\documentclass[11pt,a4paper]{article}
\usepackage{jheppub}
\usepackage{slashed}
\usepackage{amsmath}
\usepackage{amssymb}
\usepackage{xspace}
\usepackage{subcaption}
\usepackage{mathrsfs} 
\usepackage{xcolor}

\newcommand{\eq}[1]{eq.~\eqref{eq:#1}}

\renewcommand{\sec}[1]{section~\ref{sec:#1}}

\newcommand{\fig}[1]{figure~\ref{fig:#1}}

\newcommand{\nn}{\nonumber}

\newcommand{\df}{\mathrm{d}}
\newcommand{\img}{\mathrm{i}}

\newcommand{\al}{\alpha}

\newcommand{\ga}{\gamma}

\newcommand{\de}{\delta}
\newcommand{\eps}{\epsilon}

\newcommand{\w}{\omega}

\newcommand{\cJ}{{\mathcal J}}
\newcommand{\bn}{\bar{n}}

\newcommand{\MCFM}{\textsc{MCFM}\xspace}
\newcommand{\Pythia}{\textsc{Pythia}\xspace}

\allowdisplaybreaks[2]

\title{Precision boson-jet azimuthal decorrelation at hadron colliders}

\author[a]{Yang-Ting Chien,}
\author[b,c]{Rudi Rahn,}
\author[d,e]{Ding Yu Shao,}
\author[b,c]{Wouter J.~Waalewijn,}
\author[f]{Bin Wu}
\emailAdd{ytchien@gsu.edu, rudi.rahn@uva.nl, dingyu.shao@cern.ch, w.j.waalewijn@uva.nl, b.wu@cern.ch}
\affiliation[a]{Physics and Astronomy Department, Georgia State University, Atlanta, GA 30303}
\affiliation[b]{Nikhef, Theory Group,
	Science Park 105, 1098 XG, Amsterdam, The Netherlands}
\affiliation[c]{Institute for Theoretical Physics Amsterdam and Delta Institute for Theoretical Physics, University of Amsterdam, Science Park 904, 1098 XH Amsterdam, The Netherlands}
\affiliation[d]{Department of Physics and Center for Field Theory and Particle Physics, Fudan University, Shanghai, China}
\affiliation[e]{Key Laboratory of Nuclear Physics and Ion-beam Application (MOE), Fudan University, Shanghai, China}
\affiliation[f]{Instituto Galego de F\'isica de Altas Enerx\'ias IGFAE, Universidade de Santiago de Compostela, E-15782 Galicia-Spain}
\abstract{
The azimuthal angular decorrelation of a vector boson and jet is sensitive to QCD radiation, and can be used to probe the quark-gluon plasma in heavy-ion collisions. By using a recoil-free jet definition, the sensitivity to contamination from soft radiation on the measurement is reduced, and the complication of non-global logarithms is eliminated from our theoretical calculation. 
Specifically we will consider the $p_T^n$ recombination scheme, as well as the $n\to \infty$ limit, known as the winner-take-all scheme. 
 These jet definitions also significantly simplify the calculation for a track-based measurement, which is preferred due to its superior angular resolution. 
We present a detailed discussion of the factorization in Soft-Collinear Effective Theory, revealing why the transverse momentum $\vec q_T$ is more complicated than the azimuthal angle.
We show that potential glauber contributions do not spoil our factorization formalism, at least up to and including order $\alpha_s^3$. The resummation is carried out using the renormalization group, and all necessary ingredients are collected or calculated. We conclude with a detailed phenomenological study, finding an enhanced matching correction for high jet $p_T$ due to the electroweak collinear enhancement of a boson emission off di-jets. We also compare with the \Pythia event generator, finding that our observable is very robust to hadronization and the underlying event.
}
\begin{document}
\maketitle

\section{Introduction}
\label{sec:intro}

At hadron colliders, the simplest jet measurements involve a single jet, which
recoils against a color-singlet object produced alongside it. 
In the transverse plane momentum conservation enforces an (almost) back-to-back orientation of the
jet and the color singlet. 
We will focus on the case of a $Z$ boson balancing against the jet,
but our approach can also be applied to other color singlets such as a photon, $W$ or Higgs boson.
With only a single jet in the final state, this process is the minimal
extension beyond the 0-jet case at hadron colliders, and of significant
experimental interest at hadron
colliders~\cite{Chatrchyan:2013tna,CMS:2013myp,ATLAS:2016jxf,Khachatryan:2016crw,Sirunyan:2017jic,Aaboud:2017kff, CMS:2017ehl}.
In particular, in heavy-ion collisions~\cite{Sirunyan:2017jic, CMS:2017ehl} the boson provides an inert reference for the jet since it does not interact strongly with the quark-gluon plasma~\cite{Kartvelishvili:1995fr}. Therefore the process can be used to study the medium properties.

Accordingly, efforts towards precise theoretical predictions for boson-jet correlations have been made in hadron collisions: Fixed-order
results have been calculated to next-to-next-to-leading order in the QCD
corrections~\cite{Chen:2014gva,Ridder:2015dxa,Boughezal:2015dva,Boughezal:2015ded,Boughezal:2015dra,Boughezal:2015aha,Campbell:2016lzl,Gehrmann-DeRidder:2017mvr,Chen:2019zmr,Gehrmann-DeRidder:2019avi,Czakon:2020coa,Mondini:2021nck,Gauld:2021pkr},
while resummed results for the near-planar case exhibiting Sudakov logarithms have been derived to next-to-leading logarithmic (NLL) accuracy at hadron colliders for various vector (and scalar)
bosons~\cite{Banfi:2008qs,Chen:2018fqu,Sun:2018icb,Chien:2019gyf,Buonocore:2021akg,Hatta:2021jcd}. Efforts have
also been made in heavy-ion collisions to study the azimuthal decorrelation~\cite{Chen:2018fqu} as well as
the transverse momentum imbalance and other related observables (see \cite{Cao:2020wlm} for a review).

In this paper, we study the deviation from the back-to-back configuration in $V$+jet 
production, by including QCD corrections from the dominant contribution from
soft and collinear radiation. We demonstrate in detail that the azimuthal
decorrelation, in combination with a \emph{recoil-free} jet definition, has a
particularly simple theoretical description: the non-global logarithms
(NGLs)~\cite{Dasgupta:2001sh} are absent\footnote{Different recombination schemes for the azimuthal angle were first considered in ref.~\cite{Banfi:2008qs}. The H1 scheme they consider  (which corresponds to $n = 1$ in \eq{recomb}) does not have non-global logarithms of the azimuthal decorrelation. Unlike the case we study, the H1 scheme does have non-global logarithms of the jet radius parameter $R$, since the contribution of soft radiation inside the jet is suppressed compared to soft radiation outside it, due to a cancellation $\phi-\sin \phi \approx \mathcal{O}({\phi^3})$. }. This allows us to obtain resummed
predictions at next-to-next-to-leading logarithmic (NNLL) accuracy, as first reported in \cite{Chien:2020hzh}, and opens up the possibility for an N$^3$LL analysis to match to the next-to-next-to-leading order (NNLO) calculations in \cite{Chen:2014gva,Ridder:2015dxa,Boughezal:2015dva,Boughezal:2015ded,Boughezal:2015dra,Boughezal:2015aha,Campbell:2016lzl,Gehrmann-DeRidder:2017mvr,Chen:2019zmr,Gehrmann-DeRidder:2019avi,Czakon:2020coa,Mondini:2021nck,Gauld:2021pkr}, since many ingredients are already available. Also, if the jet is measured using charged particle tracks, rather than the full set of all final state particles, this can be described using the track function formalism~\cite{Chang:2013rca,Chang:2013iba} with minimal nonperturbative input (an integral over the track functions).
In contrast, the azimuthal decorrelation with the \emph{standard} jet axis suffers ---
as is typical for jet observables --- from the presence of non-global
logarithms, which has so far hindered resummation beyond the NLL accuracy. The effects of NGLs are instead power suppressed in the case we consider.
While our main focus is on the Winner-Take-All (WTA) recombination scheme~\cite{Salam:WTAUnpublished,Bertolini:2013iqa}, we also determine the effect of a different choice of recoil-free (i.e.~momentum-weighted) axis. This only leads to a minor modification in the constant term in the jet function.
We furthermore explore the other component of the $V$+jet transverse momentum
imbalance, which we refer to as the radial decorrelation. Because the factorization structure is much more
complicated, even for a recoil-free axis, we do not further pursue it here, but
include it as an illustrative foil to the azimuthal decorrelation.

The program outlined in the previous paragraph is conducted based on a
factorization formula derived using Soft-Collinear Effective Theory
(SCET)~\cite{Bauer:2000yr,Bauer:2001ct,Bauer:2001yt,Bauer:2002nz,Beneke:2002ph}. 
We calculate the necessary ingredient functions, where not yet available, and
explore potentially factorization violating effects from Glauber gluons,
identifying the order at which they may first contribute.

Our main result consists of a phenomenological study, consisting of
numerical results for the analytic resummation at NNLL accuracy. We include a discussion of the resummation,
perturbative uncertainties, nonperturbative effects, and the matching to fixed order. 
We furthermore study features of this observable using \Pythia, and compare to our NNLL result.
As promised by the recoil-free axis, we find that the observable is insensitive 
to soft effects (hadronization, underlying event), a rather small sensitivity to the jet radius, and  
negligible differences when measuring the decorrelation on tracks. 
The uncertainty band of our resummed predictions are reduced when going from NLL to NNLL. Our resummed results are consistent with \Pythia, except for the matching corrections from the NLO cross section (as we didn’t match \Pythia to NLO).
These matching corrections are substantial for high jet $p_T$, and arise from the boson being emitted from a leading order dijet configuration. Though this is formally power suppressed in the back-to-back limit, it is enhanced by an electroweak logarithm.
While we focus on the WTA axis, we also explore $p_T^n$-weighted recombination schemes, and these conclusions also hold there (if $n>1$).

The paper is structured as follows: We introduce the kinematic setup and discuss
different observables measuring the transverse momentum decorrelation of the
$V+$jet pair in section~\ref{sec:Zjet}. Section~\ref{sec:fact} establishes the
factorization formula, gathers the available ingredient functions, and includes
a brief discussion of factorization violation induced by a Glauber mode.
Section~\ref{sec:J} contains various calculations relating to the jet function:
The linearly polarized jet function appearing in our factorization formula, as
well as the changes to the jet function induced by the use of a different recoil-free jet axis or track-based
measurements, respectively. We establish our resummation
strategy in section~\ref{sec:resum} (with certain ingredients relegated to
appendix~\ref{app:anomdim}) which is then carried out to derive the results in
section~\ref{sec:result}. We conclude in section~\ref{sec:conc}.

\section{Boson-jet correlation}
\label{sec:Zjet}

\subsection{Geometry of the collision}
\label{sec:geometry}

We begin by describing the geometry of the collision and defining the observable for
which we perform the resummation. It will be instructive to
contrast our target observable, the \emph{azimuthal} decorrelation, with the
closely related \emph{radial} decorrelation. (These correspond to the two
components of the difference in transverse momentum between vector boson and
jet.) This comparison will demonstrate in detail where simplifications due to
the choice of a recoil-free recombination scheme arise and how non-global
effects are suppressed: While using a recoil-free axis removes non-global logarithms
for the azimuthal decorrelation, the radial variety still suffers from NGLs and needs
to include effects related to the technical definition of this axis. For concreteness we 
discuss the case of the Winner-Take-All (WTA) axis.

\smallskip

\begin{figure}
\centering
\includegraphics[height=0.35\textheight]{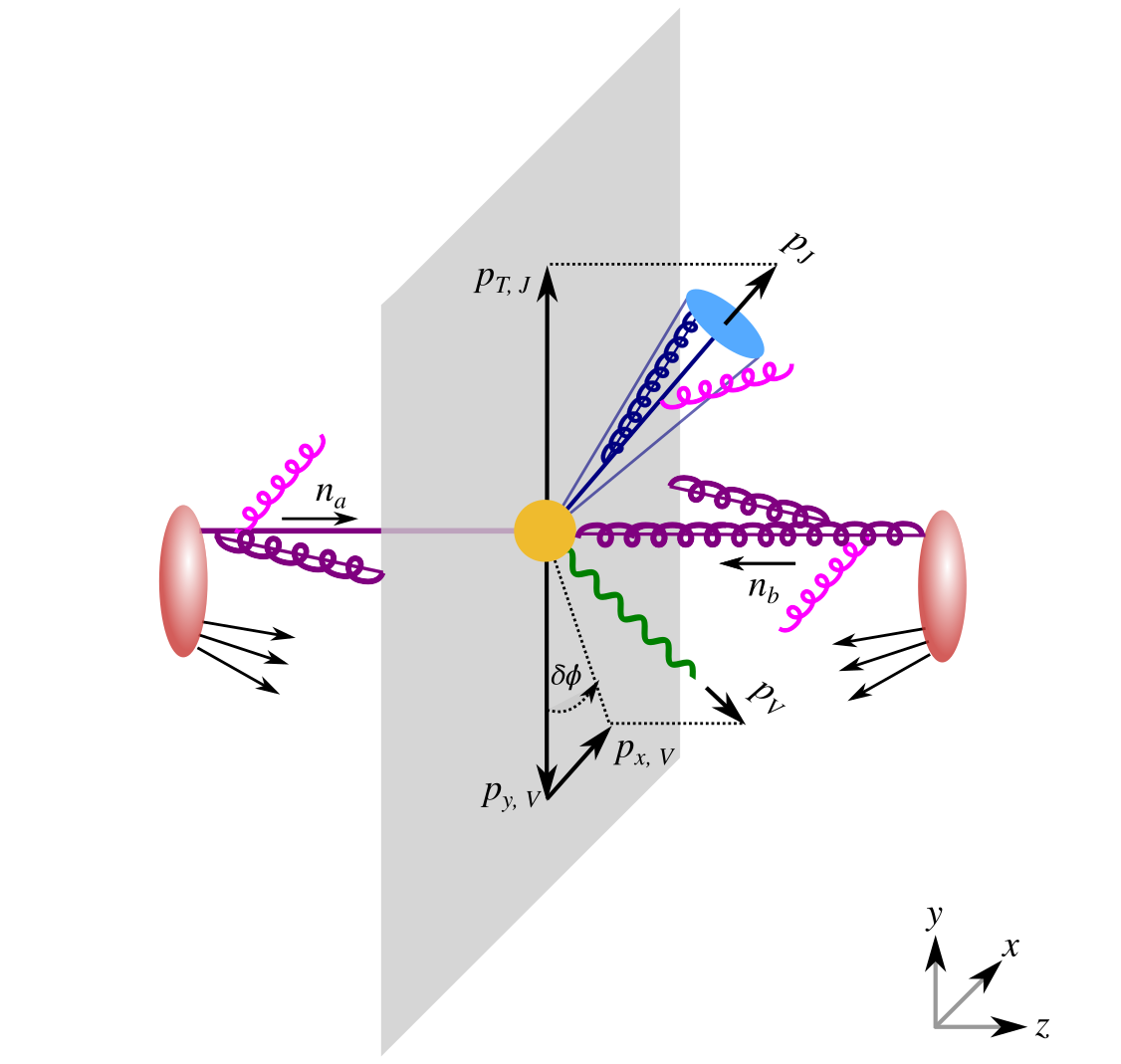}
\hfill
\includegraphics[height=0.35\textheight]{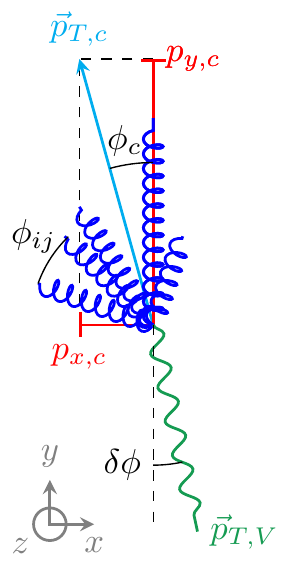}
\caption{\textbf{Left:} The azimuthal angle between the vector boson (green) and jet axis (blue) 
is related to the  momentum of the vector boson $p_{x,V}$ transverse to the 
colliding protons (red) and jet. Collinear initial  (purple) and final-state 
(blue) radiation and soft radiation (magenta) is also shown. 
\textbf{Right:} Schematic of the transverse plane: the angle $\phi_c$
between WTA axis (along $y$ axis) and collective collinear momentum, the angle $\phi_{ij}$ between two
generic collinear emissions $i$ and $j$, and the angular decorrelation
observable $\delta \phi$. These quantities are of the same parametric size.}
\label{fig:coordinate}
\end{figure}

The geometric setup is illustrated in the left panel of \fig{coordinate}, where we
 choose to align the $y$-axis with the reconstructed \emph{jet axis}
(its projection onto the transverse plane, to be precise). Our starting point is
momentum conservation in the transverse plane, which reads
\begin{equation} \label{eq:mom_cons}
\vec{p}_{T,a}+\vec{p}_{T,b}+\vec{p}_{T,S}+\vec{p}_{T,c}+\vec{p}_{T,V} = 0,
\end{equation}
and expresses that the vector boson (with transverse momentum $\vec{p}_{T,V}$)
recoils against emissions off the beams ($\vec{p}_{T,a},\vec{p}_{T,b}$), soft radiation ($\vec{p}_{T,S}$), and the \emph{total} collinear radiation in the jet ($\vec{p}_{T,c}$). Note that $\vec{p}_{T,c}$ is
non-trivial, i.e.
possesses both $x$ and $y$-components, as we chose to align our coordinate axes
with the reconstructed jet axis $\vec{p}_{T,J}$, which in the WTA case does
\emph{not} follow the collective momentum of all collinear emissions in the jet: As detailed in ref.~\cite{Bertolini:2013iqa}, the WTA axis always follows the orientation of one input particle (not necessarily the most energetic one, which would not be collinear safe), 
whereas the total momentum in general is not aligned with any individual particle
direction. 

To illustrate this point, we revisit the WTA recombination
scheme and the associated axis in detail: in sequential jet clustering
algorithms, particles are pairwise recombined and assigned a joint 4-momentum.
For the WTA recombination scheme, this joint momentum is taken massless, and
points along the direction of the ``harder" of the two particles in the pair. The criterion to determine which of two particles is the harder is typically the hierarchy in either the particle energy (``WTA-$E$-scheme'') or the transverse momentum (``WTA-$p_T$-scheme'')\footnote{See e.g.~ref.~\cite{Larkoski:2014uqa}
vs.~\cite{Bertolini:2013iqa}. Note that this scheme choice also affects the order in which particles are clustered, which in the former is determined by their energy, while in the latter it follows from their
transverse momentum.}. The latter choice is of course more suitable for
hadron colliders and shall therefore be used here.
The magnitude of the newly combined pseudo-particle is the sum of the energies
or the scalar sum of transverse momenta --- the ``winner'' absorbs the loser, so to speak. The WTA axis is then
simply the final pseudo-particle after clustering. It is in principle possible
that many soft emissions are clustered into one highly energetic pseudo particle that
emerges as the winner, but this is extremely unlikely because they would have to
be clustered together first to stand a chance against collinear emissions,
contrary to the expectation that they are spread out over the jet.
Thus the WTA axis will be aligned with one of the collinear emissions, rendering its direction free of soft recoil.

The WTA axis is a particularly powerful tool when combined with SCET, because it
can exploit the parametric hierarchy of the modes in SCET to simplify the calculation: As stated,
the magnitude of the jet axis will be a sum of transverse momenta
\begin{equation}
|\vec{p}_{T,J}| = \sum_{i} |\vec{p}_{T,i}|,
\end{equation}
where the sum runs over all emissions in the jet. One consequence of the WTA
recombination is then that soft emissions inside the jet can affect the
magnitude as participants in this sum, but never the orientation, as soft
emissions always lose against collinear emissions when determining with which
emission the axis should be aligned. The dependence of the axis on soft
emissions can be expanded away and represents a subleading effect in the
effective theory --- and in some cases even the dependence in the magnitude can
be expanded away, with more details laid out in the following two sections.

\smallskip

The azimuthal angle is now directly related to the (dimensionful) offset between
vector boson and reconstructed jet momentum, defined as
\begin{align}
\vec q_T &\equiv \vec p_{T,V}+ \vec p_{T,J} \nonumber \\&= \vec p_{T,J} -
\vec{p}_{T,c} -\vec{p}_{T,a}-\vec{p}_{T,b}-\vec{p}_{T,S}.
\end{align}
In the absence of QCD radiation, $\vec q_T$ vanishes, and the limit of small $\vec q_T$ is of
interest for the resummation.
The angular decorrelation can be e.g. written as $\delta
\phi = \arcsin q_x/|p_{T,V}| \approx q_x/|p_{T,V}|$, and is essentially a dimensionless version of
the dimensionful $\vec q_T$.

\subsection{Azimuthal and radial decorrelation}

Besides the \emph{azimuthal} decorrelation,
which represents the tangential offset $q_x$ as shown in \fig{coordinate}, we
can define a second quantity of interest here, the
\emph{radial} decorrelation, using $q_y$. Using \eq{mom_cons} these can be
written as
\begin{align} \label{eq:raddecorr}
q_x &= p_{x,V} + p_{x,J} &    q_y &= p_{y,V} + p_{y,J} 
 \\&= p_{x,V} &                              &=p_{y,J}-p_{y,a}-p_{y,b}-p_{y,S}-p_{y,c}
\nn \\&= -p_{x,a}-p_{x,b}-p_{x,S}-p_{x,c} 
\nn\end{align}
where we used that $p_{x,J}=0$ due to the alignment of the $y$-axis with the
the WTA axis. We expect the
contributions from the beams to be isotropic in the
transverse plane (and their $x$- or $y$-dependence therefore to be similar) and
the soft function to capture the $y$-dependence 
via its dependence on the geometry of the beam-jet system, so to proceed we
need to understand the relation between $ \vec{p}_{T,J}$ and $\vec{p}_{T,c}$, i.e.~the reconstructed WTA axis and the collective collinear
momentum of the jet, and their components\footnote{This is also true for $q_x$, which depends implicitly on $\vec p_{T,J}$, through our choice of coordinate system.}.
Right from the start it is clear that the radial decorrelation is more complex,
as it involves a large cancellation between the boson and jet axis magnitude.

For $\vec p_{T,c}$ and its decomposition into $x$- and $y$-components, this
yields the situation illustrated in the right panel in \fig{coordinate}. Thus
\begin{equation}
\label{eq:countazim}
p_{x,c} = |\vec{p}_{T,c}| \cdot \sin \phi_c \approx |\vec{p}_{T,c}|
\cdot\phi_c,
\end{equation}
where $\vec{p}_{T,c}$ (which is of the order of the center-of-mass energy $Q$) is the vector sum of all
collinear emissions off the parton initiating the jet, and $\phi_c$ measures the angular separation
between the WTA axis and the collective
collinear momentum flow.

To obtain $\delta \phi$ or $q_x$, we still need to include
the collinear and soft contributions in \eq{raddecorr}. 
As the $x$-direction is perpendicular to the plane containing both the jet and
the beams, we expect the appropriate SCET modes to scale --- with notation
$p^\mu=(n_b\cdot p,n_a\cdot p,p_\perp)$ --- as 
\begin{alignat}{3} \label{eq:pc}
\begin{alignedat}{3}
p_a &\sim Q(1,\phi_c^2,\phi_c),\qquad\qquad
&p_b &\sim Q(\phi_c^2,1,\phi_c),\\
p_S &\sim Q(\phi_c,\phi_c,\phi_c),
&p_c &\sim Q(1,\phi_c^2,\phi_c)_J,
\end{alignedat}
\end{alignat}
where the scaling for the jet --- with subscript $J$ ---  is to be understood
as scaling of lightcone components based on the jet direction, i.e.~$p^\mu=(\bar{n}_J\cdot p,n_J\cdot p,p_\perp)_J$. We thus find a SCET$_{\rm II}$
situation, as expected for a \emph{de
facto} transverse momentum measurement.

Finally, note that there is no special dependence on soft in-jet radiation:
It is always expanded away when the jet axis direction is determined, but it can also be
expanded away in the magnitude in this case, as the azimuthal decorrelation only
depends on the direction of the jet axis, which determines the coordinate
system we use to define it.

\subsection{Properties of the radial decorrelation}

We will now contrast this surprisingly simple observable with the radial
decorrelation $q_y$, where the details of the recombination become
explicitly relevant. We emphasize that we do not wish to actually perform
the resummation for this observable, but merely include this discussion to
highlight why the azimuthal decorrelation is so amenable to precision calculations.

A key insight is that while $p_{y,J}$ in eq.~\eqref{eq:raddecorr} includes
a sum over collinear (and soft) momenta, it makes use of a scalar sum and aligns it
with a collinear emission's direction, while $p_{y,c}$ is a component of a
vector sum. Iterating the law of cosines/binomial theorem over all relevant
momenta, the vector and scalar sum of collinear momenta can be
related, to
\begin{align}
\begin{aligned}
|\vec{p}_{T,c}|=\biggl|\sum_{i\in c} \vec{p}_{T,i} \biggr|  &= \sum_{i\in c}
|\vec p_{T,i}| \cdot \sqrt{1-
\frac{2\sum_{i<j}
|\vec p_{T,i}||\vec p_{T,j}|(1-\cos\phi_{ij})}{(\sum_{i\in c}
|\vec p_{T,i}|)^2}}.
\end{aligned}
\end{align}
Here $\phi_{ij}$ are the pairwise opening angles between emissions $i$ and $j$,
which satisfy $\phi_{ij}\ll1$ because $i$ and $j$ are both collinear. Thus we
can expand the above expression to obtain
\begin{align}
\begin{aligned}
\biggl|\sum_{i\in c} \vec p_{T,i} \biggr|  &\approx \sum_{i\in c}
|\vec p_{T,i}| -
\frac{\sum_{i<j}|\vec p_{T,i}||\vec
p_{T,j}|\phi^2_{ij}}{2\sum_{i\in c} |\vec p_{T,i}|}.
\end{aligned}
\end{align}

We can now derive the expression for the radial decorrelation: As $p_{x,J}=0$
due to our choice of conventions, $p_{y,J}$ is the scalar sum
of all momenta inside the jet, which for
$R\gg\delta\phi$ includes all collinear radiation, as well as the soft
emissions that end up in the jet: $p_{y,J} = |\vec{p}_{T,J}| =
\sum_{i\in(s\in J)} |\vec{p}_{T,i}| + \sum_{i\in c} |\vec{p}_{T,i}|$. This leads to
\begin{align}
\begin{aligned}
q_y &=
p_{y,J}-|\vec{p}_{T,c}|\cos\phi_c-p_{y,S}
-p_{y,a}-p_{y,b}
\\
&= \underbrace{-p_{y,a}-p_{y,b}}_{\text{beam functions}} +
\underbrace{\sum_{i\in(s\in J)} |p_{T,i}|-p_{y,S}}_{\text{soft function}} +
\underbrace{\sum_{i\in c} |\vec{p}_{T,i}|\frac{\phi_c^2}{2} +
\frac{\sum_{(i<j)\in c}|\vec p_{T,i}||\vec p_{T,j}|\phi^2_{ij}}{2\sum_{i\in
c}|\vec p_{T,i}|}}_{\text{jet function}},
\label{eq:sumqy}
\end{aligned}
\end{align}
where the leading scalar sum of collinear momenta cancelled, and the
corresponding functions in the factorization are indicated. The expected power
counting of the modes involved is also non-trivial:
The jet power counting is determined by its $\perp$-component, which must scale
as $\mathcal{O}(Q \phi_c)$, as the angle between WTA axis (i.e.~one of the
emissions) and the collective collinear momentum has to be parametrically
similar to the opening angle of collinear splittings.
The transverse components $p_{y,a}$ and $p_{y,b}$ for the beams must
scale as $\mathcal{O}(Q \phi_c^2)$ to contribute at equal footing with the jet
contributions in~\eqref{eq:sumqy}. The soft contributions are isotropic and a
similar argument applies. It thus follows that
\begin{alignat}{3}
\begin{alignedat}{3}
p_a &\sim Q(1,\phi_c^4,\phi_c^2),\qquad\qquad
&p_b &\sim Q(\phi_c^4,1,\phi_c^2),\\
p_S &\sim Q(\phi_c^2,\phi_c^2,\phi_c^2),
&p_J &\sim Q(1,\phi_c^2,\phi_c)_J,
\end{alignedat}
\end{alignat}
which is an interesting combination of SCET$_{\rm I}$ and SCET$_{\rm II}$
characteristics.
For the beams this is a transverse momentum measurement, and so they share the
virtuality of the soft emissions, as typical in SCET$_{\rm II}$. The jet, however,
measures essentially an invariant-mass-like observable, a SCET$_{\rm I}$ situation.

We also note that the differing treatment of soft radiation inside and
outside the jet means this observable exhibits non-global
logarithms. This is a persistent feature: while it is e.g.~possible
to remove the term depending on pairs of collinear emissions from \eq{sumqy} by
determining the magnitude of $\vec p_{T,J}$ using a vector-sum-based recombination scheme, 
the inclusion of soft in-jet emissions will also be present in such a case.

From the above discussion we can now understand why the azimuthal
decorrelation is so simple, by comparison: The non-trivial effects present in its radial sibling, including
the appearance of non-global logarithms, are $\mathcal{O}(\phi_c^2)$
effects, and are therefore power suppressed compared to the dominant
contributions to azimuthal decorrelation in \eqref{eq:countazim}, which are
$\mathcal{O}(\phi_c)$.

\subsection{Extensions}

We conclude this section, by discussing other recoil-free axes, double differential measurements in the azimuthal and radial decorrelation and tracks for the radial decorrelation:

First, we note that the choice of a recoil-free axis other than the WTA-axis (in
essence the \texttt{FastJet}~\cite{Cacciari:2011ma} recombination scheme
of equation (4) with $w_i=p_{Ti}^n$ and
$1<n<\infty$, see also \eq{recomb}) constitutes a small change for the azimuthal recombination: as
long as it is recoil free, only the jet function will be changed, and we will
calculate the corresponding jet function in sec.~\ref{subsec:recoscheme}. (For the radial decorrelation this may be a more intricate affair, as subleading effects are elevated by virtue of a large cancellation.)

Second, we point out that measuring any quantity in the $\vec q$-plane other
than the elementary $q_x$ and $q_y$ is a \emph{de facto} double-differential
measurement, and requires the specification of the relative scaling of the $q_x$
and $q_y$ components, to cleanly establish which contributions to the observable
from the different regions (or ingredient functions) are subleading and can be
neglected. This ultimately is a consequence of the different scaling of $q_x$
and $q_y$ themselves, in terms of power counting. In particular, one could consider $q_x \sim q_y$ or $q_x^2 \sim Q q_y$, which correspond to making the factorization of the azimuthal or radial decorrelation differential in both $q_x$ and $q_y$,  or something in between (see e.g.~ref.~\cite{Procura:2014cba} for a factorization description of double differential measurements).

Finally, we point out that using track-based measurements is a possibility for the
azimuthal decorrelation, and will be discussed in~\ref{subsec:tracks}, but is
firmly ruled out for the radial decorrelation. The reason is simply that the vector boson will always
be fully reconstructed with all of its transverse momentum, but the
jet axis is only established based on the particles included in the jet recombination. 
If all collinear particles are clustered into the jet, the jet axis magnitude is roughly equal to the
total collinear transverse momentum, and the cancellation against the
vector boson transverse momentum establishes the radial decorrelation as a small
quantity, which vanishes in the singular limit. A resummation program can then be set up. If
tracks are used, however, only the subset of charged particles in the final
state contributes to the jet axis. The difference to the vector boson $p_T$ is
therefore a large quantity, finite even in the singular limit, and given by the
transverse momentum of the neutral particles in the jet. The azimuthal
decorrelation is unaffected, as it does not involve a large cancellation.

\section{Factorization}
\label{sec:fact}

In the previous section we performed an analysis of the contributions to the
transverse momentum $\vec q$ of the boson+jet system from soft and collinear
emissions, indicating that these contributions are dominant at leading power. In
the limit where $q_x$ is small, the infrared structure of QCD results in large
logarithms of $\ln (Q/q_x)$ caused by soft and collinear emissions from initial
state beam partons and final state jet partons (and similarly for $q_y$). To
obtain reliable predictions in this limit, we will derive a factorization
formula in sec.~\ref{subsec:SCET}, that enables us to resum these large
logarithms, which will be discussed in sec.~\ref{sec:resum}. Resummation to
next-to-next-to-leading logarithmic accuracy requires the ingredients of the
factorization theorem at one-loop order, given in sec.~\ref{subsec:oneloop} or
calculated in~sec. \ref{sec:J}\footnote{The calculation of the
one-loop gluon jet function is detailed in sec.~\ref{sec:gluon_jet}, showing for
the first time how to obtain the linearly-polarized contribution. Generalizing
the WTA recombination scheme to the recoil-free $p_T^n$ scheme (with $n>1$) only
results in a modification of the jet function, which is calculated in
sec.~\ref{subsec:recoscheme}. Switching to track-based measurements also only
affects the jet function constant, this calculation is outlined
in~\ref{subsec:tracks}.}, as well as their anomalous dimensions at two-loop
order (and the three-loop cusp anomalous dimension), collected in
sec.~\ref{subsec:RGE}. Finally, in sec.~\ref{subsec:glauber} we discuss the
contribution from Glauber gluons, showing that they may first appear at
$\mathcal{O}(\alpha_s^4)$.

\subsection{Factorization formula for azimuthal decorrelation}
\label{subsec:SCET}

As the sum of momenta in \eqref{eq:raddecorr} shows, the component $q_x$ of the
vector boson perpendicular to the jet axis receives contributions from a linear
sum of four terms. Three of these terms represent the characteristic emissions
from one of the hadronic directions, and by choosing the WTA axis, the other
term is the momentum component of \emph{all} soft radiation in the collision.
We will consider the case where the azimuthal decorrelation $\delta \phi=\arcsin
(q_x/p_{T,V})\approx
q_x/p_{T,V} \ll R$, so that contributions from
out-of-jet emissions are suppressed by powers of $\delta \phi/R$. This implies
that the observable $\de \phi$ is not sensitive to the non-global correlation between radiation inside
and outside the jet.
Starting with these assumptions, we can factorize the cross section, which
allows us to calculate the $\de \phi$ distribution at high logarithmic accuracy.
The corrections to this factorization are suppressed by powers of $\de \phi$,
and can be included for the region where $\delta \phi$ is not small by matching
to the fixed-order prediction for the cross section, as discussed in
sec.~\ref{subsec:fixed}.

A factorization formula for the boson-jet transverse momentum imbalance $q_T$
was derived by some of us in ref.~\cite{Chien:2019gyf}, for the case where jets
are defined using the anti-$k_t$ algorithm with the standard recombination
scheme. In that case, the jet axis is along the direction of the total jet
momentum, and \emph{is} sensitive to recoil from soft radiation enclosed within
the jet boundary. As a consequence, the cross section receives nonglobal
contributions that involve soft radiation inside and outside the jet, preventing
a \emph{simple} factorization of collinear and soft contributions (though there
has been substantial progress in resumming nonglobal logarithms, see
e.g.~\cite{Dasgupta:2001sh, Banfi:2002hw, Weigert:2003mm, Avsar:2009yb,
Hatta:2013iba, Khelifa-Kerfa:2015mma, Caron-Huot:2015bja, Becher:2015hka,
Larkoski:2015zka, AngelesMartinez:2018cfz, Banfi:2021owj}).
Here we instead use the WTA recombination scheme for jet clustering, which is
only sensitive to the total amount soft emissions. The WTA axis is sensitive to
the distribution of collinear radiation in the jet, but the contributions from
collinear emissions off the beams are the same as for the standard jet axis
case.

We start from a general differential momentum distribution of all the particles
in an event and project it onto the observable $q_x$,
\begin{equation}
\frac{\df\sigma}{\df q_x \,\df p_{T,V}\,\df y_V \,\df \eta_J}=\int
\df \Phi_N \,\frac{\df\sigma}{\df \Phi_N\,\df p_{T,V}\,\df y_V}\,\delta(q_x- \hat q_x)
\,\delta(\eta_J- \hat \eta_J)\;,
\end{equation}
where $\df \Phi_N = \prod^N \df p_i \, \theta(p_i^0) \, \de(p_i^2)$ is the
complete phase space of final state particles except the boson, whose transverse
momentum and rapidity are denoted by $p_{T,V}$ and $y_V$, respectively\footnote{As the jet and vector boson recoil in the transverse plane, we could
choose to keep $p_{T,J}$ differential, instead of $p_{T,V}$.
However, $p_{T,J}$ is the reconstructed jet momentum, and as such loses its
sensitivity to the hard scattering kinematics if we measure the
angular decorrelation using tracks: the neutral particles' contribution will
then be missing, and $p_{T,J}$ will no longer be parametrically similar to
$p_{T,V}$. We therefore pick the more robust $p_{T,V}$.}.
Similarly, the transverse momentum and pseudo-rapidity of the jet are labelled $p_{T,J}$ and $\eta_J$. The delta function restricts the value of the observable $q_x$, through the measurement function $\hat q_x$, which is a function of final state particle momenta $\{p_i\}$. The leading power region of soft and collinear emissions, denoted as $X_{\rm IR}$, thus constitutes the relevant degrees of freedom in the SCET of the observable $q_x$,
\begin{equation}
	\df \Phi_N \rightarrow \df X_{\rm IR}= \df X_a\,\df X_b\,\df
	X_c\,\df X_S\;,
\end{equation}
where $\df X_a$, $\df X_b$, $\df X_c$ and $\df X_S$ correspond to
the differential phase space of collinear emissions of beam $a$, beam $b$ and
jet $c$, as well as soft emissions from these collinear directions.

From (\ref{eq:raddecorr}) we see that the measurement $\hat q_x$
simplifies, as
\begin{equation}
	q_x = - p_{x,a} - p_{x,b} - p_{x,S} - p_{x,c}\;, 
\end{equation}
with the contribution from each factorized IR sector given by\footnote{Remember
that $p_{x,c}$ sensitively depends on the WTA scheme, since the $y$-axis is
along the jet direction.}
\begin{equation}
	p_{x,a} = p_{x,X_a},\quad p_{x,b} = p_{x,X_b},\quad p_{x,S} = p_{x,X_S},\quad p_{x,c} =
	p_{x,X_c}\;.
\end{equation}
Therefore, at leading power, the differential cross section can be organized as
follows,
\begin{align}
\frac{\df\sigma}{\df q_x \,\df p_{T,V}\,\df y_V \,\df \eta_J}
&= \int \df p_{x,a} \,\df p_{x,b} \,\df p_{x,c}\,\df p_{x,S}\, 
\delta\bigl(q_x+p_{x,a}+p_{x,b}+p_{x,c}+p_{x,S}\bigr)
\nn \\ & \quad \times
\frac{\df\sigma}{\df p_{x,a} \,\df p_{x,b} \,\df p_{x,c} \,\df p_{x,S} \,\df
p_{T,V}\,\df y_V \,\df \eta_J} \;.
\end{align}

In the next section we will discuss how the soft-collinear decoupling in the
SCET Lagrangian allows us to factorize the multi-differential cross section
$\df\sigma/(\prod_i \df p_{x,i} \,\df p_{T,V}\,\df y_V\,\df \eta_J)$ to obtain
\begin{align}
\label{eq:FactThm}
&\frac{\df\sigma}{\df q_x \,\df p_{T,V}\,\df y_V \,\df \eta_J}
=\int \df p_{x,a} \,\df p_{x,b} \,\df p_{x,c} \,\df p_{x,S}\, 
\delta\bigl(q_x+p_{x,a}+p_{x,b}+p_{x,c}+p_{x,S}\bigr)
 \\
 &\quad \quad\quad\times
\sum_{ijk}\mathcal{H}_{ij\rightarrow Vk}(p_{T,V},y_V-\eta_J) \tilde{B}_i(x_a,
p_{x,a})\tilde{B}_j(x_b,p_{x,b}) \tilde{\mathscr{J}}_k(p_{x,c}) \tilde{S}_{ijk}(p_{x,S},\eta_J)\;.
\nn \end{align}
Here the indices $i,j,k$ label the partonic channels of the hard scattering
processes, encoded in the hard function $\mathcal{H}$, producing a
high-transverse momentum boson $V$ recoiling against the jet. The contribution
to $q_x$ of collinear initial and final-state radiation is described by the beam
functions $\tilde{B}$ and jet function $\tilde{\mathscr{J}}$, and the soft
function $\tilde S$ accounts for the contribution from soft radiation.
(The tilde signifies that these functions are defined in momentum space, not impact parameter space.)
The treatment of Lorentz and color indices in deriving \eq{FactThm} will be
discussed in the next section, and leads to a linearly-polarized contribution
from gluon beams and jets, for which there is a corresponding change to the hard
function $\mathcal{H}$. The Bjorken variables $x_a$ and $x_b$ are determined by the boson and jet kinematics,
\begin{align}
x_a = \frac{1}{\sqrt{s}} \big( e^{\eta_J} p_{T,V} + e^{y_V} \sqrt{p_{T,V}^2 +
m_V^2} \big),~~~x_b = \frac{1}{\sqrt{s}} \big( e^{-\eta_J}  p_{T,V} + e^{-y_V} \sqrt{p_{T,V}^2 + m_V^2} \big).
\end{align}
We can eliminate the convolution in the above factorization formula by switching
to the impact parameter variable $b_x$, which is the Fourier conjugate of $q_x$,
\begin{align}
\label{eq:FactThmFourier}
&\frac{\df\sigma}{\df q_x \,\df p_{T,V}\,\df y_V \,\df \eta_J} =\nn \\
&\int \frac{\df b_x}{2\pi} e^{\img b_x q_x}
\sum_{ijk}\mathcal{H}_{ij\rightarrow Vk}(p_{T,V},y_V-\eta_J) B_i(x_a,b_{x}) B_j(x_b,b_{x}) \mathscr{J}_k(b_{x}) S_{ijk}(b_{x},\eta_J)
\end{align}
The factorization formula in impact
parameter space thus has a product form. In the next section we will present
some details of the derivation of the above factorization formula, and discuss
each factorized component.

\subsection{Derivation of factorization formula}
\label{subsec:fact}

We will derive the factorization theorem using Soft-Collinear Effective Theory
(SCET)~\cite{Bauer:2000yr,Bauer:2001ct,Bauer:2001yt,Bauer:2002nz,Beneke:2002ph}.
For an introduction to SCET, see  e.g.~refs.~\cite{Stewart:notes,Becher:2014oda}. 
The leading operators describing the hard scattering for $Z$+jet production
are~\cite{Becher:2009th}
\begin{align} \label{eq:ops}
	&{\cal O}^{S\nu}_{q\bar q}(b,t_a,t_b,t_c)=\nn
	\\&\quad\quad\bar\chi_{n_b}(b+t_b\bar n_b)S_{n_b}^\dagger(b_x)
	S_{n_c}(b_x){\cal B}_{{n_c}\perp}^{\nu,A} (b+t_c\bar
	n_c) t^A S_{n_c}^\dagger(b_x) S_{n_a}(b_x) \chi_{n_a}(b+t_a\bar n_a)\nn
	\\
	&{\cal O}^{T\nu}_{q\bar q}(b,t_a,t_b,t_c)=\nn
	\\&\quad\quad\bar\chi_{n_b}(b+t_b\bar	n_b)S_{n_b}^\dagger(b_x) 
	S_{n_c}(b_x)i\sigma^\nu_{\rho}
	{\cal B}_{{n_c}\perp}^{\rho,A}(b+t_c\bar
	n_c)t^A S_{n_c}^\dagger(b_x) S_{n_a}(b_x)  \chi_{n_a}(b+t_a\bar n_a) \,.
\end{align}
The superscripts $S$ and $T$ on the operators label different quark spin
structures. The four-vector $b$ denotes the position of the operator and the $t_i$ will be integrated over in the matching in \eq{match} below. The lightcone directions $n_a$, $n_b$ and $n_c$ are along the two
beams and jet directions,
and $\chi_{n_i} = W_{n_i}^\dagger \xi_{n_i}$ ($\bar \chi_{n_i} = \bar \xi_{n_i} W_{n_i}$) is the collinear
quark field and $\mathcal{B}_{n_i\perp}^\mu
=\frac{1}{g}W_{n_i}^\dagger \img D_{n_i\perp}^\mu W_{n_i}$ the collinear gluon field, which include collinear Wilson
lines $W_{n_i}$ for gauge invariance. They appear in combination with soft
Wilson lines $S_{n_i}$ in the pattern outlined by the square brackets. Multipole
expansions simplify the coordinate dependence of the soft Wilson lines, and are
implied for the $b$-dependence of the collinear ones, as well. We have here labelled the fields
according to the $q\bar q \rightarrow Zg$ channel, and the other channels can be
obtained by a permutation of the lightcone directions. To keep the notation
compact in the following, we will denote these SCET operators as ${\cal
O}^{\nu}_{j} $, with the index $j$ labeling the partonic channels \emph{and}
quark spin structures, and denote the collinear fields collectively as
$[\phi_{n_i}]^{\alpha_i}_{a_i}$ with Lorentz and color indices $\alpha_i$ and
$a_i$. The contribution from operators with additional collinear fields is power
suppressed by ${\cal O}(\de \phi)$.

The SCET operators are matched to the full electroweak current which produces the boson,
\begin{equation} \label{eq:match}
	J_{\rm EW}^\nu(b^\mu)\rightarrow\sum_j\int \df t_a\,\df t_b\,\df t_c\,
	C_j(t_a,t_b,t_c){\cal O}_j^\nu(b^\mu, t_a,t_b,t_c)\;,
\end{equation}
where $C_j$'s are the Wilson coefficients. In the case of direct photon production, the cross section differential in the
infrared degrees of freedom has the following form after performing the
integrals over the $t_i$: 
\begin{align}
	\frac{\df\sigma}{\df X_{\rm IR}}&=\frac{2\pi\alpha_{em} e_q^2}{ E_{\rm
	CM}}\frac{\df^3 p_V}{(2\pi)^3 2E_V} \sum_{j,k} \epsilon^{p*}_\mu \epsilon^q_\nu \,\langle P_a P_b | {\cal O}^{\mu\dagger}_j
	(b) | X_{\rm IR}\rangle \langle X_{\rm IR} | {\cal O}^{\nu}_k (0) |  P_a P_b
	\rangle\nonumber\\&\quad\times\tilde
	C^*_j(\bar{n}_a\cdot P_a,\bar{n}_b\cdot P_b,\bar{n}_c\cdot P_c) \tilde
	C_k(\bar{n}_a\cdot P_a,\bar{n}_b\cdot P_b,\bar{n}_c\cdot P_c)\;,
\end{align}
where $\epsilon^p_\mu$ is the polarization vector of the boson with the index
$p$ labeling the polarization states, and $|P_a\rangle$ and $|P_b\rangle$ are
the initial hadron states with momenta $P_a$ and $P_b$. 
Here $\al_{em}$ is the electromagnetic coupling constant, $e_q$ is the electric charge of the quark (in terms of elementary charge units), and for $Z$ production one replaces $e_q^2$ by
\begin{align}\label{eq:emdef}
e _ { q } ^ { 2 } \rightarrow \frac { \left( 1 - 2 \left| e _ { q } \right| \sin ^ { 2 } \theta _ { W } \right) ^ { 2 } + 4 e _ { q } ^ { 2 } \sin ^ { 4 } \theta _ { W } } { 8 \sin ^ { 2 } \theta _ { W } \cos ^ { 2 } \theta _ { W } }.
\end{align}

With the capability of tagging the boson polarization, one will be able to
assign a specific polarization tensor $\epsilon^{p*}_\mu \epsilon^{q}_\nu$. For
observables inclusive in boson polarization, the polarization sum and the Ward
identity give
\begin{align}
	\frac{\df\sigma}{\df X_{\rm IR}\,\df p_{T,V}\,\df y_V}&=-\frac{\alpha_{em}e_q^2~
	p_{T,V}}{2(2\pi) E_{\rm CM}} \sum_{j,k} \langle P_a P_b | {\cal O}^{\mu\dagger}_j
	(b) | X_{\rm IR}\rangle \langle X_{\rm IR} | {\cal O}_{\mu k} (0) |  P_a P_b
	\rangle\nonumber\\&\quad\times\tilde
	C^*_j(\bar{n}_a\cdot P_a,\bar{n}_b\cdot P_b,\bar{n}_c\cdot P_c) \tilde
	C_k(\bar{n}_a\cdot P_a,\bar{n}_b\cdot P_b,\bar{n}_c\cdot P_c)\;.
\end{align}
Together with the factorization of infrared degrees of freedom, $|X_{\rm
IR}\rangle = |X_{a}\rangle |X_{b}\rangle |X_{c}\rangle |X_{S}\rangle$, the above matrix element factorizes
into a product of three collinear matrix elements and one soft matrix element.

For particular quark spins or gluon polarizations either in the initial or final
state\footnote{Polarized beams can induce preferred quark spins and gluon
polarizations. Even for the case of unpolarized beams considered here, there is
a contribution from linearly-polarized
gluons~\cite{Mulders:2000sh,Nadolsky:2007ba,Mantry:2009qz,Catani:2010pd}.}, one
needs to project the fields onto the corresponding components.
The collinear matrix elements associated with the incoming protons give the bare
quark and gluon beam functions,
\begin{align}\label{eq:beam}
  &
  \langle P_i|[\phi_{n_i}^{f\dagger}]^{\alpha'_i}_{a'_i}(b+t'_i \bar n_i)[\phi^f_{n_i}]^{\alpha_i}_{a_i}(t_i \bar n_i)|P_i\rangle \notag\\
  &\qquad=\frac{\delta_{a'_i a_i}}{d_i}\int_0^{1} \frac{\df x_i}{x_i}~
  \sum\limits_j [{P}_{n_i}^{\alpha_i \alpha'_i}]_{f}^{j} B^j_{f}(x_i,\vec
  b_T,\epsilon)e^{\img x_i \bar n_i \cdot P_i\left( \frac{n_i\cdot b}{2}+t_i'-t_i \right)}\;.
\end{align}
Here  $[{P}_{n_i}^{\alpha_i \alpha'_i}]_{f}^{j}$ is the Dirac or Lorentz
structure of the beam function for parton flavor $f$ and spin or polarization
structure $j$, $d_i$ is the dimension of the color representation of a
generic collinear field $\phi_{n_i}$, and $\eps = (4-d)/2$ is the dimensional
regulator.

For the quark beam function (i.e.~$i=a,b$, $f=q, \bar q$), only a single Dirac structure contributes at leading power,
\begin{equation}
	[{P}_{n_i}^{\alpha_i \alpha'_i}]_{q} = \frac{1}{2} (\slashed{ n}_i)^{\alpha_i
	\alpha'_i}\,x_i \,\bar n_i \cdot P_i\;.
\end{equation}
Another possible Dirac structure, $\gamma^\mu_\perp$, would need to be combined with a transverse momentum, and is therefore power suppressed. For the gluon beam function ($i=a,b$, $f=g$), there are two independent Lorentz structures: transverse ($T$)  and linear ($L$) polarization, for which the projectors are~\cite{Catani:2010pd, Becher:2012yn},
\begin{align} \label{eq:proj_B_g}
	[{P}_{n_i}^{\alpha_i \alpha'_i}]^T_{g}&=-\frac{1}{d-2}g_T^{\alpha_i \alpha'_i}
	=-\frac{1}{d-2}\Big[g^{\alpha_i \alpha'_i}-\frac{1}{2}(n^{\alpha_i}_an^{\alpha'_i}_b + n^{\alpha_i}_b n^{\alpha'_i}_a)\Big]\,, \nn \\
	[{P}_{n_i}^{\alpha_i \alpha'_i}]^L_{g}&=
	\frac{1}{d-2}g_T^{\alpha_i \alpha'_i}+\frac{b_T^{\alpha_i}b_T^{\alpha'_i}}{\vec b_T^{\,2}}
	.
\end{align}
The linear polarization contributes, because the collinear gluon splitting is
intrinsically polarized. 

The collinear matrix elements associated with the outgoing jet are the quark and gluon jet functions. Before introducing any jet definition, one has
\begin{align}
    &e^{\img bp_{x,V}}\langle 0|[\phi_{n_J}^{f}]^{\alpha'_J}_{a'_J}(b+t'_J \bar
    n_J)[\phi^{f\dagger}_{n_J}]^{\alpha_J}_{a_J}(t_J \bar n_J)|0\rangle\notag\\
    &=e^{\img bp_{x,V}}\langle
    0|[\phi_{n_J}^{f}]^{\alpha'_J}_{a'_J}(0)e^{-\img p_{x,c}\cdot(b+t'_J \bar n_J-t_J \bar n_J)}[\phi^{f\dagger}_{n_J}]^{\alpha_J}_{a_J}(0)|0\rangle.
\end{align}
One can, then, make the following decomposition
\begin{align}
    e^{-\img p_{c}\cdot(b+t'_J \bar n_J-t_J \bar
    n_J)}=e^{-\img \bar{n}_J\cdotp_{c}(\frac{n_J\cdot n_a}{4}\bar{n}_a\cdot b+\frac{n_J\cdot n_b}{4}\bar{n}_b\cdot b+t'_J-t_J)} e^{\img p_{x,c} b}.
\end{align}
Below, we shall drop the first phase factor on the right hand side of this equation, which specifies the jet momentum that enters the hard function and the conservation of $p^\pm$ with respect to the beam directions, respectively.
As discussed in sec.~\ref{sec:geometry}, the jet definition respects factorization, meaning that the jet functions are only dependent on $n_J$-collinear modes. Effectively, a jet definition simply defines the transverse momentum and rapidity of the jet:
 \begin{align} \label{eq:J_def}
     &e^{\img b_x p_{x,V}}\langle
     0|[\phi^f_{n_J}]^{\alpha_J'}_{a_J'}(0)e^{\img p_{x,c}
     b_x}\delta^{(2)}\bigl(\vec{p}_{T,J}-\hat{\vec{p}}_{T,J}\bigr)\delta(\eta_J-\hat{\eta}_J)[\phi_{n_J}^{f\dagger}]^{\alpha_J}_{a_J}(0)|0\rangle\;
     \nn \\
  &\qquad=e^{\img b_x q_{x}}\langle
  0|[\phi^f_{n_J}]^{\alpha_J'}_{a_J'}(0)e^{\img \delta_x
  b_x}\delta^{(2)}\bigl(\vec{p}_{T,J}-\hat{\vec{p}}_{T,J}\bigr)\delta(\eta_J-\hat{\eta}_J)[\phi_{n_J}^{f\dagger}]^{\alpha_J}_{a_J}(0)|0\rangle\;
  \nn \\
  & \qquad \equiv
    \frac{\delta_{a_J' a_J}}{2(2\pi)^{d-1}} e^{\img q_x b_x}\sum\limits_j
    [P_{n_J}^{\alpha_J' \alpha_J}]^j_f \cJ^{j}_f(b_x, p_{T,J}, \eta_J,\epsilon) \,
  \end{align}
with\footnote{As shown in sec. \ref{sec:gluon_jet}, it is more convenient to evaluate jet functions in a frame in which $p_{x,J}\neq 0$. Therefore, we keep the dependence of $p_{x,J}$ explicit here.
}
\begin{align}
\delta_x\equiv p_{x,c}-p_{x,J},\qquad
q_x = p_{x,V}+p_{x,J}.
\end{align}
Here the momentum $p_{x,c}$ corresponds to the $x$-component of the momenta of the jet constituents, and the operators $\hat{\vec{p}}_{T,J}$ and $\hat{y}_J$ give the jet momentum transverse to the beam and jet rapidity, respectively.
The decomposition of the Dirac and Lorentz structures is similar to the beam functions (with a different normalization), 
\begin{align} \label{eq:proj}
	[{P}_{n_J}^{\alpha'_J \alpha_J}]_{q} &= \frac{1}{2} (\slashed{ n}_J)^{\alpha'_J \alpha_J} \bar n_J \cdot P_J\,, \nn \\
	[{P}_{n_J}^{\alpha'_J \alpha_J}]^T_{g}&= -g_\perp^{\alpha'_J \alpha_J}
	= - g^{\alpha'_J \alpha_J}+\frac{1}{2}(n^{\alpha'_J}_J \bar n^{\alpha_J}_J + \bar n^{\alpha'_J}_J n^{\alpha_J}_J)
	\,, \nn \\
	[{P}_{n_J}^{\alpha'_J \alpha_J}]^L_{g}&=
	g_\perp^{\alpha'_J \alpha_J}+(d-2)\frac{b_\perp^{\alpha'_J}b_\perp^{\alpha_J}}{\vec b_\perp^2}	,
\end{align}
where this is now perpendicular to the jet (not beam) direction, as indicated by the $\perp$ (instead of $T$).

The soft matrix element associated with the soft operator ${\cal
O}^s_j$ give the soft
function,
\begin{equation}
	S_j(b_x)=\frac{2}{N_c^2-1}{\rm Tr}\langle 0|\bar{\bf T}[{\cal O}^{s,a\dagger}_j(b_x)] {\bf T}[{\cal O}^{s,a}_j(0)] | 0\rangle\;,
\end{equation}
where ${\bf T}$ ($\bar{\bf T}$) denote (anti-)time ordering. The overall factor is chosen such that at tree-level the soft function equals 1.
The $q\bar q g$ color space is one-dimensional, but there is a different soft function for each partonic channel $j$, because it matters whether the jet or beam are in the adjoint representation for the gluon. E.g.~for a gluon jet ${\cal O}^s_j$ is $S_{n_b}^\dagger S_{n_c}t^a S_{n_c}^\dagger S_{n_a}$.
 In the remainder of this paper we will label the soft function with the full partonic channel $ijk$, where $i,j,k$ are now parton flavors.

The hard functions are the square of the matching coefficients of SCET operators and observable independent. These are at tree-level given by the corresponding QCD matrix-elements $M$ and including factors that account for averaging over spin/color states of incoming partons (indicated by the bar in $\overline M$) and the flux factor,
\begin{equation}
	\mathcal{H}_{ij\rightarrow Vk}=\frac{x_a x_b p_{T,V}}{8\pi\hat s^2}|\overline M(ij\rightarrow Vk)|^2\;.
\end{equation}

At one-loop order, this equation still holds after renormalization and dropping infrared divergences (which cancel in the matching between QCD and SCET).
Note that the indices of the projectors in eqs.~\eqref{eq:proj_B_g} and \eqref{eq:proj} will be contracted with the matrix elements of the hard scattering. If an observable is not sensitive to the spin or polarizations states, the typical spin-averaged hard function is sufficient. However, in our case we will get a different expression when the initial or final gluon is linearly polarized, and we need to sum all polarizations in \eq{FactThmFourier}, which affects both the gluon beam/jet function and the hard function.

\subsection{One-loop ingredients}
\label{subsec:oneloop}

Here we provide one-loop expressions of the beam, jet, soft and hard functions. Our observable leads to rapidity divergences in the beam, jet and soft function that are not regulated by dimensional regularization. We use the $\eta$-regulator~\cite{Chiu:2011qc,Chiu:2012ir} to regularize rapidity divergences, and the resulting evolution in the rapidity scale $\nu$ can be used to resum the corresponding rapidity logarithms. This is discussed in sec.~\ref{subsec:RGE}, where also the  anomalous dimensions needed for next-to-next-to-leading logarithmic resummation are collected.

The beam functions are matched to collinear parton distribution functions (PDFs) with perturbatively calculable matching coefficients~\cite{Collins:1981uw,Collins:1984kg,Becher:2010tm,Chiu:2012ir}
\begin{align}
  B_i(x, b_x,\mu,\nu) = \sum_j \int \frac{\df x'}{x'}\,\mathcal{I}_{i
  j}\Bigl(\frac{x}{x'}, b_x, \mu, \nu\Bigr) f_j(x',\mu) \bigl[1 + \mathcal{O}(\Lambda_{\rm QCD}^2 \vec b_T^{\,2})] \,.\end{align}
The matching coefficients $\mathcal{I}$ have been calculated up to three-loop order~\cite{Catani:2011kr,Catani:2012qa,Gehrmann:2012ze,Gehrmann:2014yya,Luebbert:2016itl,Echevarria:2016scs,Luo:2019hmp,Luo:2019bmw,Behring:2019quf,Luo:2019szz,Ebert:2020yqt}, and the linearly-polarized contribution at two-loop order~\cite{Gutierrez-Reyes:2019rug}.
Up to one-loop order, they are given by
\begin{align}
\mathcal{I}_{q  q}(z, b_x, \mu, \nu)&=\delta(1\!-\!z)+\frac{\alpha_{s}}{4 \pi}\left[C_{F} L_{b}\left(3+4 \ln \frac{\nu}{\omega}\right) \delta(1\!-\!z)-2P_{qq}(z)  L_{b}+2 C_{F}(1\!-\!z)\right] 
\nn \\ & \quad
+\mathcal{O}(\alpha_{s}^{2}), \nn\\ 
\mathcal{I}^T_{g  g}(z, b_x, \mu, \nu)&=\delta(1-z)+\frac{\alpha_{s}}{4 \pi}\left[L_{b}\left(\beta_{0}+4 C_{A} \ln \frac{\nu}{\omega}\right) \delta(1-z)-2 P_{gg}(z) L_{b}\right]+\mathcal{O}(\alpha_{s}^{2}), \nn\\ 
\mathcal{I}_{q  g}(z, b_x, \mu, \nu)&=\frac{\alpha_{s}}{4 \pi}\left[-2P_{qg}(z)  L_{b}+4T_F z(1-z)\right]+\mathcal{O}(\alpha_{s}^{2}), \nn\\ 
\mathcal{I}^T_{g  q}(z, b_x, \mu, \nu)&=\frac{\alpha_{s}}{4 \pi}\left[-2P_{gq}(z) L_{b}+2 C_{F} z\right]+\mathcal{O}(\alpha_{s}^{2}), \nn\\
\mathcal{I}^L_{g  g}(z, b_x, \mu, \nu) &= -\frac{\alpha_{s}}{4 \pi} C_A \frac{4(1-z)}{z} +\mathcal{O}(\alpha_{s}^{2}), \nn\\
\mathcal{I}^L_{g  q}(z, b_x, \mu, \nu) &= -\frac{\alpha_{s}}{4 \pi} C_F \frac{4(1-z)}{z}+\mathcal{O}(\alpha_{s}^{2}),
\end{align}
with 
\begin{align} \label{eq:wLb}
   \omega = \bar n_i\cdot p_i = x_i E_{\rm cm} \,, \qquad L_b = \ln \frac{\mu^2 b_x^{\,2}}{4 e^{-2\gamma_{\rm E}}}
\,,\qquad \beta_0 &= \frac{11}{3} C_A - \frac{4}{3} T_F n_f \,,
\end{align}
and the splitting functions
\begin{align} \label{eq:splitfun}
P_{qq}(z)&=C_{F}\left[\frac{1+z^{2}}{(1-z)_{+}}+\frac{3}{2} \delta(1-z)\right], \nn \\
P_{gg}(z)&=2 C_{A}\left[\frac{z}{(1-z)_{+}}+\frac{1-z}{z}+z(1-z)\right]+ \frac{\beta_{0}}{2} \delta(1-z)\,,\nn \\
P_{qg}(z)&= T_{F}\left[z^{2}+(1-z)^{2}\right], \nn\\
P_{gq}(z)&= C_{F} \frac{1+(1-z)^{2}}{z}
\,.\end{align}
The one-loop jet functions for the WTA recombination scheme are~\cite{Gutierrez-Reyes:2018qez,Gutierrez-Reyes:2019vbx,Chien:2020hzh}
\begin{align}
    \mathscr{J}_q(b_x,\mu,\nu) & = 1 + \frac{\al_s}{4\pi} C_F \left[ L_b \left( 3 + 4\ln\frac{\nu}{\omega} \right) + 7 -\frac{2\pi^2}{3} - 6\ln 2 \right] + \mathcal{O}(\al_s^2) \nn \,,\\
    \mathscr{J}_{g}^{T}(b_x,\mu,\nu) & = 1+  \frac{\al_s}{4\pi} \biggl[
    L_{b}\Bigl(\beta_{0}+4 C_{A} \ln \frac{\nu}{\omega}\Bigr)
     + C_A\Bigl(\frac{131}{18} -\frac{2\pi^2}{3} - \frac{22}{3}\ln 2 \Bigr) \nn \\ & \quad + T_F n_f \Bigl(- \frac{17}{9} + \frac{8}{3}\ln 2\Bigr)\biggr] + \mathcal{O}(\al_s^2) \nn \,, \\
     \mathscr{J}_{g}^{L}(b_x,\mu,\nu) & = \frac{\al_s}{4\pi} \left[- \frac{1}{3}C_A + \frac{2}{3} T_F n_f \right]+ \mathcal{O}(\al_s^2)\,.
\end{align}
The result for the linearly-polarized gluon jet function was first quoted in our letter~\cite{Chien:2020hzh}, and we therefore provide a calculation of the gluon jet functions in \sec{gluon_jet}. There we also calculate jet functions for other recoil-free axes and for track-based measurements.

Up to order $\al_s^2$, the soft function can be obtained~\cite{Chien:2020hzh} from the standard TMD soft function, which is known up to three loop order~\cite{Echevarria:2015byo,Luebbert:2016itl,Li:2016ctv}. The contribution involving exchanges between three Wilson lines vanishes due to color conservation~\cite{Becher:2012za}. For exchanges involving only two Wilson lines, we can perform a boost to make them back-to-back. Since our observable is perpendicular to the boost, only the rapidity regulator is affected, which can be taken into account in a straightforward manner~\cite{Gao:2019ojf}. The resulting one-loop soft functions are
\begin{align}
\frac{\alpha_s}{4\pi}
S_{i j k}^{(1)}(b_x, \mu, \nu)=\frac{\alpha_s}{4\pi} \Bigl[\left(C_{i}+C_{j}-C_{k}\right) \frac{\omega_{i j}}{2}+\left(C_{i}+C_{k}-C_{j}\right) \frac{\omega_{i k}}{2}+\left(C_{j}+C_{k}-C_{i}\right) \frac{\omega_{j k}}{2}\Bigr]
\,,\end{align}
where the color factor $C_i$ is $C_F$ if parton $i$ is an (anti-)quark and $C_A$ if it is a gluon, and 
\begin{align}
    \omega_{i j}=-2 L_{b}^{2}+4 L_{b}\left(\ln \frac{\mu^{2}}{\nu^{2}}-\ln \frac{n_i\cdot n_j}{2}\right)-\frac{\pi^{2}}{3}
\,.\end{align}

The hard function for the process $ij\to  V k$ is given by
\begin{align}
     \mathcal{H}_{ij\to  V k} = \frac{x_1 x_2 p_{T,V}^2}{8\pi \hat s^2} |\overline{\mathcal{M}}_{ij\to  V k}|^2
\end{align}
for which the tree-level matrix elements are
\begin{align}
|\overline{\mathcal{M}}_{q\bar q\to  V g}|^{2} &=\frac{16 \pi^{2} \alpha_{s} \alpha_{e m} e_{q}^{2}\left(N_{c}^{2}-1\right)}{N_{c}^{2}} \frac{\hat{t}^{2}+\hat{u}^{2}+2 \hat{s} m_{V}^{2}}{\hat{t} \hat{u}}\,, \nn\\
|\overline{\mathcal{M}}_{qg\to  V q}|^{2} &=-\frac{16 \pi^{2} \alpha_{s} \alpha_{e m} e_{q}^{2} }{N_{c}} \frac{\hat{s}^{2}+\hat{t}^{2}+2 \hat{u} m_{V}^{2}}{\hat{s} \hat{t}}, \nn \\
    |\overline{\mathcal{M}}_{q\bar q\to V g_L }|^2 &= - \frac{32\pi^2 \alpha_{s} \alpha_{e m}e_q^2(N_c^2-1)}{N_c^2} \frac{\hat s m_V^2}{\hat u \hat t}
    , \nn \\ 
    |\overline{\mathcal{M}}_{qg_L\to V q }|^2 &=  \frac{32\pi^2 \alpha_{s} \alpha_{e m} e_q^2}{N_c} \frac{\hat u m_V^2}{\hat s \hat t} 
\,,\end{align}
and $g_L$ denotes a linearly-polarized gluon. For $Z$ production one replaces
$e_q$ in analogy to~\eqref{eq:emdef}, and the partonic Mandelstam variables are
\begin{align} \label{eq:mandelstam}
  \hat s &= m_V^2 + 2 p_{T,V}^2 + 2 p_{T,V} \sqrt{m_V^2 + p_{T,V}^2} \cosh (\eta_J - y_V)\,, \nn \\
  \hat t &= - p_{T,V}^2 - p_{T,V} \sqrt{m_V^2 + p_{T,V}^2} \exp(\eta_J - y_V)\,, \nn \\
  \hat u &= - p_{T,V}^2 - p_{T,V} \sqrt{m_V^2 + p_{T,V}^2} \exp(y_V - \eta_J)\,.
\end{align}
The loop corrections that enter the hard function have been calculated at one-loop order in refs.~\cite{Aurenche:1983ws,Aurenche:1987fs,Gordon:1993qc,Ellis:1981hk,Arnold:1988dp,Gonsalves:1989ar}, and we give their expressions in the appendix \ref{app:oneloophard}. The resulting one-loop hard function for a transversely polarized gluon can be obtained from the appendices of \cite{Becher:2009th,Becher:2012xr,Moult:2015aoa}. Since the beam and jet function for a linearly-polarized gluon only start at one-loop, the tree-level hard function suffices in this case.

\subsection{Glauber interaction and factorization breaking}
\label{subsec:glauber}

We conclude this section by briefly commenting on the appearance of a Glauber mode and
its impact on the validity of the factorization theorem in~\eqref{eq:FactThm}
and~\eqref{eq:FactThmFourier}, following the framework laid out
in ref.~\cite{Rothstein:2016bsq}, supplemented by the results
of refs.~\cite{Schwartz:2017nmr,Schwartz:2018obd}.

Ref.~\cite{Rothstein:2016bsq} explains in detail that Glauber rungs
connecting active lines to other active or to spectator lines can be
absorbed\footnote{Alternatively, the
Glauber mode is then a true subset of the corresponding soft or collinear
mode, so a dedicated Glauber contribution does not add anything
that is not already covered by the naive soft or collinear mode.} into the soft (for active-active) or collinear Wilson lines (for
active-spectator), and only the pure spectator-spectator Glauber exchanges
require special attention. 
Before we discuss our specific application, we first summarize and clarify a few points:

First, we note that in line with the expectation that ``spectator gluons or
quarks may be created by collinear radiation from active lines'', according
to ref.~\cite{Rothstein:2016bsq}, we take only the parton lines directly connected to
the hard scattering vertex as ``active''. This implies that collinear splittings
in the initial state, as well as the branching of the jet progenitor parton
generate spectators, and that thus there are three collinear sectors containing
spectators (two proton remnants, one jet sector). 

Second, we note that for single-scale observables,
refs.~\cite{Schwartz:2017nmr,Schwartz:2018obd} demonstrate that pure Glauber
exchanges (i.e. without soft emissions off Lipatov vertices) do not lead to
factorization violation. 

Third, we resolve the ambiguity of the Glauber mode for three distinct rapidity
sectors by observing that Glauber modes only ever connect two of them, and that
the frames in which these are pairwise back-to-back are connected by ---
generically --- $\mathcal{O}(1)$ boosts. Together with the collapse rule this
means that the different Glauber dipoles communicate only via soft emissions from Lipatov vertices, whose power
counting is isotropic, and therefore unchanged by the boosts relating the
different Glauber frames.

Lastly, we follow the diagrammatic conventions
of ref.~\cite{Rothstein:2016bsq} for the discussion below, where auxiliary interactions are introduced that
create active and spectator lines from incoming hadron fields, to be able to
focus on the topology of the appearing diagrams. 

\medskip

\begin{figure}
     \centering
     \begin{subfigure}[b]{0.215\textwidth}
         \centering
         \includegraphics[width=\textwidth]{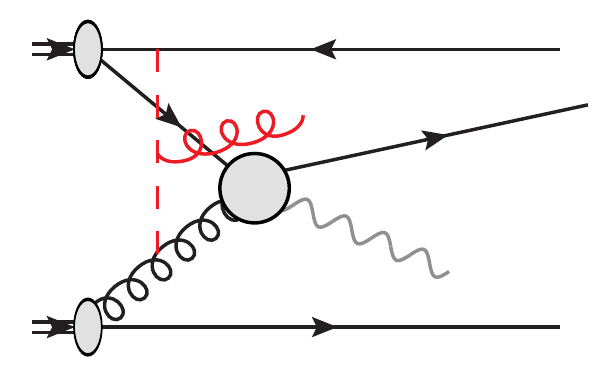}
         \caption{Collinear overlap}
         \label{fig:LipAS}
     \end{subfigure}
     \hfill
     \begin{subfigure}[b]{0.215\textwidth}
         \centering
         \includegraphics[width=\textwidth]{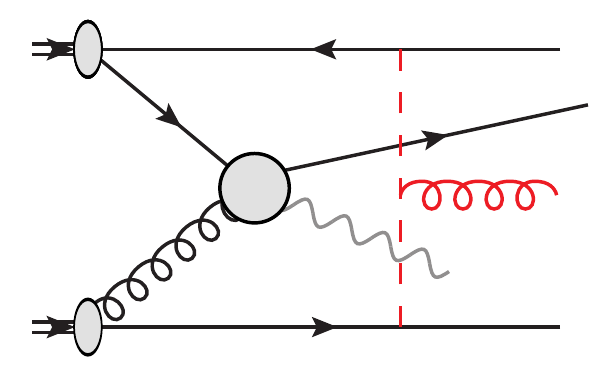}
         \caption{Remnant recoil}
         \label{fig:LipRemRem}
     \end{subfigure}
     \hfill
     \begin{subfigure}[b]{0.252\textwidth}
         \centering
         \includegraphics[width=\textwidth]{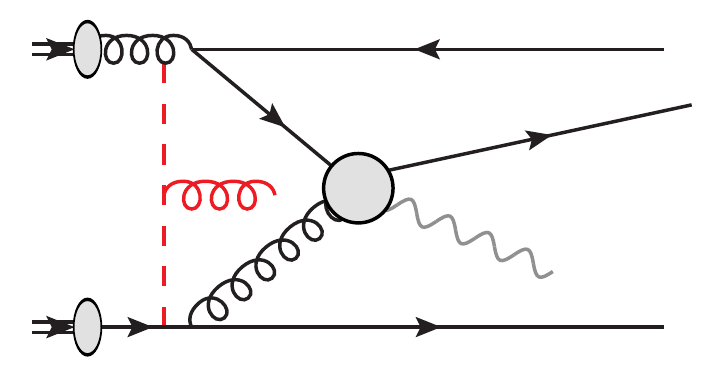}
         \caption{IS splittings}
         \label{fig:LipPertSplit}
     \end{subfigure}
     \hfill
     \begin{subfigure}[b]{0.215\textwidth}
         \centering
         \includegraphics[width=\textwidth]{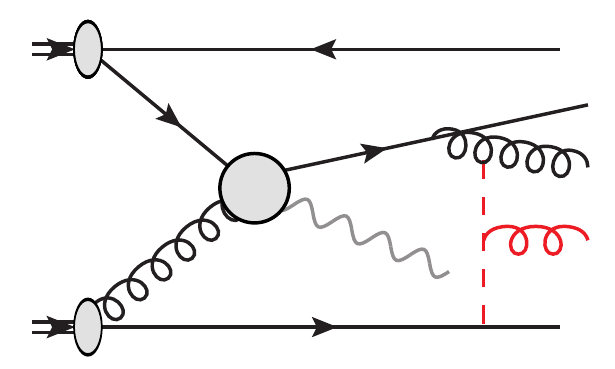}
         \caption{Jet spectators}
         \label{fig:LipJet}
     \end{subfigure}
        \caption{Glauber topologies. Glauber and Lipatov emissions are colored
        in red, Glauber emissions are dashed. The diagrams describe (a) the
        collinear overlap of active-spectator interactions, (b) rescattering of the proton
        remnants, (c) Glauber bursts after perturbative initial state (IS) splittings,
        and (d) spectator-spectator exchanges involving the jet, respectively.}
        \label{fig:three graphs}
\end{figure}

Moving on to the discussion of our case: We begin with Glauber exchanges involving Lipatov vertices
off Glauber rungs connecting to at least one active parton, as e.g. in
\fig{LipAS}. Following ref.~\cite{Rothstein:2016bsq}, the Glauber exchange
is an element of the soft or collinear sector, the Lipatov vertex therefore
represents soft radiation off either another soft, or off a collinear emission.
In the former case (the soft overlap for active-active Glauber exchange), this
simply is the soft dynamics encoded in the soft Wilson line structure and Lagrangian. For the
latter case, it corresponds to a soft emission off a collinear field, which does not appear in the SCET${_{\rm II}}$ Lagrangian. Instead it has been integrated out and is thus accounted for in the matching of QCD to SCET, where a soft Wilson line is introduced for every
collinear Wilson line. This soft Wilson line then accounts for the Lipatov
emissions off the Glauber subset of the corresponding collinear Wilson line
(besides other soft effects).
This leaves us with Glauber exchanges between spectators only, which also have
at least one Lipatov emission. It is clear that the lowest order that such diagrams could appear in the perturbative expansions is $\mathcal{O}(\alpha_s^2)$.

Focusing on Glauber exchanges between the proton remnants, we naively expect
that a contribution of the form of \fig{LipRemRem} could appear at
$\mathcal{O}(\alpha_s^2)$, when interfering with a suitable tree-level diagram.
However, such interference is prohibited by momentum conservation: The proton
remnants have transverse momenta of $\mathcal{O}(\Lambda_{\rm QCD})$, the typical
scale of intra-proton dynamics, while the Glauber exchange represents a recoil between
the involved spectators, causing them to have transverse momentum of
$\mathcal{O}(q_x)$. This does not match up with any tree level conjugate
diagram. Accordingly, the conjugate amplitude also requires some perturbative
exchange (Glauber or otherwise), which pushes the contribution to
$\mathcal{O}(\alpha_s^3)$. In addition, such diagrams do not contribute to our
observable, as we only measure properties of the jet: Any recoil of beam
remnants is never measured, and accompanying soft emissions are not relevant, because we use the WTA axis.
For a diagram to have any effect it must 
impart recoil on the active partons or the jet constituents: The WTA axis
follows a collinear emission in the jet, which can only pick up Glauber effects
by direct recoil, or by inheriting recoil from the partons initiating the hard
scattering.

An example for an allowed diagram would be
\fig{LipPertSplit}, where a perturbative splitting of an initial
state gluon creates spectators that already have transverse momentum of $\mathcal{O}(q_T)$, such that a Glauber exchange
does not fall prey to momentum conservation. Such diagrams appear, by virtue
of the initial state branching, at the earliest at $\mathcal{O}(\alpha_s^4)$.
Lastly, we turn to diagrams involving the jet, like e.g.~\fig{LipJet}.
As stated above, spectators in the jet can arise from perturbative splittings of the parton initiating the jet, which means such
topologies start at $\mathcal{O}(\alpha_s^3)$ at the lowest. Momentum
conservation does not pose an obstacle here, as the soft emission across the cut
can attach to the beam spectator in the conjugate amplitude, pushing its
transverse momentum to the right scaling. 
However, an explicit calculation of the loop diagram in \fig{LipJet}
shows that it evaluates to zero, as may have been expected due to its similarity with deep-inelastic scattering.
Explicitly, as only two rapidity sectors are involved, we
can boost to a frame in which they are back-to-back. Of the four non-Glauber
propagators involving the loop momentum $k$, two will then depend only on $n
\cdot k$, and two only on $\bar{n}\cdot k$, after expansions around the momentum scaling for Glauber and collinear
modes. These two subsets correspond to the two collinear sectors, i.e. two
propagators arise from the jet, the other from the beam remnant and active line.
The two propagators arising from the jet have poles on the same side of the real
line, and the loop integral thus evaluates to zero by residues\footnote{See
in particular the discussions surrounding equations (11.9) and (B.6)
in~\cite{Rothstein:2016bsq}.}.
Numerator factors and the rapidity regulator do not change this outcome.

We thus conclude that effects of perturbative Glauber exchanges are suppressed
by at least $\mathcal{O}(\alpha_s^3)$ compared to the accuracy achieved in this
paper.

\section{Jet function calculations}
\label{sec:J}

\subsection{Gluon jet function calculation at order \texorpdfstring{$\alpha_s$}{alpha(s)}}
\label{sec:gluon_jet}

Next we present the calculation of the one-loop gluon jet functions.
According to \eq{J_def}, they are defined through
\begin{align} \label{eq:gJ_deffull}
    \mathscr{J}_g^{\mu\nu}&\equiv\sum\limits_j [P_{n}^{\mu\nu}]^j_g \mathscr{J}^{j}_g(b_x, \vec{p}_{T,J}, \eta_J,\epsilon) 
    \nn\\
    &=\frac{2(2\pi)^{d-1}}{N_c^2-1}
     \langle 0|\mathcal{B}_{n\perp}^{a\mu}(0)\left(0\right)e^{\img \delta_xb_x}\delta^{(2)}(\vec{p}_{T,J}-\hat{\vec{p}}_{T,J})\delta(y_J-\hat{y}_J)\mathcal{B}_{n\perp}^{a\nu}(0)|0\rangle\;
\end{align}
where the transverse momentum of the jet with respect to the initial parton is encoded in
\begin{align}
     \delta_x=p_{x,c}-p_{x,J}.
\end{align}
In our original coordinates  $p_{x,J}=0$, but we will instead perform our calculation in a frame where  the total collinear momentum $p_{x,c} =0$.
In this subsection we drop the subscript $J$ on light-cone coordinates for brevity.
As discussed in sec.~\ref{sec:geometry}, the difference between $\vec{p}_{T,J}$ and $\vec{p}_{T,c}$ is $O(\lambda)$, so we can replace $\hat{\vec{p}}_{T,J} \to \vec{p}_{T,c}$, and similarly we can replace $y_J$ by $y_c$. 
Then, following the same steps as for the standard jet axis in ref.~\cite{Chien:2019gyf}, we can switch to coordinates along the momentum of the initial parton, to obtain
\begin{align} \label{eq:J_g_def}
\mathscr{J}_g^{\mu\nu}=&\frac{2(2\pi)^{d-1}}{N_c^2-1}\bar{n}\cdot p_J
 \langle0|\mathcal{B}_{n\perp}^{a\mu}(0)e^{\img \delta_x {b}_x}\delta(\bar{n}\cdot p_J-\bar{n}\cdot p_c)\delta^{(d-2)}(\vec{p}_{\perp,c})\mathcal{B}_{n\perp}^{a\nu}(0)|0\rangle.
\end{align}
Using the projectors in \eq{proj}, one finds at tree-level
\begin{align}
    &\mathscr{J}_{g}^T=\frac{1}{d-2}[{P}_{n}^{\mu\nu}]^T_{g}\mathscr{J}_{g\mu\nu} = 1+\mathcal{O}(\alpha_s),\notag\\
    &\mathscr{J}_{g}^L=\frac{1}{(d-2)(d-3)}[{P}_{n}^{\mu\nu}]^L_{g}\mathscr{J}_{g\mu\nu}
    =\mathcal{O}(\alpha_s).
\end{align}

\begin{figure}
\begin{center}
\includegraphics[width=0.6\textwidth]{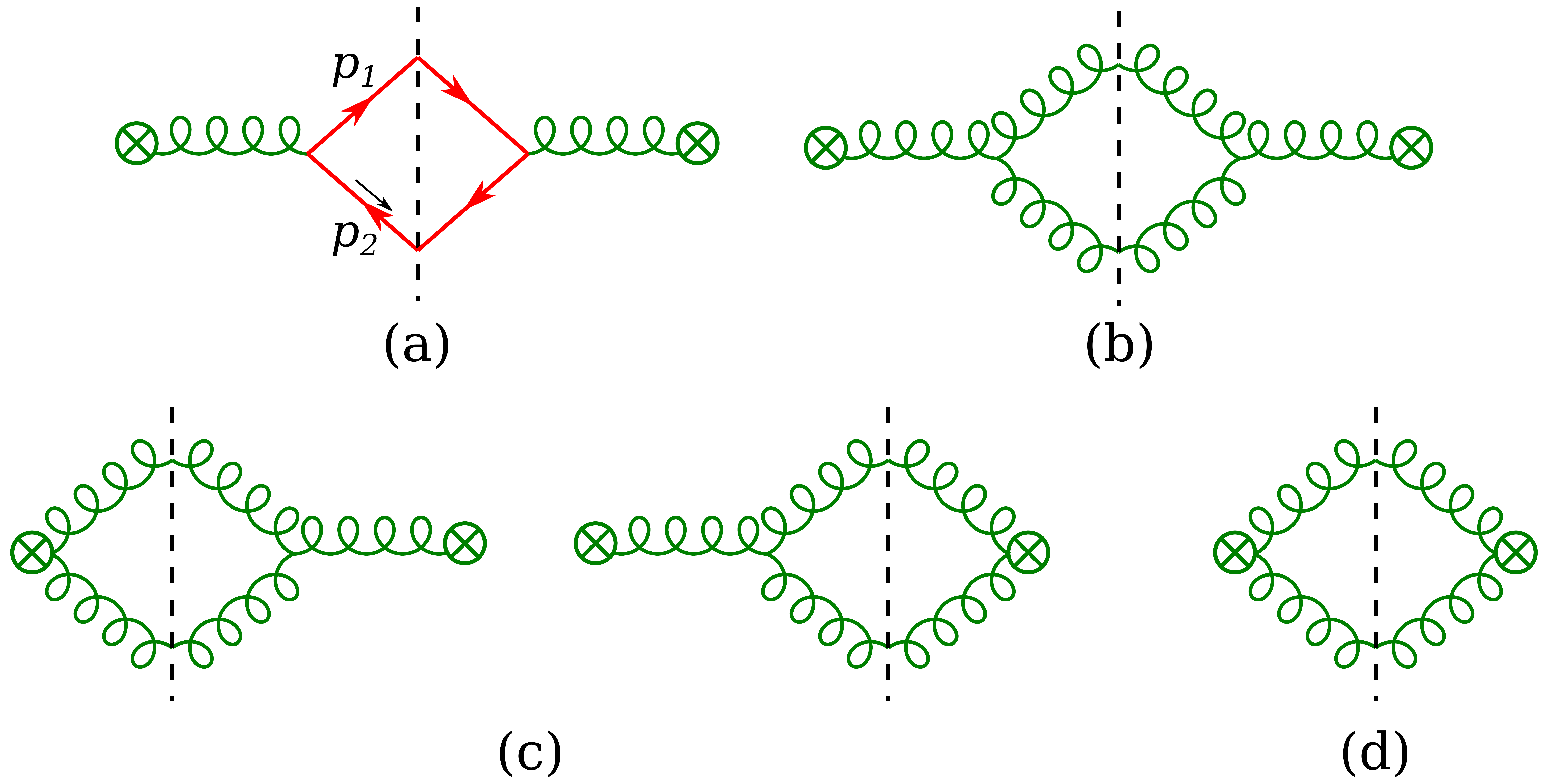}
\end{center}
\caption{One-loop diagrams contributing to the gluon jet functions. The $\otimes$ denotes the collinear gluon field $\mathcal{B}_{n\perp}^\mu$ (dressed with Wilson lines).
\label{fig:gJets}}
\end{figure} 

All one-loop diagrams for the gluon jet functions are shown in \fig{gJets}. 
The relevant Feynman rules for $\mathcal{B}^{ \mu}_{n\perp}=\frac{1}{g}W_{n}^\dagger \img D_{n\perp}^\mu W_{n}$ are given by
\begin{align} \label{eq:B_feyn}
    \begin{array}{c}
         \includegraphics[width=0.12\textwidth]{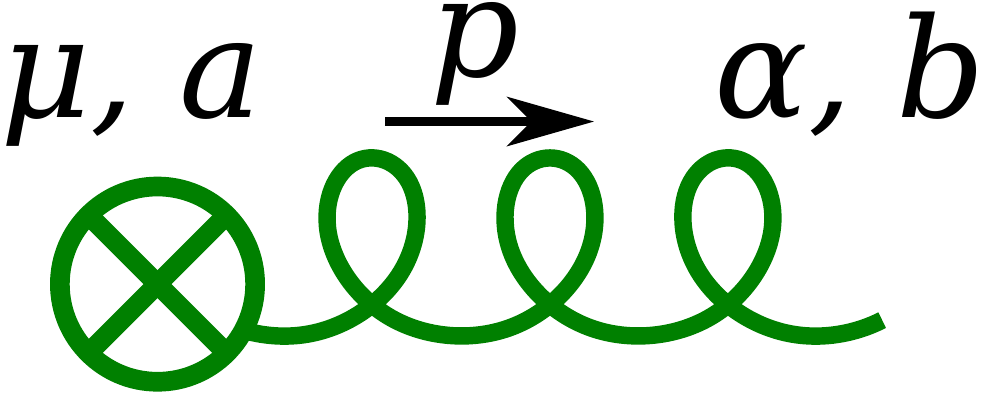}
    \end{array}&=\frac{-\img\delta^{ab}}{p^2+\img0} \bigg(g_\perp^{\alpha\mu}-\frac{\bar{n}^\alpha p_\perp^\mu}{\bar{n}\cdot p-\img0}\bigg),
\notag\\
    \begin{array}{c}
         \includegraphics[width=0.15\textwidth]{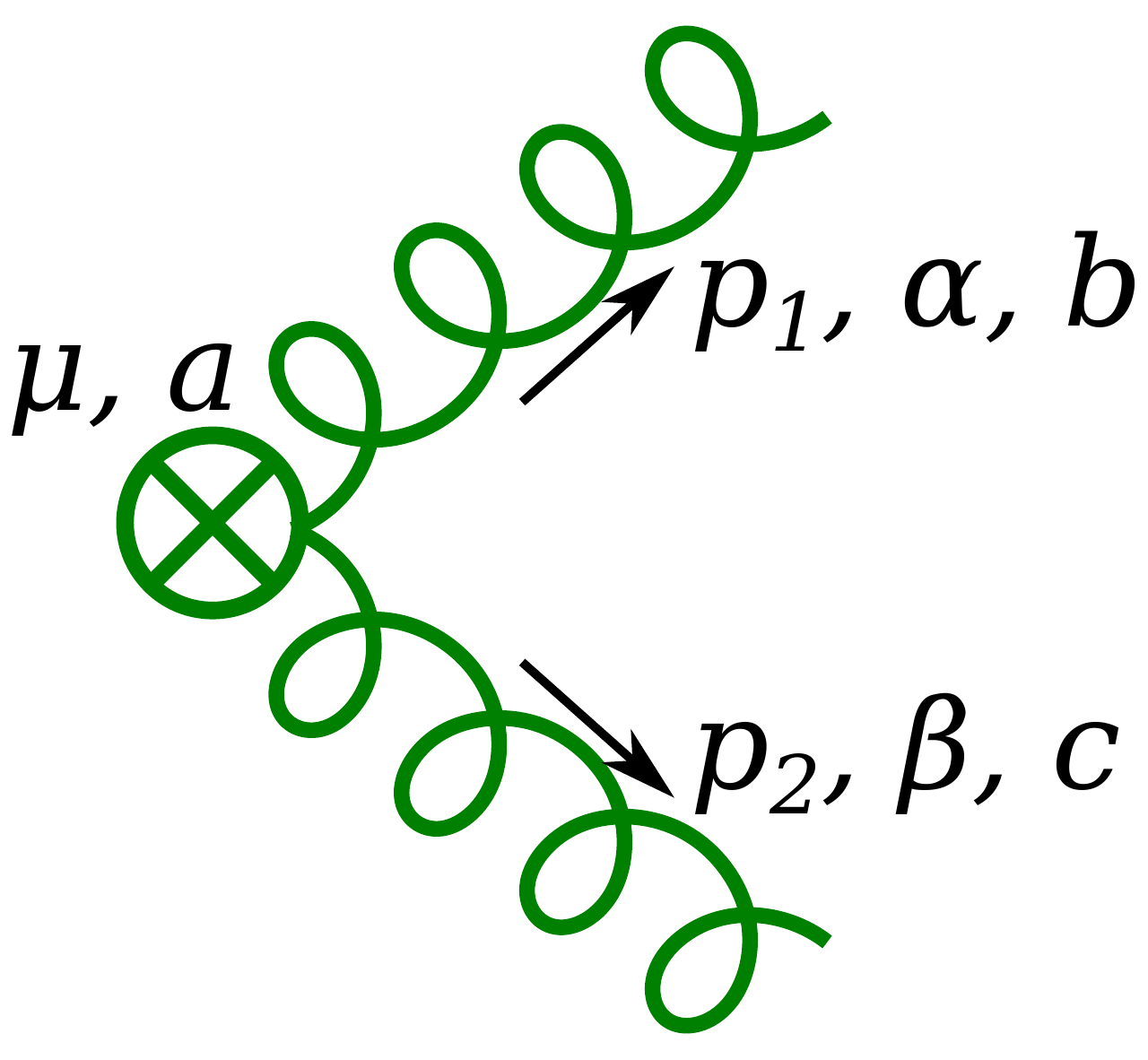}
    \end{array}&=\img g f^{abc} \tilde \mu^\eps w^2 \nu^\eta \bigg[g_\perp^{\alpha\mu}\frac{\bar{n}^\beta}{(\bar{n}\cdot p_2)^{1+\eta}}-g_\perp^{\beta\mu}\frac{\bar{n}^\alpha}{(\bar{n}\cdot p_1)^{1+\eta}}\bigg],
\end{align}
where $g$ is the renormalized coupling with $\tilde \mu^2 = \mu^2e^{\gamma_E}/(4\pi)$,
 and we use the $\eta$-regulator~\cite{Chiu:2011qc,Chiu:2012ir} to regularize rapidity divergences.
The $w$ in \eq{B_feyn} is an (artificial) coupling used to derive the corresponding rapidity evolution. 
Diagram (d) vanishes in Feynman gauge, and the others contribute to the jet function according to, 
\begin{align} \label{eq:Jmunu}
    \frac{\al_s}{4\pi}\,\mathscr{J}_g^{\mu\nu(1)}=\int {\rm d} \Pi_{2}\, e^{\img \delta_x b_x}\Bigl[n_f M_a^{\mu\nu}+\frac{1}{2}M_b^{\mu\nu}+\frac{1}{2}M_{c}^{\mu\nu}\Bigr]
\end{align}
where the $1/2$ is an identical particle factor and 
\begin{align}\label{eq:MgJets}
    M_a^{\mu\nu} &= \frac{g^2 T_F}{(p_1\cdot p_2)^2}\tilde \mu^{2\epsilon}\big(-2p_\perp^\mu p_\perp^\nu-g_\perp^{\mu\nu} p_1\cdot p_2\big),\notag\\
    M_b^{\mu\nu}&=\frac{g^2 C_A}{(p_1\cdot p_2)^2}\tilde \mu^{2\epsilon}\big[(d-2)p_\perp^\mu p_\perp^\nu+2 g_\perp^{\mu\nu} p_1\cdot p_2\big],\notag\\
    M_{c}^{\mu\nu}&=-\frac{g^2 w^2 C_A}{p_1\cdot p_2}\tilde \mu^{2\epsilon}\Bigl(\frac{\nu}{\omega_J}\Bigr)^\eta g_\perp^{\mu\nu}\bigg[\frac{1+z}{(1-z)^{1+\eta}}+\frac{2-z}{z^{1+\eta}}\bigg]\,.
\end{align}
with $p_1 \cdot p_2 = \vec p_\perp^{\,2}/(2z(1-z))$.
The two-body \emph{collinear} phase space entering in  \eq{Jmunu}, is defined as 
\begin{align}\label{eq:dPi2}
    \int {\rm d}\Pi_{2} &\equiv 2 (2\pi)^{d-1} \omega_J\prod\limits_{i=1}^2\int\frac{{\rm d}p_i^- {\rm d}^{d-2}p_{i\perp}}{2p_i^- (2\pi)^{d-1}}\,\delta(p^-_J-p_1^--p_2^-)\delta^{(d-2)}(\vec{p}_{1\perp}+\vec{p}_{2\perp})\notag\\
    &=\frac{1}{4\pi}\int\frac{{\rm d}^{d-2}p_\perp}{(2\pi)^{d-2}}\int_0^1\frac{{\rm d}z}{z(1-z)}
\end{align}
which we rewrite in terms of the transverse momentum $\vec p_\perp$ and momentum fraction $z$,
\begin{align} \label{eq:jet_kin}
    \vec{p}_\perp=\vec{p}_{1\perp}=-\vec{p}_{2\perp},\qquad z = \frac{p_{1}^-}{p^-_J}=1-\frac{p_{2}^-}{p^-_J}.
\end{align}
The amplitudes in \eq{MgJets} are also expressed in terms of these variables. For the WTA scheme, one has
\begin{align}\label{eq:deltaxWTA}
    \delta_x = \left\{\begin{array}{ll}
    -\frac{p_x}{1-z} &  \text{for $\frac{1}{2}>z>0$}\\
    \frac{p_x}{z} &\text{for $1>z>\frac{1}{2}$} 
    \end{array}\right.
    .
\end{align}

The evaluation of $\mathscr{J}_{g}^i$ involves the following two integrals
\begin{align}\label{eq:Is}
    I_{1}(b_x)&\equiv\int\frac{\df^{d-2}p_\perp}{(2\pi)^{d-2}}\frac{e^{\img b_x p_x}}{\vec{p}_\perp^{\,2}}=\frac{1}{4\pi} (\pi b_x^2)^{\epsilon } \Gamma (-\epsilon ),\notag\\
    I_{2}(b_x)&\equiv\int\frac{\df^{d-2}p_\perp}{(2\pi)^{d-2}}\frac{p_x^2e^{\img b_x p_x}}{(\vec{p}_\perp^{\,2})^2}=\frac{1}{8\pi} (2 \epsilon +1) (\pi b_x^{2})^{ \epsilon} \Gamma (-\epsilon ).
\end{align}
For $\mathscr{J}_{g}^T$, one has 
\begin{align}
    \mathscr{J}_{g}^{T}&=1-\frac{g_{\perp\mu\nu}}{d-2}\int \df \Pi_{2}\,e^{\img \delta_x b_x}\bigg[n_f M_a^{\mu\nu}+\frac{1}{2}M_b^{\mu\nu}+\frac{1}{2}M_{c}^{\mu\nu}\bigg].
\end{align}
By using \eq{dPi2} and \eq{Is}, this leads to 
\begin{align}\label{eq:JgT}
    \mathscr{J}_{g}^T&= 1+\alpha_s\tilde{\mu}^{2\epsilon}\int_0^1 \df z\, I_{1}\bigg(\frac{b_x}{\hat z}\bigg)\bigg\{w^2 C_A\bigg(\frac{\nu}{\omega_J}\bigg)^\eta\bigg[\frac{1+z}{(1-z)^{1+\eta}}+\frac{2-z}{z^{1+\eta}}\bigg]\notag\\
    &\qquad + 2 C_A[z(1-z)-1]+2 T_F n_f \bigg(1-\frac{4}{d-2}z(1-z)\bigg)\bigg\}\notag\\
    &=Z_{g} + \frac{\alpha_s }{4\pi}\bigg[ C_A \bigg(4 L_b \ln\frac{\nu}{\omega_J}+\frac{11}{3} L_b-\frac{2}{3}\pi ^2 +\frac{131}{18}-\frac{22}{3} \ln2\bigg)+ \notag\\
    &\quad - T_F n_f \bigg(\frac{4}{3}L_b+\frac{17}{9}-\frac{8}{3} \ln2 \bigg)\bigg]\,,
\end{align}
where 
\begin{align}
    \hat z = \max\{z,1-z\},\qquad L_b=\log\Bigl(\frac{b_x^2\mu^2}{4e^{-2\gamma_E}}\Bigr),\qquad \omega_J = 2 p_{T,J}\cosh\eta_J,
\end{align}
and the jet function renormalization is
\begin{align}\label{eq:Zg1}
    Z_{g}=1-\frac{\alpha_s C_A}{\pi\eta}w^2 e^{(L_b-\gamma_E)\epsilon}\Gamma(-\epsilon)+\frac{\alpha_s }{4\pi\epsilon}\bigg[C_A\bigg(4\ln\frac{\nu}{\omega_J}+\frac{11}{3}\bigg)-\frac{4}{3} n_fT_F\bigg].
\end{align}
From this, one can easily obtain the one-loop anomalous dimensions 
\begin{align}\label{eq:gjetanomdim1}
    &\Gamma_\nu^{\mathscr{J}_{g}} = \frac{\alpha_s C_A}{\pi}L_b,\notag\\
    &\Gamma_\mu^{\mathscr{J}_{g}} = \frac{\alpha_s}{2\pi}\bigg[C_A\bigg(4\ln\frac{\nu}{\omega_J}+\frac{11}{3}\bigg)-\frac{4}{3} n_fT_F\bigg]=2 C_A \Gamma_{\text{cusp}}\ln\frac{\nu}{\omega_J}-2 \gamma^g
\,,\end{align}
which agree with their all-order form in  eq.~\eqref{Bad}.

For $\mathscr{J}_{g}^L$, one only needs to include the terms $\propto p_\perp^\mu p_\perp^\nu$ in eq.~(\ref{eq:MgJets}). It can then easily be evaluated by using the two integrals in \eq{Is}, yielding
\begin{align}
    \mathscr{J}_{g}^L&=\frac{g^2}{2}\tilde \mu^{2\epsilon}\frac{(d-2){C_A}-4T_F n_f}{(d-2)(d-3)}\int \df\Pi_{2}\, \frac{e^{\img \delta_x b_x}}{(p_1\cdot p_2)^2}\,[{P}_{n}^{\mu\nu}]^L_{g}\, p_\perp^\mu p_\perp^\nu\\
    &=2\alpha_s\tilde \mu^{2\epsilon}\frac{(d-2){C_A}-4T_F n_f}{(d-2)(d-3)}\int_0^1 \df z\, z(1-z)\bigg[(d-2)I_{2}\bigg(\frac{b_x}{\hat{z}}\bigg)-I_{1}\bigg(\frac{b_x}{\hat{z}}\bigg)\bigg].\notag
\end{align}
Since
\begin{align}\label{eq:I2mI1}
    (d-2)I_{2}\bigg(\frac{b_x}{\hat{z}}\bigg)-I_{1}\bigg(\frac{b_x}{\hat{z}}\bigg)=-\frac{1}{4\pi}+\mathcal{O}(\epsilon),
\end{align}
$\mathscr{J}_{g}^L$ is finite at this order (as required, since it vanishes at tree-level). It is given by
\begin{align}\label{eq:JgL}
    \mathscr{J}_{g}^L=-\frac{\alpha_s}{4\pi}(2C_A-4T_F)\int_0^1 z(1-z)=\frac{\alpha_s}{4\pi}\Bigl(-\frac{1}{3}C_A+ \frac{2}{3}T_Fn_f\Bigr).
\end{align}

\subsection{Recombination scheme dependence}
\label{subsec:recoscheme}

The WTA algorithm is not the only recombination scheme that can be used to construct a recoil-free jet axis. In this subsection we will employ a more general recombination scheme. It turns out that this only changes the finite part of the jet function, and at the end of this section we give explicit results.
 
A recombination scheme dictates how particles are merged during the clustering
procedure. The simplest is to add the momentum four-vectors, which is known as
the $E$-scheme. In the rest of this paper we focus on the WTA-scheme, specifically the WTA-$p_T$-scheme,
described in \sec{geometry}. Here, we consider a generalization in which the
momenta $p_i$ and $p_j$ of two particles (or pseudojets) are recombined into
$p_r$ as follows
\begin{align} \label{eq:recomb}
p_{T,r} &=p_{T,i}+p_{T,j}, \nonumber \\
\phi_r &= (p_{T,i}^n \phi_i +p_{T,j}^n \phi_j)/(p_{T,i}^n +p_{T,j}^n ), \nonumber \\
y_r &= (p_{T,i}^n y_i +p_{T,j}^n y_j)/(p_{T,i}^n +p_{T,j}^n ),
\end{align}
in terms for the transverse momentum $p_T$, azimuthal angle $\phi$ and rapidity $y$.
The azimuthal angle and rapidity are combined in a way that favors the direction
of the harder particle, with the weight of the factors $p_{T,i}^n$ controlled by the power of $n\geq 1$. For $n > 1$ this recombination scheme is recoil free, implying that the same factorization theorem holds as for the WTA scheme (which corresponds to $n \to \infty$). Since only the jet function depends on $n$, consistency of the factorization implies that only the constant term of the jet function depends on it, which is borne out by an explicit calculation.

Now, let us calculate gluon jet functions at one loop using the general scheme in \eq{recomb}. 
For two collinear partons, using coordinates along the parton that initiates the jet, one has $\phi_i \approx p_{x,i}/p_{T,i}$ and $p_x = p_{x,1} = -p_{x,2}$ such that~\footnote{
Here, it does not matter whether one uses $p_{T,i}$ or $p_i^-$ as long as the jet radius is not too large. Our final result, however, shows explicit breaking of boost invariance along the beam axis, resulting from the rapidity regulator.}
\begin{align}
    \delta_x &= p_{T,J}\frac{p_{T,1}^n \phi_1 - p_{T,2}^n  \phi_2}{p_{T,1}^n +p_{T,2}^n}
    = p_{J}^-\frac{(p_{1}^-)^n \frac{p_x}{p_{1}^-} - (p_{2}^-)^n \frac{p_x}{p_2^-}}{({p_{1}^-})^n +({p_{2}^-})^n}=\frac{z^{n-1}  - (1-z)^{n-1} }{z^n + (1-z)^n }\,p_x
\end{align}
in terms of the kinematic variables in \eq{jet_kin}.
As a consistancy check, one can easily see that $\delta_x$ reduces to the expression for the WTA axis in \eq{deltaxWTA} in the limit $n\to \infty$. With the following replacement 
\begin{align}\label{eq:zhatn}
    \hat{z}\to \hat{z}_n=\bigg|\frac{z^n + (1-z)^n }{z^{n-1}  - (1-z)^{n-1} }\bigg|,
\end{align}
one can straightforwardly evaluate the gluon jet functions for the general recombination scheme by extending the calculation in the previous section. 

\begin{figure}[t!]
\centering
\includegraphics[width=0.4\textwidth]{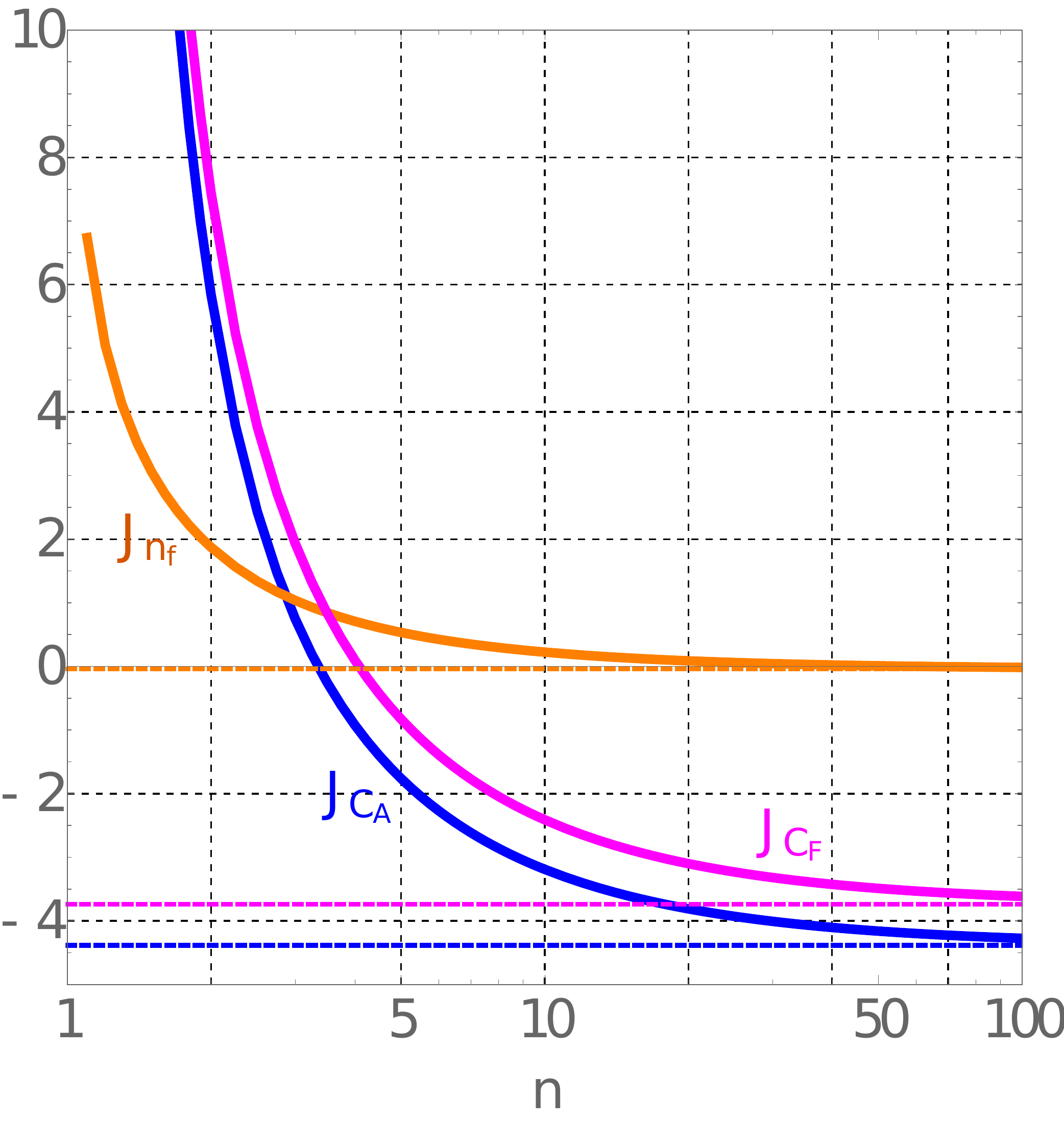}
\caption{
Finite part (i.e.~$L_b=0$) of the contributions from gluons ($J_{C_A}$) and quarks ($J_{n_f}$) to the gluon and to the quark ($J_{C_F}$) jet function for the recombination scheme in \eq{recomb} with $n>1$. The results for the WTA scheme (black dashed lines) are approached for $n \to \infty$. $J_{C_A}$, $J_{n_f}$ and $J_{C_A}$ are defined in \eq{JgTn} and \eq{Jqn}, respectively.  \label{fig:jet_const}}
\end{figure} 
For $\mathscr{J}_{g}^L$, one can easily see that \eq{JgL} is valid for all the values of $n$ because the dependence on $\hat z_n$ drops out in \eq{I2mI1}. $\mathscr{J}_{g}^T$ is given by the first two lines of \eq{JgT} with the replacement for $\hat{z}_n$ in \eq{zhatn}. The $1/\eta$ pole arises only from the following expansions
\begin{align}
    \frac{1}{(1-z)^{1+\eta}}\to -\frac{1}{\eta}\delta(1-z),\text{ and } \frac{1}{z^{1+\eta}}\to -\frac{1}{\eta}\delta(z)
\end{align}
in the integrand of \eq{JgT}. Since
\begin{align}
    I_{1}\bigg(\frac{b_x}{\hat{z}_n}\bigg)\bigg|_{z= \text{0 or 1}}=I_{1}(b_x),
\end{align}
the $\hat{z}_n$-dependence drops out and
the $1/\eta$ term is the same as for the WTA Scheme. Similarly, the only source for the $1/\epsilon$ pole in $\mathscr{J}_{g}^T$ comes from the pole in $I_{1}$ with
\begin{align}\label{eq:I1exp}
    \Bigl(\frac{\mu^2e^{\gamma_E}}{4\pi}\Bigr)^\epsilon I_{1}\bigg(\frac{b_x}{\hat{z}_n}\bigg) = -\frac{1}{4 \pi}\bigg[\frac{1}{\epsilon }+L_b-\ln(\hat{z}_n^2)\bigg] +\mathcal{O}(\epsilon).
\end{align}
Hence, the jet function renormalization in \eq{Zg1} and the anomalous dimensions in \eq{gjetanomdim1} are valid for any value of $n>1$, as required by consistency of the factorization. By using the expansion in \eq{I1exp}, we obtain the constant term of $\mathscr{J}_g^{T}$ (i.e.~taking $L_b=0$) 
\begin{align} \label{eq:JgTn}
  \mathscr{J}_g^{T}\bigr|_{L_b = 0} &= 1+ \frac{\alpha_s}{2\pi}\bigg\{C_A\int_0^1\df z\,\ln(\hat{z}_n^2)\bigg[\frac{z}{1-z}+\frac{1-z}{z}+z(1-z)\bigg]\notag\\
  &\quad +T_F n_f\bigg[\frac{1}{3}+\int_0^1\df z\,\ln(\hat{z}_n^2)((1-z)^2+z^2)\bigg]
  \bigg\}
  \nn\\ &
  \equiv1+\frac{\alpha_s}{4\pi}(C_A J_{C_A}+T_F n_f J_{n_f})
\end{align}
For $n\to \infty$, it reproduces that for the WTA scheme:
\begin{align}
     J_{C_A}|_{n\to\infty} =\frac{131}{18}-\frac{22 \ln 2 }{3}- \frac{2\pi ^2}{3}\,, \quad
    J_{n_f}|_{n\to\infty} = \frac{8 \ln 2 }{3}-\frac{17}{9}.
\end{align}
One can also find analytic results for specific values of $n$. E.g.~for the $p_t^2$ scheme ($n=2$), we have 
\begin{align}
    J_{C_A}|_{n=2} =\frac{58}{9}-\frac{10 \pi }{3}+\pi ^2\,, \quad
    J_{n_f}|_{n=2} = \frac{2 \pi }{3}-\frac{2}{9}
\,.\end{align}
The two constants $J_{C_A}$ and $J_{n_f}$ are plotted as function of $n$ in \fig{jet_const}. The analytic result for the WTA scheme is also shown, and one can see that it is approached in the limit $n\to \infty$. For $n=1$ the jet is no longer recoil-free. This is reflected in a diverging constant in the $n\to 1$ limit, indicating that the poles of the jet function, and thus the entire factorization are different for $n=1$.
Similarly, the finite part of $\mathscr{J}_q$ in general depends on $n$ and takes the form 
\begin{align} \label{eq:Jqn}
    \mathscr{J}_q\bigr|_{L_b=0} =&
    1+ \frac{\alpha_sC_F}{4\pi}\bigg[1+2\int_0^1\df z\,\ln(\hat{z}_n^2)\frac{1+z^2}{1-z}\bigg]
    \equiv 1+ \frac{\alpha_sC_F}{4\pi}J_{C_F}\,.
\end{align}
In \fig{jet_const}, $J_{C_F}$ is also shown.
For the $p_t^2$ scheme and for the WTA scheme,
\begin{align}
   J_{C_F}\bigr|_{n=2} = 7-3 \pi +\pi ^2\,, \qquad
   J_{C_F}\bigr|_{n\to\infty} = 7-\frac{2 \pi ^2}{3}-6 \ln 2\,.
\end{align}

\subsection{Track-based jet function}
\label{subsec:tracks}

Next, we will consider the case where the jet is measured using only charged particles, exploiting the superior angular resolution of the tracking system. Since this does not modify the effect of soft radiation on the measurement, which still contributes through the total recoil, this only affects the jet function. This is similar to the different recombination schemes considered in sec.~\ref{subsec:recoscheme}, and we can reuse much of the calculation, though we will restrict ourselves to the effect of a track-based measurement for WTA scheme. In particular, the jet function anomalous dimensions are not modified but only the constant, which should be contrasted with the complicated jet function encountered for a track-based measurement of thrust in ref.~\cite{Chang:2013iba}. 

We will account for the conversions of the partons to charged particles using
the track function formalism~\cite{Chang:2013rca,Chang:2013iba}. At one-loop
order, the jet consists of (at most) two partons and for $q_x \gg \Lambda_{\rm
QCD}$ they can be treated as fragmenting independently into charged hadrons
moving in the same direction as the original partons. Denoting the total
momentum fractions of charged hadrons produced by each of the two partons by
$z_1$ and $z_2$, the only change due to a track-based measurement is that the
condition which parton ``wins'' gets modified:
\begin{align}
  \hat z = \max\{z,1-z\} \quad \to \quad
  \hat z_{\rm ch} = \left\{ \begin{tabular}{ll} $z$ & $z_1 z>z_2(1-z)$ \\
   $1-z$ & $z_1 z < z_2(1-z)$ \end{tabular} \right.
   \,.\end{align}
We also need to take into account the nonperturbative distribution of $z_1$ (and
$z_2$) which is described by a track function $T_f(z_1,\mu)$, where $f$ is the flavor of the parton.
For example, for the gluon jet function in \eq{Jmunu}, the corresponding track-based measurement (indicated by the bar) is
\begin{align}
    \frac{\al_s}{4\pi}\,\bar {\mathscr{J}}_g^{\mu\nu(1)}&=\int\! \df z_1 \df z_2    
    \int {\rm d} \Pi_{2}\, e^{\img \delta_{x,{\rm ch}} b_x} \Bigl[\sum_q T_q(z_1) T_q(z_2) M_a^{\mu\nu} + T_g(z_1) T_g(z_2)  \frac{1}{2}M_b^{\mu\nu} 
    \nn \\ & \quad
    + T_g(z_1) T_g(z_2) \frac{1}{2}M_{c}^{\mu\nu}\Bigr]
\,,\end{align}
where $|\delta_{x,{\rm ch}}| = |p_x| / \hat z_{\rm ch}$, and the track functions depend on the flavor of the partons in the final state. Note that different quark flavors have different track functions, but $T_q = T_{\bar q}$ due to charge conjugation invariance.

The subsequent steps directly parallel those in sec.~\ref{subsec:recoscheme}, so we find again that the one-loop jet function for the linearly-polarized gluon $\bar {\mathscr{J}}_g^L$ is not modified, while for the transversely-polarized gluon and quark we have
\begin{align} 
  \bar{\mathscr{J}}_g^{T}\bigr|_{L_\perp = 0} &= 1+
   \int\! \df z_1 \df z_2\, 
  \frac{\alpha_s}{2\pi}\bigg\{T_g(z_1) T_g(z_2)\,C_A\int_0^1\df z\,\ln(\hat{z}_{\rm ch}^2)\bigg[\frac{z}{1-z}+\frac{1-z}{z}+z(1-z)\bigg]\notag\\
  &\quad +\sum_q T_q(z_1) T_q(z_2) T_F \bigg[\frac{1}{3}+\int_0^1\df z\,\ln(\hat{z}_{\rm ch}^2)((1-z)^2+z^2)\bigg]
  \bigg\}
  \,,\nn \\
    \bar {\mathscr{J}}_q\bigr|_{L\perp=0} &=
   1+\int\! \df z_1 \df z_2\, T_q(z_1) T_g(z_2)
    \frac{\alpha_sC_F}{4\pi}\bigg[1+2\int_0^1\df z\,\ln(\hat{z}_{\rm ch}^2)\frac{1+z^2}{1-z}\bigg]
\,,\end{align}
in direct analogy to eqs.~\eqref{eq:JgTn} and \eqref{eq:Jqn}.

\section{Resummation and matching}
\label{sec:resum}

\subsection{Renormalization group evolution}
\label{subsec:RGE}

In  section \ref{sec:fact} we have given the derivation of the factorization
formula and the explicit expressions for the one-loop ingredients. Within the
EFT framework one then uses the renormalization group (RG) evolution equations
to resum the large logarithms between different scales. In addition to the
standard UV divergences regularized by the dimensional regularization, the jet,
beam and soft functions also involve rapidity divergences, as these modes are
not separated in invariant mass but rapidity, see \eq{pc}. In order to resum the
corresponding rapidity logarithms we apply the rapidity RG method developed in
ref.~\cite{Chiu:2011qc,Chiu:2012ir}. Different regulator choices (e.g.~\cite{Collins:2011zzd,Chiu:2009yx,Becher:2011dz,Echevarria:2015byo,Li:2016axz,Ebert:2018gsn})
or resummation via the collinear anomaly framework~\cite{Becher:2011pf,Becher:2010tm} are also possible and (up to the possibility of scale variation) equivalent. To achieve
next-to-next-to-leading logarithmic resummation, we need the one-loop fixed order ingredients in sec.~\ref{subsec:oneloop} and the two-loop anomalous dimensions (except for the cusp term in the anomalous dimensions, which is required at three loop order).

Generally, the RG equations for a function $F$ are given by 
\begin{align}\label{eq:RGE}
    \frac{\df}{\df\ln \mu} F(\mu) = \Gamma^F_{\mu} F(\mu), \quad \frac{\df}{\df \ln \nu} F(\mu,\nu)= \Gamma^F_{\nu} F(\mu,\nu).
\end{align}
where $\Gamma_\mu^F$ and $\Gamma_\nu^F$ denote the standard and rapidity
anomalous dimensions, respectively. For the beam, jet and soft function, this
multiplicative form of the evolution equation only holds in impact parameter
space, and the anomalous dimension depends on $b_x$. In addition the anomalous
dimension may e.g.~depend on the hard kinematics or the jet direction, which we omit.

 The anomalous dimension of the hard function are
\begin{equation}\label{Had}
\Gamma_\mu^{{\cal H}_{ij\to Vk}} = \Gamma_{\rm cusp}(\alpha_s) \Big( C_i 
\ln\frac{\hat u^2}{p_{T,V}^2\mu^2} + C_j \ln\frac{\hat
t^2}{p_{T,V}^2\mu^2} +C_k \ln\frac{p_{T,V}^2}{\mu^2}
\Big) + 2\big(\gamma_\mu^{i}+\gamma_\mu^{j}+\gamma_\mu^{k}\big)(\alpha_s)
\,,\end{equation} 
where $C_i$ and $\gamma_\mu^i$ are the color factor and non-cusp anomalous dimension for parton $i$, i.e.~$C_A$ and $\gamma_\mu^g$ for a gluon, and
$C_F$ and $\gamma_\mu^q$ for an (anti-)quark. As we always have (up
to permutation) two quarks and a gluon in our process, the non-cusp anomalous part simplifies 
to $\sum_{a}\gamma_\mu^a=2\gamma_\mu^q+\gamma_\mu^g$.
The expressions for the
partonic Mandelstam variables in terms of the kinematics of the boson and jet are collected in \eq{mandelstam}.
The perturbative expansion for $\Gamma_{\rm cusp}$ and $\gamma^{a}_\mu$ have been collected in appendix~\ref{app:anomdim}.

The anomalous dimensions of the beam functions are 
\begin{align}\label{Bad}
&\Gamma_\mu^{B_i}(\alpha_s) = 2 C_i \Gamma_{\rm cusp}(\alpha_s)
\ln\frac{\nu}{\omega_i} + \gamma_\mu^{B_i}(\alpha_s)\,,\notag \\
&\Gamma_\nu^{B_i}(\alpha_s) = 2 C_i A_{\Gamma_{\rm cusp}}(\mu,\mu_B) - 
C_i \frac{\gamma_{\nu}(\alpha_s)}{2}  
\end{align} 
where $\omega_i$ is explicitly defined in \eq{wLb},
$\mu_B=2/e^{\gamma_E} b_x$ (so $L_b=\ln
\mu^2/\mu_B^2$), and  the cusp anomalous dimension
$\Gamma_{\rm cusp}$, non-cusp anomalous dimension $\gamma^{B_i}_\mu$, and
rapidity anomalous dimension $\gamma_\nu$ in
appendix~\ref{app:anomdim}. The function $A_{\Gamma_{\rm cusp}}$ is obtained by replacing $\ga^i \to \Gamma_{\rm cusp}$ in $A_{\gamma^i}$ in \eq{S_and_A}.
The jet function $\mathscr{J}$ satisfies the same anomalous dimension as the beam function, where now $\omega_J = \bn_J \cdot p_J = 2 p_{T,J} \cosh\eta_J$. 

The anomalous dimensions of the soft function are
\begin{align}\label{Sad}
\Gamma_\mu^{S_{ijk}}(\alpha_s) &= \Gamma_{\rm cusp}(\alpha_s) \biggl[
\left(C_i+C_j+C_k\right)\ln\frac{\mu^2}{\nu^2}  - C_i \ln\frac{\alpha_{ji}\alpha_{ik}}{\alpha_{jk}} - C_j \ln\frac{\alpha_{ij} \alpha_{jk}}{\alpha_{ik}}  \\
& \quad - C_k \ln\frac{\alpha_{ik}\alpha_{kj}}{\alpha_{ij}}\biggr] +
(C_i+C_j+C_k)\gamma_\mu^S(\alpha_s) \nn \\ 
\Gamma_\nu^{S_{ijk}}(\alpha_s) &= -2(C_i\!+\!C_j\!+\!C_k) A_{\Gamma_{\rm cusp}}(\mu,\mu_B)
+ (C_i\!+\!C_j\!+\!C_k) \frac{\gamma_{\nu}(\alpha_s)}{2}\nn
\end{align} 
with
\begin{align}
    \alpha_{ij}\equiv\frac{n_i\cdot n_j}{2}.
\end{align}

From eqs.~\eqref{Had}, \eqref{Bad}, \eqref{Sad} and the non-cusp anomalous dimensions in appendix~\ref{app:anomdim}, it is straightforward to verify that the factorized cross section in \eqref{eq:FactThmFourier} is independent of the renormalization scale $\mu$ and rapidity renormalization scale $\nu$:
\begin{align}
\gamma_\mu^{{\cal H}_{ij\to Vk}}(\alpha_s) + \gamma_\mu^{S_{ijk}}(\alpha_s) + \gamma_\mu^{B_{i}}(\alpha_s) + \gamma_\mu^{B_{j}}(\alpha_s) + \gamma_\mu^{\mathscr{J}_{k}}(\alpha_s) &=0\,,\notag\\
\gamma_\nu^{S_{ijk}}(\alpha_s) + \gamma_\nu^{B_{i}}(\alpha_s) + \gamma_\nu^{B_{j}}(\alpha_s) + \gamma_\nu^{\mathscr{J}_{k}}(\alpha_s) &=0\,. 
\end{align}

\begin{figure}[t!]
\centering
\includegraphics[width=0.35\textwidth]{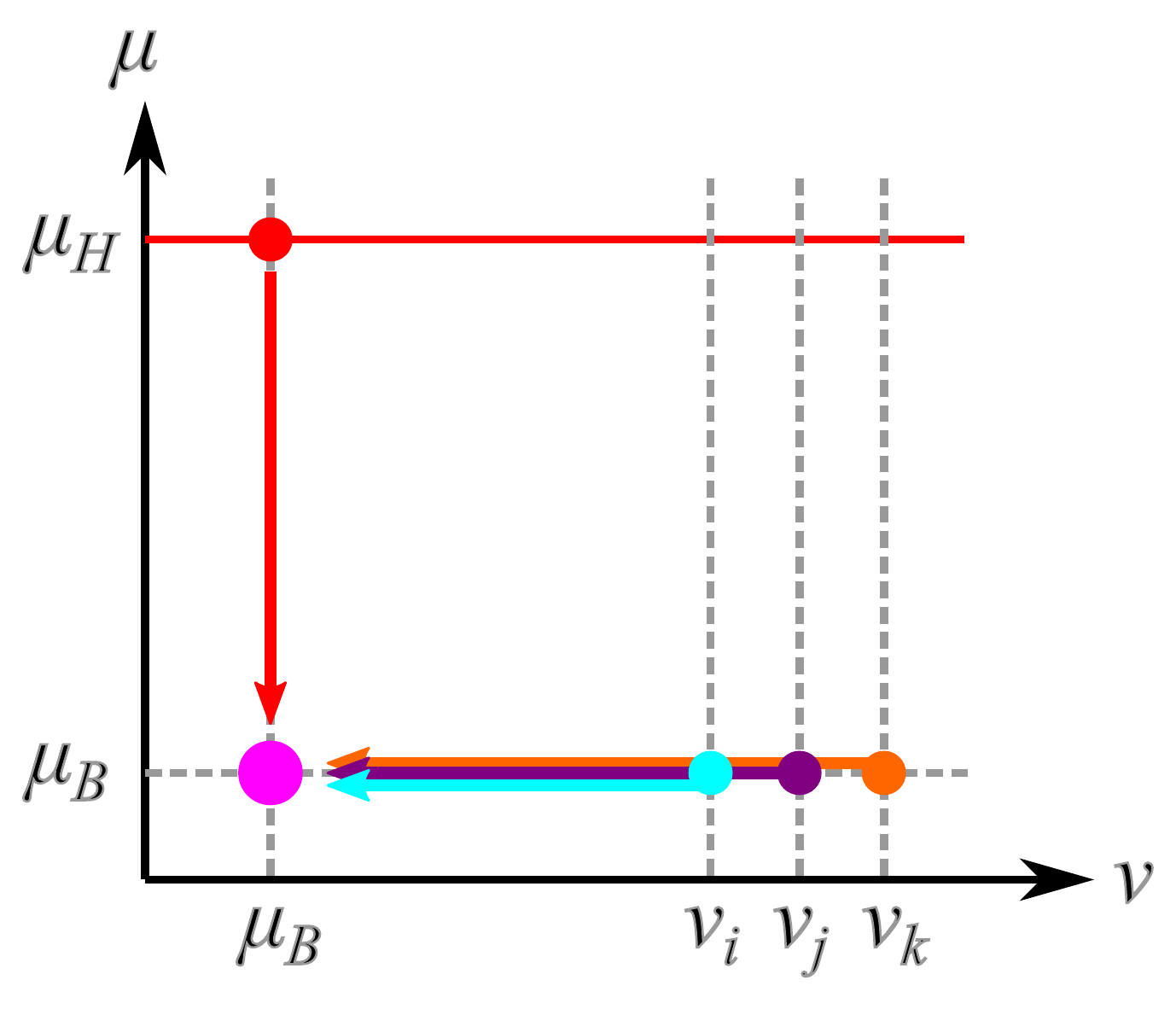}\caption{We evolve all ingredients from their natural $(\mu,\nu)$ scale to the scale of the soft function along the indicated paths. We take the different $\nu_i$ into account though they are of the same parametric size.\label{fig:pathinmunu}}
\end{figure} 

Using the renormalization group equations to evolve all ingredients in \eq{FactThmFourier} from their natural $\mu$ and $\nu$ scales to a common scale, the all-order resummation formula can be written as
\begin{align}\label{eq:resum}
     \frac{\df \sigma_{\text {resum }}}{\df q_x\, \df p_{T,V} \df y_V} &=
     \sum_{i j k} \int_0^\infty \frac{\df b_x}{\pi}\, \cos(b_x q_x) \prod_{a=i j
     k}\Bigl(\frac{\nu_{S}}{\nu_{a}}\Bigr)^{\Gamma_\nu^{B_a}(\mu_B)}
     \exp\biggl(\int^{\mu_B}_{\mu_H} \frac{\df \mu}{\mu} \Gamma_\mu^{\mathcal{H}_{i j} \rightarrow Vk}(\alpha_{s}) \biggr)   \notag\\ & \quad \times \mathcal{H}_{i j \rightarrow k V}(p_{T,V}, y_V - \eta_J,\mu_H) \mathcal{B}_{i}(x_{1}, b_x, \mu_B, \nu_{i}) \mathcal{B}_{j}(x_{2}, b_x, \mu_B, \nu_{j}) \nn \\ & \quad \times \mathscr{J}_{ k}( b_x, \mu_B, \nu_{k}) S_{i j k}(b_x, \mu_B, \nu_{S})\,, 
\end{align}
where we use $i$ to label the parton flavor but also the rapidity scales of the
beams and jets, and understand $\Gamma_\nu^{B_k}$ to refer to the jet function
rapidity anomalous dimension.
We chose to evolve the beam and jet function from their natural rapidity scales $\nu_i$ to the rapidity scale $\nu_S$ of the soft function at the common invariant mass scale $\mu_B$, and evolve the hard function from $\mu_H$ down to the scale $\mu_B$, as summarized in \fig{pathinmunu}.
The evolution factor can be evaluated analytically as
\begin{align}\label{eq:RG-kernel}
 &\exp\biggl(\int_{\mu_H}^{\mu_B} \frac{\df \mu}{\mu} \Gamma_\mu^{\mathcal{H}_{i j} \rightarrow Vk}(\alpha_{s}) \biggr) 
 \nn \\ &\quad
 =\biggl(\frac{\hat{u}^{2}}{p_{T, V}^{2} \mu_H^{2}}\biggr)^{-C_{i} A_{\Gamma_{\rm cusp}} (\mu_H, \mu_B)}\biggl(\frac{\hat{t}^{2}}{p_{T, V}^{2} \mu_H^{2}}\biggr)^{-C_{j} A_{\Gamma_{\rm cusp}}(\mu_H, \mu_B)}
 \biggl(\frac{p_{T, V}^{2}}{\mu_H^{2}}\biggr)^{-C_{k} A_{\Gamma_{\rm cusp}} (\mu_H, \mu_B)} \notag \\ 
& \qquad \times \exp \Bigl[2(C_{i}+C_{j}+C_{k}) S(\mu_H, \mu_B)-2 \sum_{a=i j k} A_{\gamma^{a}}(\mu_H, \mu_B)\Bigr]. 
\end{align}
The functions $A$ and $S$ are given in appendix~\ref{app:anomdim}. 
The natural scales for the various ingredients in the resummation formula are
taken to be
\begin{align}\label{eq:scale}
    \mu_H = \sqrt{m_V^2+p_{T,V}^2}\,,~~~\mu_B = \nu_S = 2 e^{-\ga_E}/b_*,~~~\nu_a = \w_a = \bar n_a\cdot p_a\,,
\end{align}
where we avoid unphysical results in the large-$b$ region, by applying the $b_*$-prescription \cite{Collins:1984kg}
\begin{align}
b_{*}=|b_x|/\sqrt{1+b_x^2/b_{\rm max}^2}
\,,\end{align}
with $b_{\rm max} = 1.5\,{\rm GeV}^{-1} $. 

\begin{figure}
    \centering
    \includegraphics[width=0.48\linewidth]{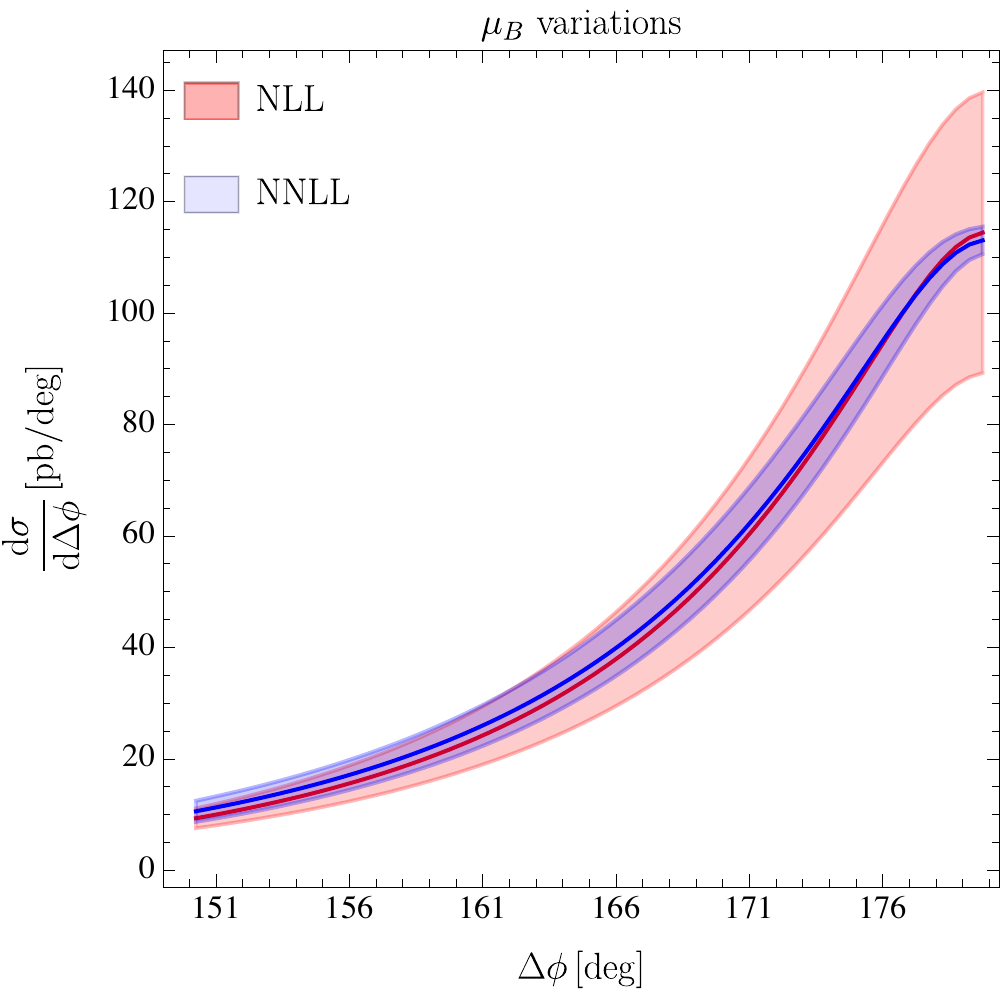}\quad
    \includegraphics[width=0.48\linewidth]{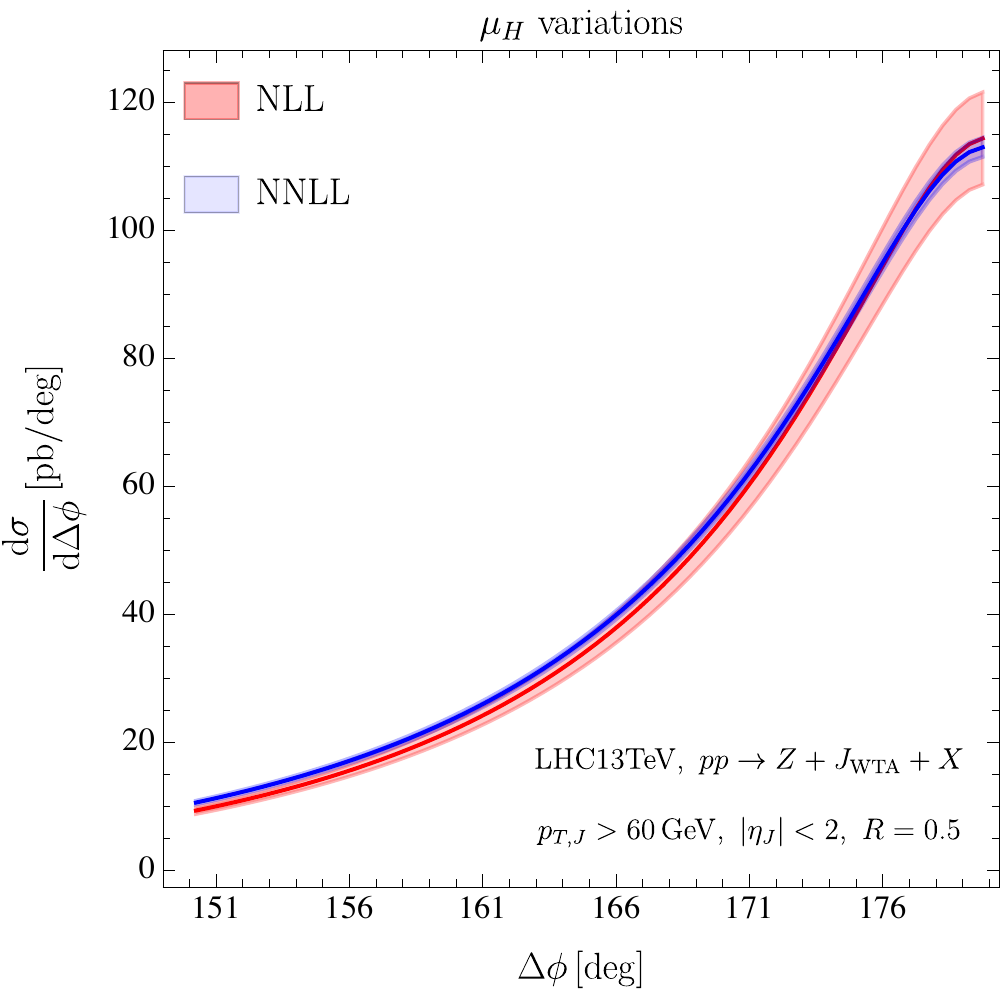}
    \caption{Uncertainties estimated by varying the renormalization scale $\mu_B$ (left) and $\mu_H$ (right) up and down by a factor of 2. The rapidity scale variation is not shown, as it is negligible for our central scale choice for $\mu_B$.}
    \label{fig:scale_muh_mub}
\end{figure}

In \fig{scale_muh_mub} we show the resummation results at NLL and NNLL accuracy, separately displaying the various contributions to the perturbative uncertainties. Specifically, we assess the uncertainties by varying $\mu_H$, $\mu_B$ and $\nu_S$ up and down by a factor of two around their default values in \eqref{eq:scale}. The fact that the uncertainty band is smaller at NNLL than at NLL, and that the bands overlap over almost the entire range, suggest that this is a reasonable estimate. In our full result we will combine these uncertainties by taking the envelope.

\begin{figure}
    \centering
    \includegraphics[width=0.48\linewidth]{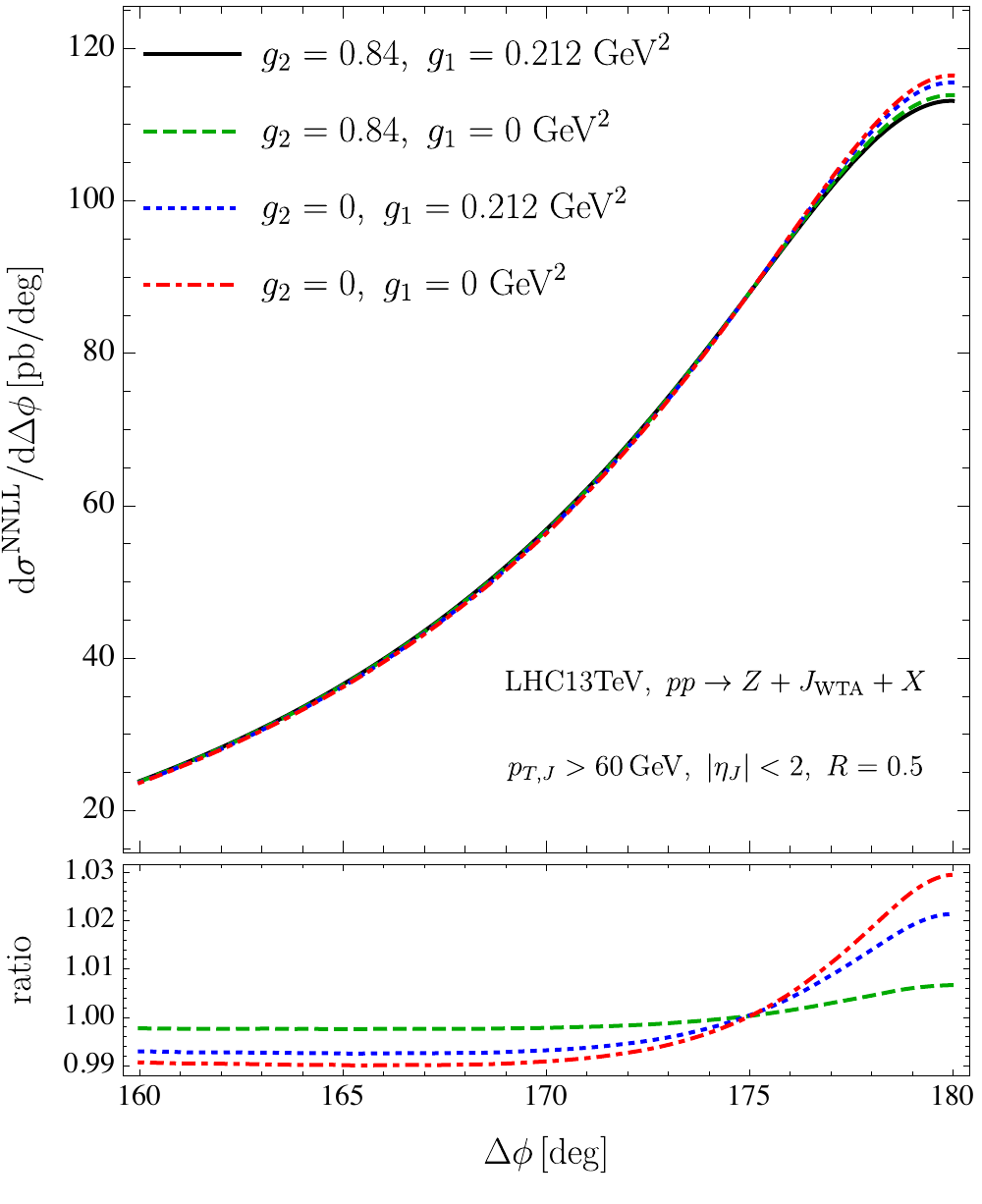}\quad
    \includegraphics[width=0.48\linewidth]{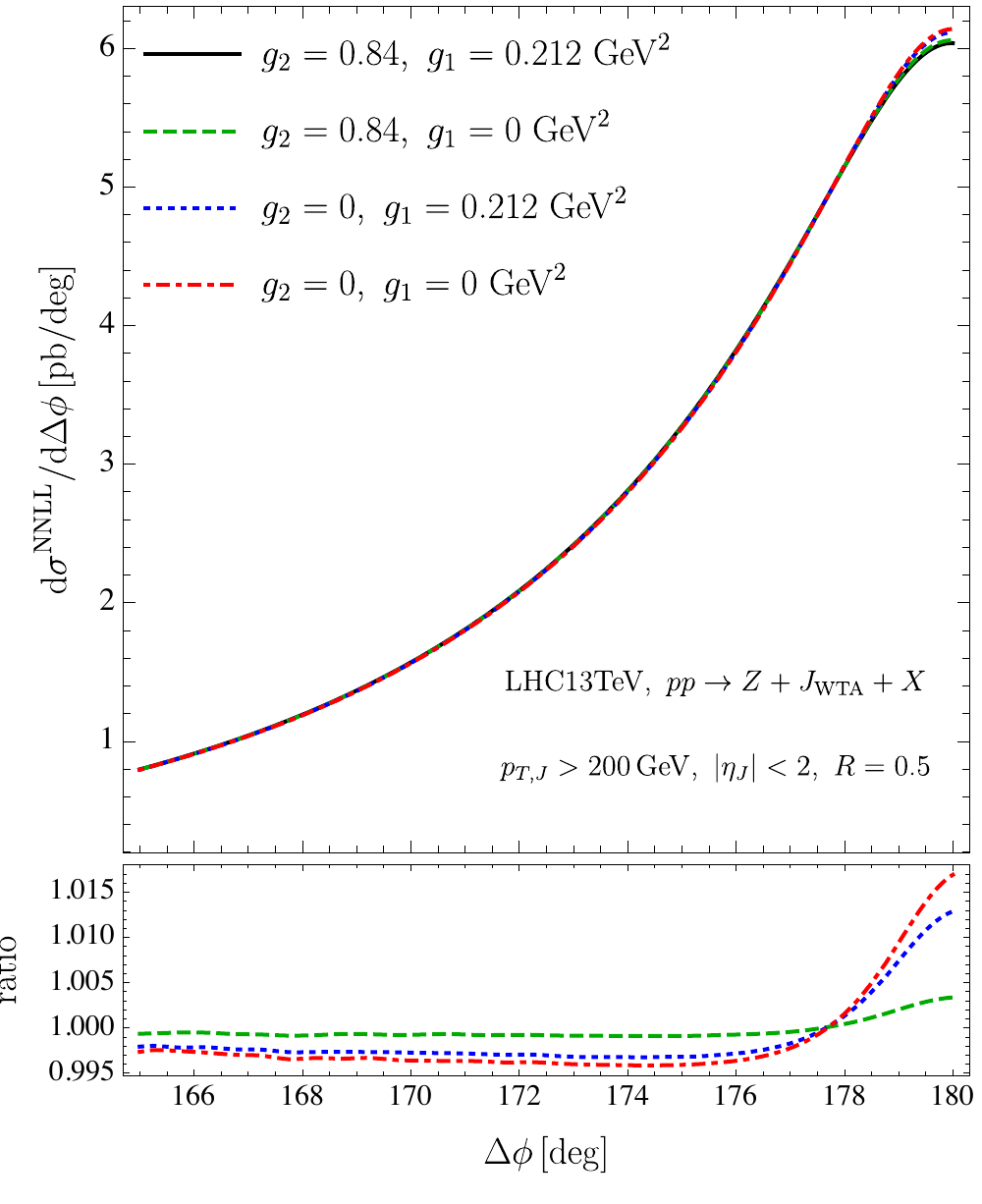}
    \caption{Variations of the nonperturbative parameters $g_1$ and $g_2$ in \eqref{eq:NPfun} for different jet transverse momentum cuts: $p_{T,J}>60\,{\rm GeV}$ (left) and $p_{T,J}>200\,{\rm GeV}$ (right).
    }
    \label{fig:NPfunction}
\end{figure}

To model non-perturbative corrections, we furthermore will include the multiplicative function $e^{-S_{\rm NP}(b)}$ 
\begin{align}\label{eq:NPfun}
    e^{-S_{\rm NP}(b_x)} = e^{-g_1 b_x^2}\prod_{a=ijk}\exp\Bigl(- \frac{C_a}{C_F} \frac{g_2}{2}\ln\frac{|b_x|}{b_*}\ln\frac{\omega_a}{Q_0} \Bigr)\,.
\end{align}
We take the same nonperturbative function $S_{\rm NP}$ as for transverse momentum distributions in~\cite{Su:2014wpa}, assuming that the nonperturbative contribution to the rapidity anomalous dimension (with coefficient $g_2$) can be obtained for a gluon beam or jet function by Casimir scaling. It is not clear whether this should also be the case for the nonperturbative model (with parameter $g_1$) and we take it the same for quark and gluon beams/jets, finding that it has a negligible effect on numerics anyway. We will use the results for the nonperturbative parameters obtained in ref.~\cite{Su:2014wpa}:
$Q_0^2=2.4~{\rm GeV}^2$, $g_1=0.212~{\rm GeV}^2$ and $g_2=0.84$. 
The sensitivity of our predictions to these nonperturbative parameters is explored in \fig{NPfunction}, finding minimal sensitivity to $g_1$ but sensitivity to $g_2$ at the percent level in the region of $\Delta\phi \sim \pi$.

\subsection{Matching to fixed-order MCFM}
\label{subsec:fixed}

In the back-to-back region where $\delta \phi \to 0$ ($\Delta \phi \to \pi$), the resummation formula will cure the divergence behavior of fixed-order results. However, if $\delta \phi$ is not small, the factorization formula receives large corrections of powers of $\delta \phi$. In this region, the resummation should be switched off, since $\ln \delta \phi$ is no longer large, and we therefore need to use fixed-order calculations that include these power corrections.

We use an additive matching scheme, in which the ``naive" matched result of 
NNLL resummed prediction and the fixed-order can be obtained by the following relation
\begin{align} \label{eq:naive}
    \df \sigma_{\rm naive}(\mathrm{NLO}+\mathrm{NNLL})=\df \sigma(\mathrm{NNLL})+ \underbrace{\df\sigma(\mathrm{NLO})-\df \sigma(\mathrm{NLO} \text { singular})}_{{\rm d}\sigma({\rm NLO}~{\rm non-singular})},
\end{align}
where we use \MCFM~\cite{Campbell:2002tg,Campbell:2003hd} to calculate the NLO results. The NLO singular distribution removes the overlap between the first two terms and can be obtained by expanding the resummation formula \eqref{eq:resum} to $\mathcal{O}(\alpha_s)$. The NLO non-singular distribution is given by the difference between NLO and NLO singular results, as indicated in the above definition. 

\begin{figure}
    \centering
    \includegraphics[width=0.95\linewidth]{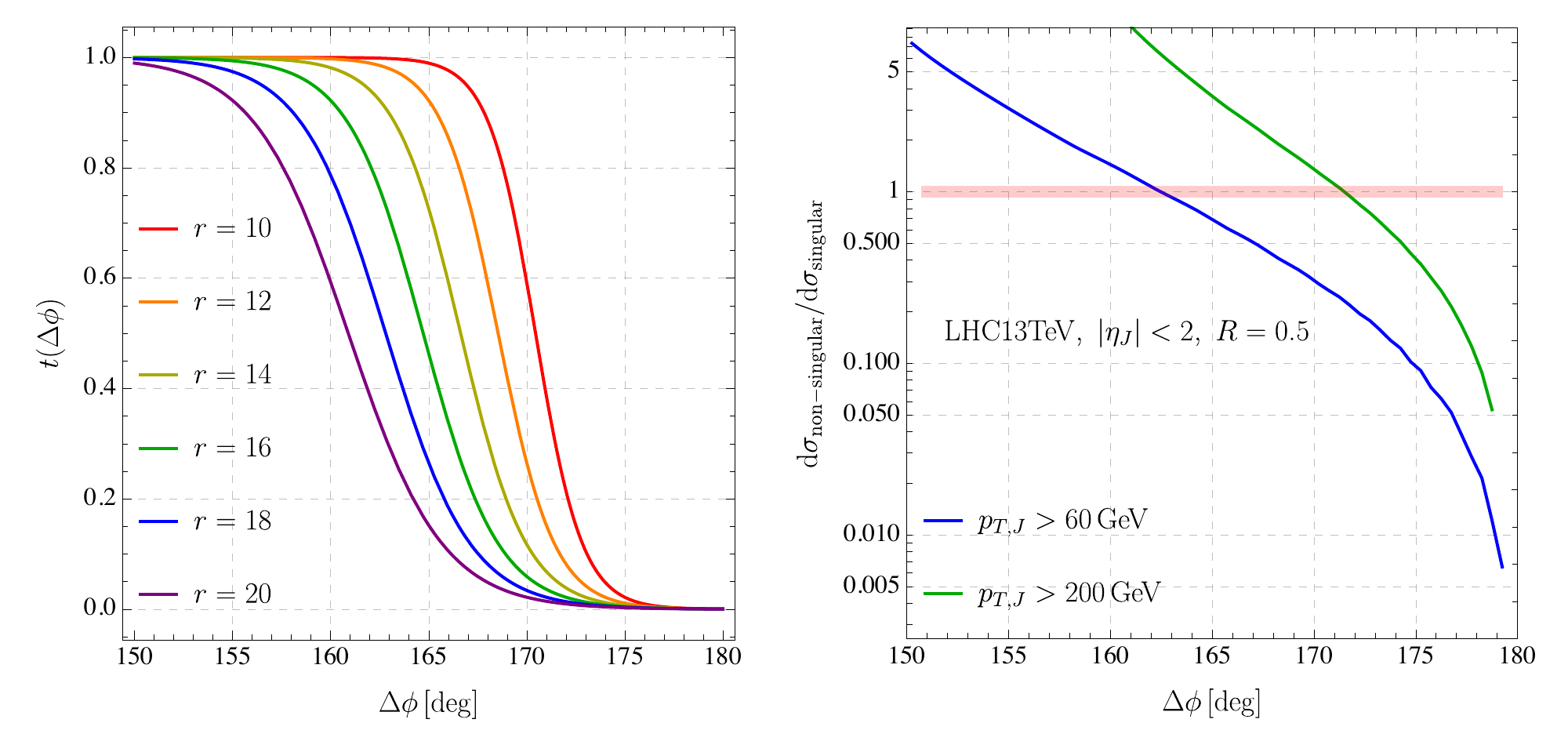}
    \caption{
     Left: the transition function defined in \eqref{eq:transfun} with transition points at approximately $180 - r$ degrees with $r=10$, $12$, $14$, $16$, $18$ and $20$. Right: the NLO non-singular divided by the NLO singular, indicating the appropriate choice of transition point.}
    \label{fig:transfun}
\end{figure}

\begin{figure}[t]
    \centering
    \includegraphics[width=0.45\linewidth]{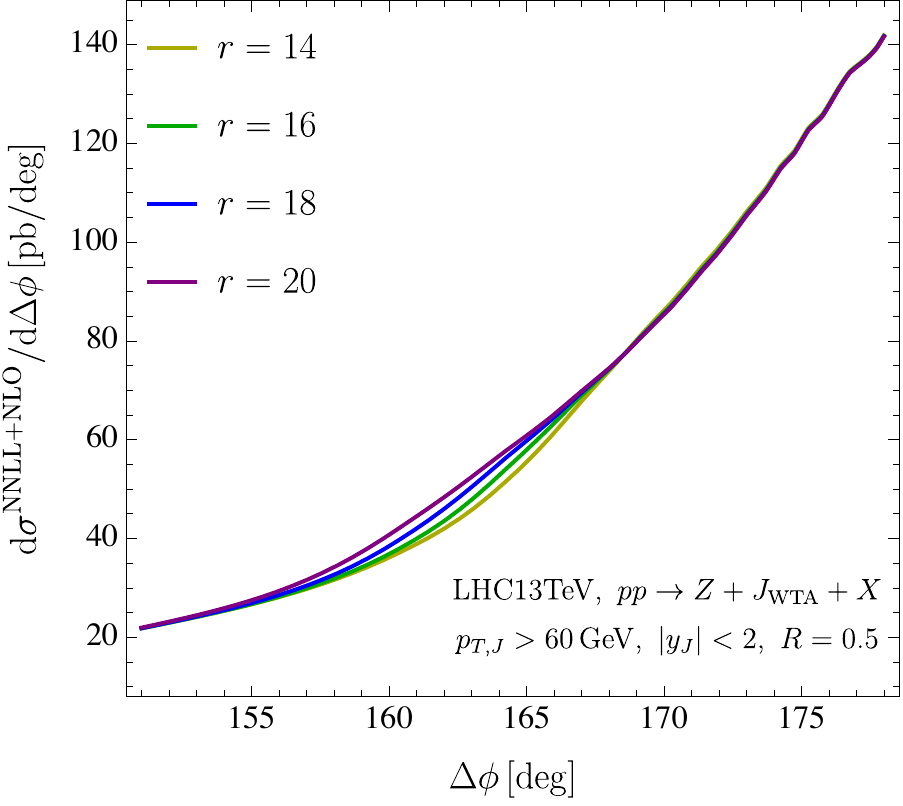} \quad 
    \includegraphics[width=0.435\linewidth]{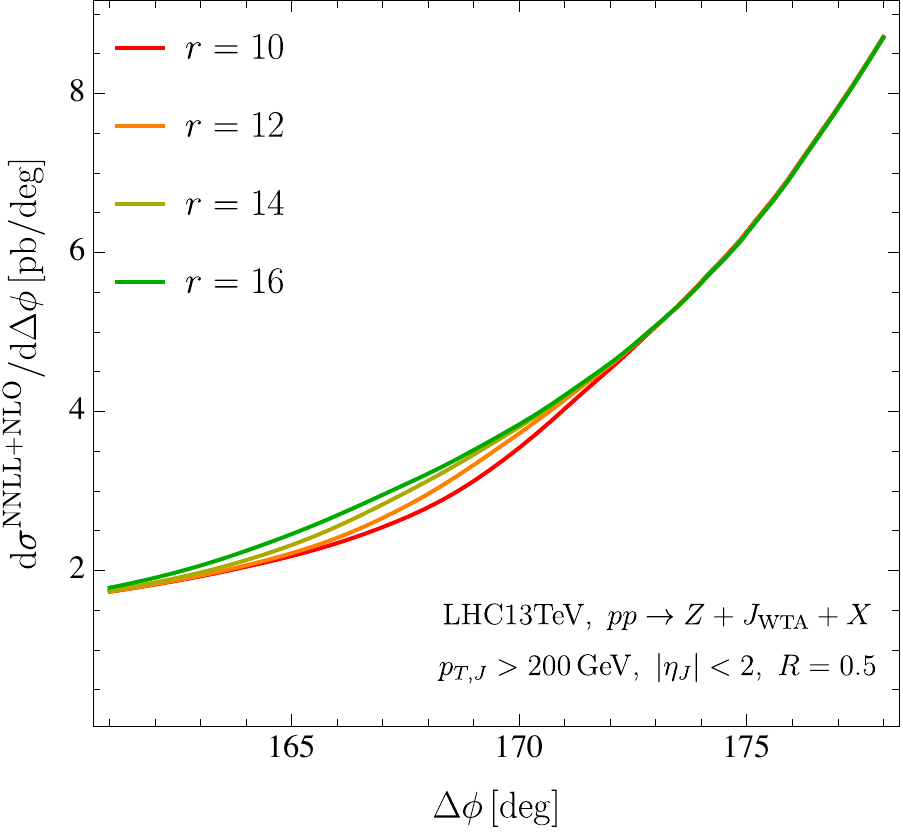}  
    \caption{Varying the transition point in matching our NNLL resummation to the NLO for $p_{T,J}>60$ GeV (left) and 200 GeV (right).}
    \label{fig:nonsing}
\end{figure}

In principle, as $\de \phi$ becomes large, the NNLL resumed cross section reduces to the NLO singular, leading to a cancellation between the first and third term in \eq{naive}. 
However, as is clear from \eqref{eq:resum}, the numerical Fourier transformation from $b$-space to the momentum space rapidly oscillates when the resummation turns off and the evolution factor approaches $1$. To avoid the corresponding numerical instability, we apply a transition function $t(\Delta \phi)$ as follows:
\begin{align}\label{eq:transfun}
    \df \sigma(\mathrm{NLO}+\mathrm{NNLL}) = [1-t(\Delta \phi)] \df \sigma_{\rm naive}(\mathrm{NLO}+\mathrm{NNLL}) + t(\Delta \phi) \df \sigma(\mathrm{NLO})
\,\end{align}
The transition function we use is defined as
\begin{align}
    t(\Delta \phi) = \frac{1}{2} - \frac{1}{2} \tanh \Bigl[ 4 - \frac{240(\pi-\Delta\phi)}{r}  \Bigr],
\end{align}
where the parameter $r$ fixes the transition point to be at approximately $180 - r$ degrees. Different choices of $r$ are illustrated in the left panel of \fig{transfun}. In the right panel we show the difference of the NLO and NLO singular cross section, divided by the NLO singular. This indicates that the power corrections to our factorization theorem are order one around $\Delta \phi$ of $160^{\rm o}$ and $170^{\rm o}$ for $p_{T,J}>60$ GeV (blue curves) and $p_{T,J}>200$ GeV (green curves), respectively, which leads us to choose $r = 20$ and $10$ in these two kinematic regions. The dependence on the choice of transition point $r$  is shown in \fig{nonsing}. The nonsingular is in principle much smaller than the singular, but because the singular is resummed, there is less of a difference between them. Therefore we can see a sizable effect on how the nonsingular correction is treated in the resummation region, particularly when the jet $p_T$ is large. We will discuss the origin of this large nonsingular correction in sec.~\ref{subsec:analytic}, arguing that it should not be Sudakov suppressed in the back-to-back limit, which is why we use an additive rather than multiplicative matching.

We conclude this section by presenting expressions for the NLO singular at the level of the integrated cross section, 
\begin{align}
    \Sigma_{\rm singular}(\delta\phi^{\rm cut}) = \int_0^{\delta\phi^{\rm cut}}\!\! \df(\delta\phi) \frac{\df \sigma}{\df(\delta\phi)\, \df p_{T,V} \df y_V}
\,.\end{align}
At order $\alpha_s$ we have
\begin{align}\label{eq:sing_int}
    \Sigma^{(1)}_{\rm singular}(\delta \phi^{\rm cut}) &= \sum_{ab} \int_{x_1}^1 \frac{\df z_1}{z_1} f_{a/p}(x_1/z_1,\mu)\int_{x_2}^1 \frac{\df z_2}{z_2} f_{b/p}(x_2/z_2,\mu) 
    \nn \\ & \quad \times
    \sum_{ijk} \Bigl(
     \mathcal{H}_{ij\to Vk}^{(0)} C_{ij\leftarrow ab}^{(1)} +
 \mathcal{H}_{ij\to Vk}^{L,(0)}  C_{ij\leftarrow ab}^{L,(1)}\Bigr)\,.
\end{align}
The first term on the second line is the contribution from the unpolarized gluon beam/jet functions, while the second term is the linearly-polarized contribution, as indicated by the superscript $L$. The one-loop coefficients are given by 
\begin{align}
    C^{(1)}_{ij\leftarrow ab} &= A_{ij} \delta_{ia}\delta_{jb}\delta(1-z_1)\delta(1-z_2) + \delta(1-z_2)\delta_{jb} \left[ 4P_{ia}(z_1) \left(L - \ln\frac{\mu}{2p_T}\right) + R_{i a}(z_1)\right] \notag \\
    &\quad + \delta(1-z_1)\delta_{ia} \left[ 4P_{jb}(z_2) \left(L - \ln\frac{\mu}{2p_T}\right) + R_{j b}(z_2)\right]  \\
    C^{(1),L}_{ij\leftarrow ab} &= A^L_{ij} \delta_{ia}\delta_{jb}\delta(1-z_1)\delta(1-z_2) + \delta(1-z_2)\delta_{jb}   L_{i a}(z_1) + \delta(1-z_1)\delta_{ia}  L_{j b}(z_2),
\nn \end{align}
where $L=\ln\delta\phi^{\rm cut}$, the splitting functions are given in \eq{splitfun}, and 
\begin{align}
 R_{q q}(z) &= 2C_F(1-z),~~ R_{ g g}(z)=0,~~R_{ q g}(z)=2z(1-z),~~ R_{ g q}(z)=2C_F z,
 \nn 
\\
 & L_{g g}(z) = - C_A \frac{4(1-z)}{z},~~~ L_{g q}(z) = - C_F \frac{4(1-z)}{z} 
\,.\end{align}
For the different partonic channels, the coefficients $A_{ij}$ are given by
\begin{align}
    A_{q\bar q} &= C_F \left[ -8L^2 + L\left( -12 + 8\ln\frac{\hat s}{4p_T^2} \right)  - \pi^2 \right] + C_A \left( - 4L^2 - 8\ln 2 L + \frac{25}{12} - \frac{7\pi^2}{6}\right) \notag \\
    & \quad + \beta_0 \left( - 2L + 2\ln\frac{\mu}{4p_T} + \frac{17}{12}\right)+ H^{(1)}_{q\bar q\to gV}(\mu_h=2p_T) \,, \nn \\
     A_{qg} &= C_F \left[ -8L^2 + L\left( -12 + 8\ln\frac{-\hat u}{4p_T^2} \right) + 7  - \frac{5\pi^2}{3} - 6\ln 2 \right] 
     \nn \\ & \quad
     + C_A \left( - 4L^2 +4 L \ln\frac{\hat s \hat t}{4p_T^2\hat u}  -\frac{\pi^2}{2}\right) 
    + \beta_0 \left( - 2L + 2\ln\frac{\mu}{2p_T} \right)+H^{(1)}_{q g\to qV}(\mu_h=2p_T), \notag \\
    A^{L}_{q\bar q} &=- \frac{C_A}{3} + \frac{2T_F n_f}{3} , ~~~ A^{L}_{qg}=0
\,,\end{align}
where the partonic Mandelstam variables are given in terms of the kinematics of the hard scattering in \eq{mandelstam}.

\section{Results}
\label{sec:result}

We start in sec.~\ref{subsec:MC} with a detailed study of the azimuthal angle between a recoil-free jet and $Z$ boson using the \Pythia Monte Carlo parton shower. This allows us to investigate the effect of hadronization and underlying event, and corroborate conclusions of our factorization analysis with regards to the dependence on the jet radius, recombination scheme and track-based measurements. Our resummed predictions are shown in sec.~\ref{subsec:analytic}. We also explain sizable non-singular corrections, particularly at large $p_{T,J}$, which may be largely removed by boson isolation cuts.

\subsection{Monte Carlo analysis}
\label{subsec:MC}

In this subsection we present a phenomenological study of recoil-free boson-jet correlation using the \Pythia 8.3 \cite{Sjostrand:2014zea} Monte Carlo parton shower. The $Z$+jet events in 13 TeV proton-proton collisions at the LHC are simulated with the decay of the $Z$ boson turned off. In experiments the clean, leptonic decay channels of $Z$ boson are reconstructed with suitable cuts on the lepton kinematics. In these Monte Carlo studies we sum over all the $Z$ boson polarization states to match our analytic calculation.

In all events, jets are reconstructed using the anti-$k_t$ algorithm \cite{Cacciari:2008gp} with $R = 0.5$ (also $R = 0.8$ or $R = 1.0$ when studying the jet radius dependence) using \textsc{FastJet} 3~\cite{Cacciari:2011ma} and $|\eta_J|<2$. 
 The azimuthal angle is defined as the one between the $Z$ boson and the leading jet in each event\footnote{Note that in some studies this angle is instead defined for an inclusive jet sample \cite{Sirunyan:2017jic, ATLAS:2018dgb}. In our factorization analysis the contribution from additional jets is power suppressed, assuming $\de \phi \ll R$.}. We consider two kinematic regions: $p_{T,J}>60$ GeV and $p_{T,J}>200$ GeV, to study the dependence of this observable on the hard energy scale. Two million events for each region are simulated, providing sufficient statistics to obtain smooth distributions.

\begin{figure}
    \centering
    \includegraphics[height=0.5\linewidth]{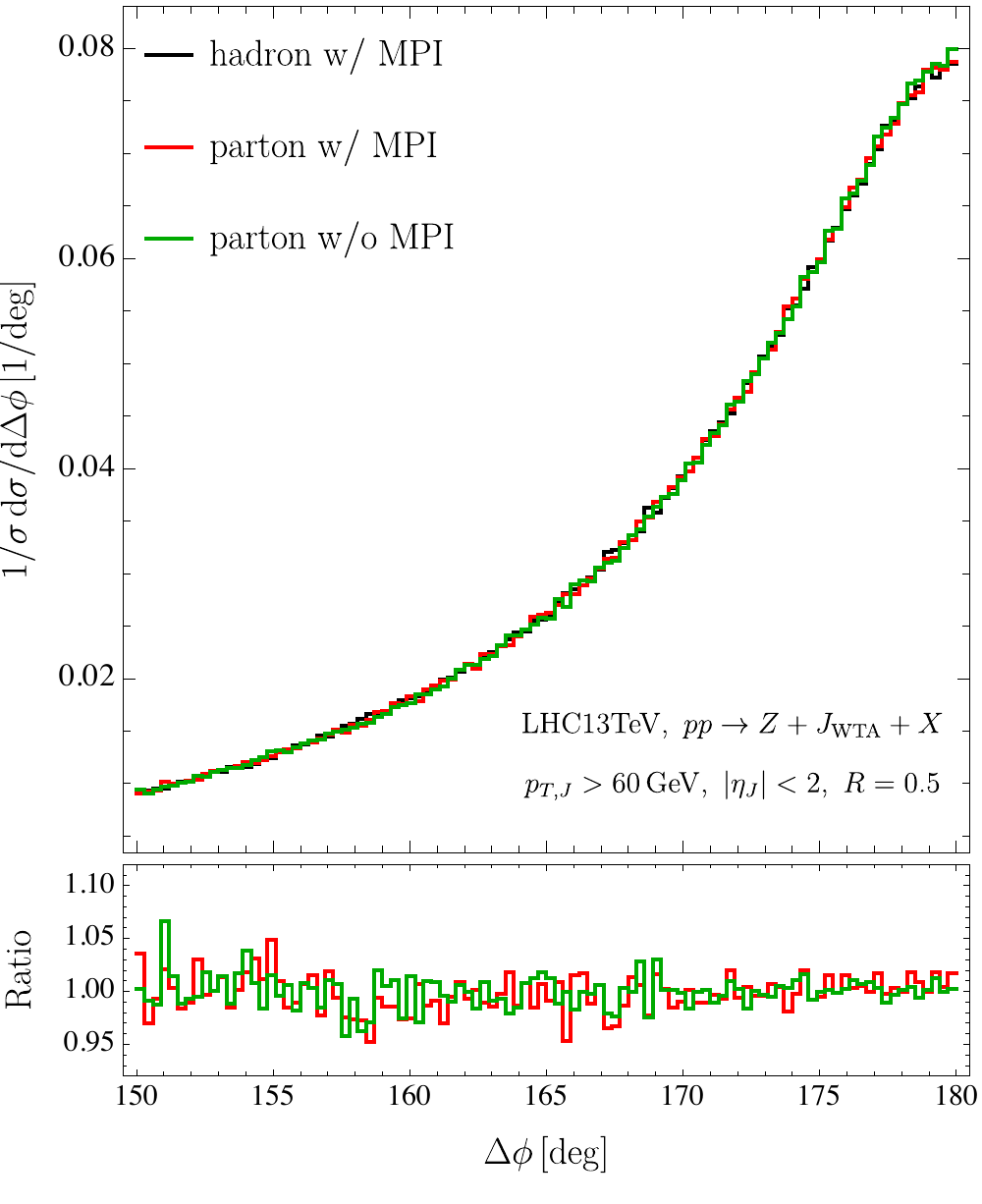}~~~~~~
    \includegraphics[height=0.5\linewidth]{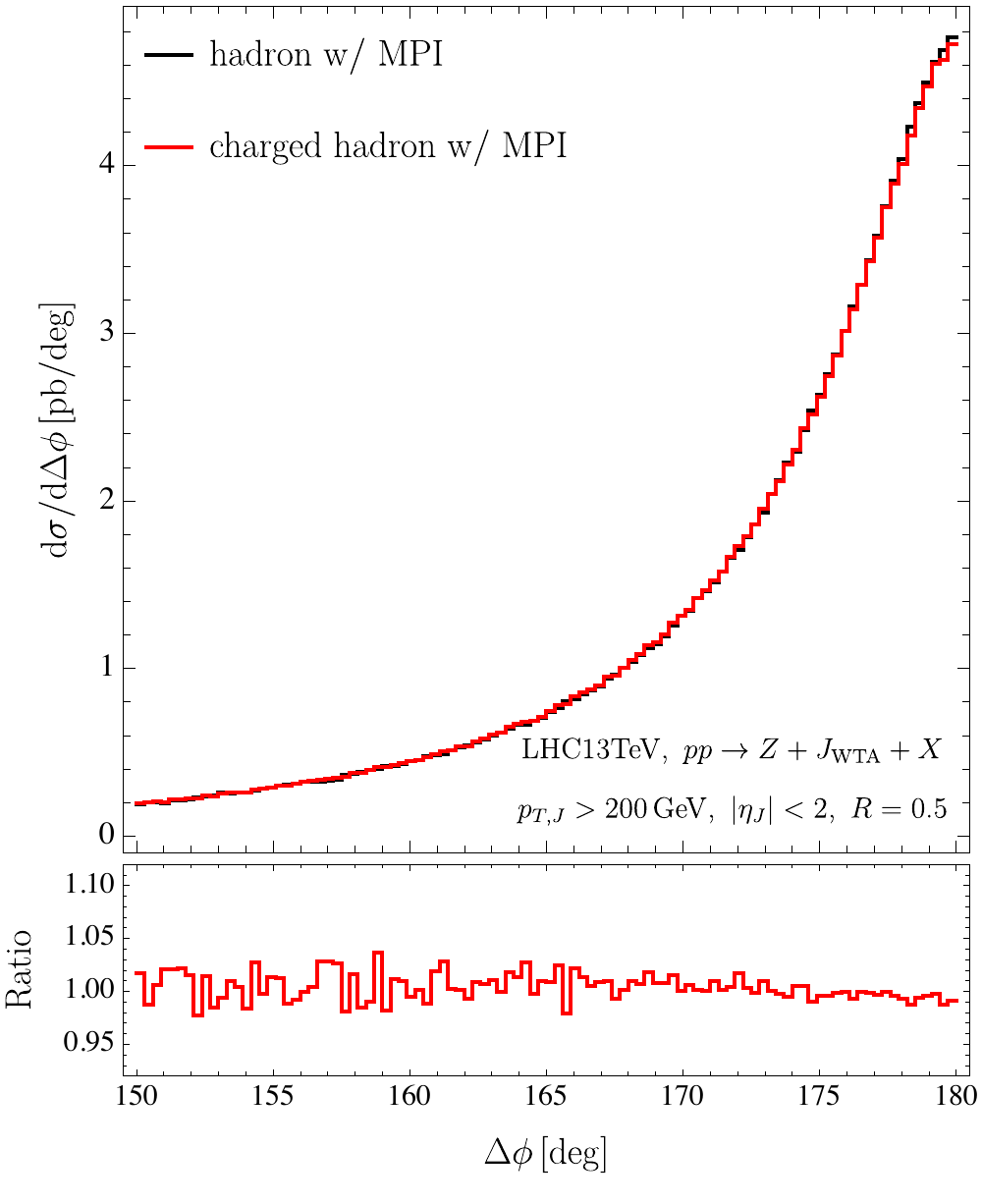}
    \caption{Normalized $\Delta \phi$ distribution for $Z$+jet in \Pythia. Left: at the parton level with or without MPI contributions, as well as at the hadron level (including MPI). Right: at the hadron level using all or only charged particles for WTA axis definition.}
    \label{fig:PHtrack}
\end{figure}

We first examine the sensitivity of the azimuthal decorrelation to hadronization and underlying event in the left panel of \fig{PHtrack}, which  shows the $\Delta \phi$ distributions with or without hadronization or underlying event contributions. In \Pythia the underlying event is modeled as multi-parton interactions (MPI). We see that the shape of the $\Delta \phi$ distribution is remarkably insensitive to hadronization and MPI, which suggests that it is dominated by perturbative contributions. This is expected: due to our recoil-free jet definition, these soft contributions do not interfere with the jet finding, and only provide a total recoil of the $V+$jet system. Since the azimuthal angle is a vector quantity, the net effect of this recoil tends to be (close to) zero. There is a change in the normalization of the absolute cross section, because the additional radiation affects the number of events having a jet with sufficient transverse momentum.

In order to exploit the high angular resolution of charged particle tracking, we study the case where the recoil-free axis is determined only by charged tracks, and the difference compared to using all charged and neutral particles is examined. This is shown in the right panel of \fig{PHtrack}. As can be seen, the azimuthal decorrelation distributions using the recoil-free axis determined by charged particles or all the particles within jets are almost identical, in line with the our conclusion in sec.~\ref{subsec:tracks} that this difference is beyond NLL accuracy. 

\begin{figure}[t]
    \centering
    \includegraphics[width=0.45\linewidth]{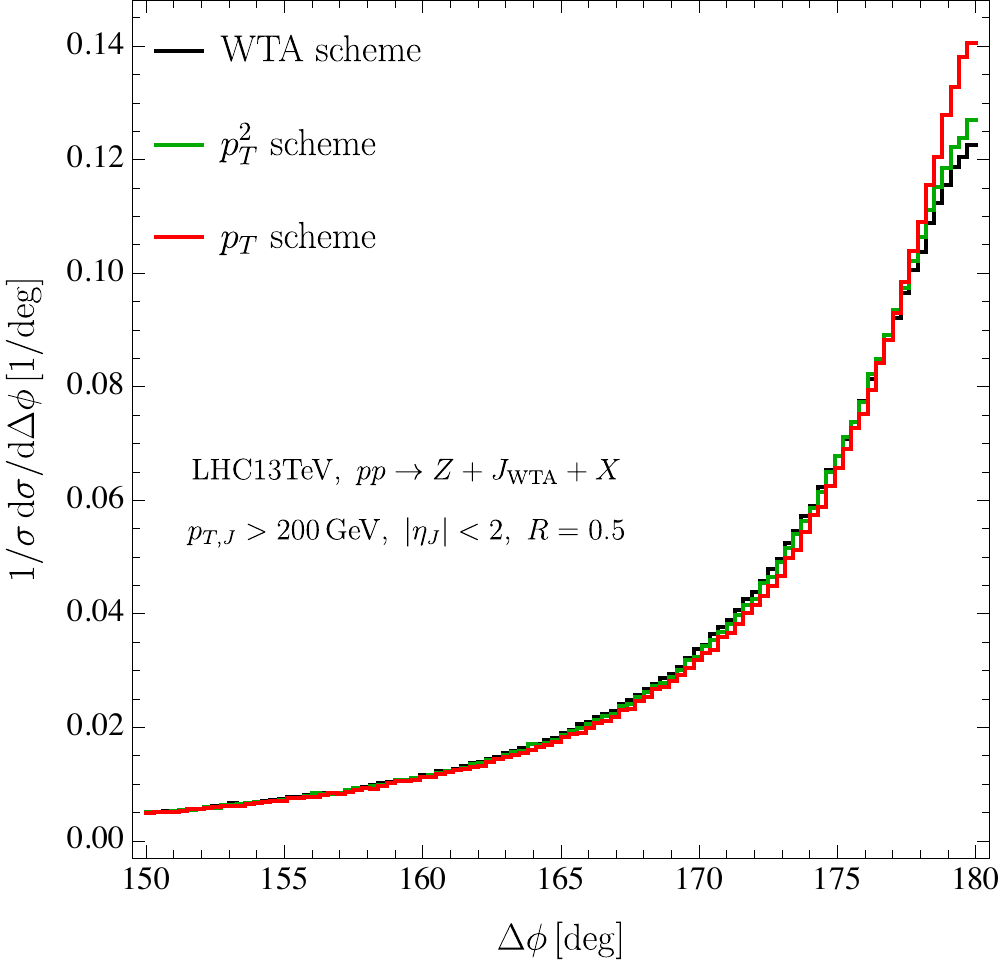}~~~~~~
    \includegraphics[width=0.45\linewidth]{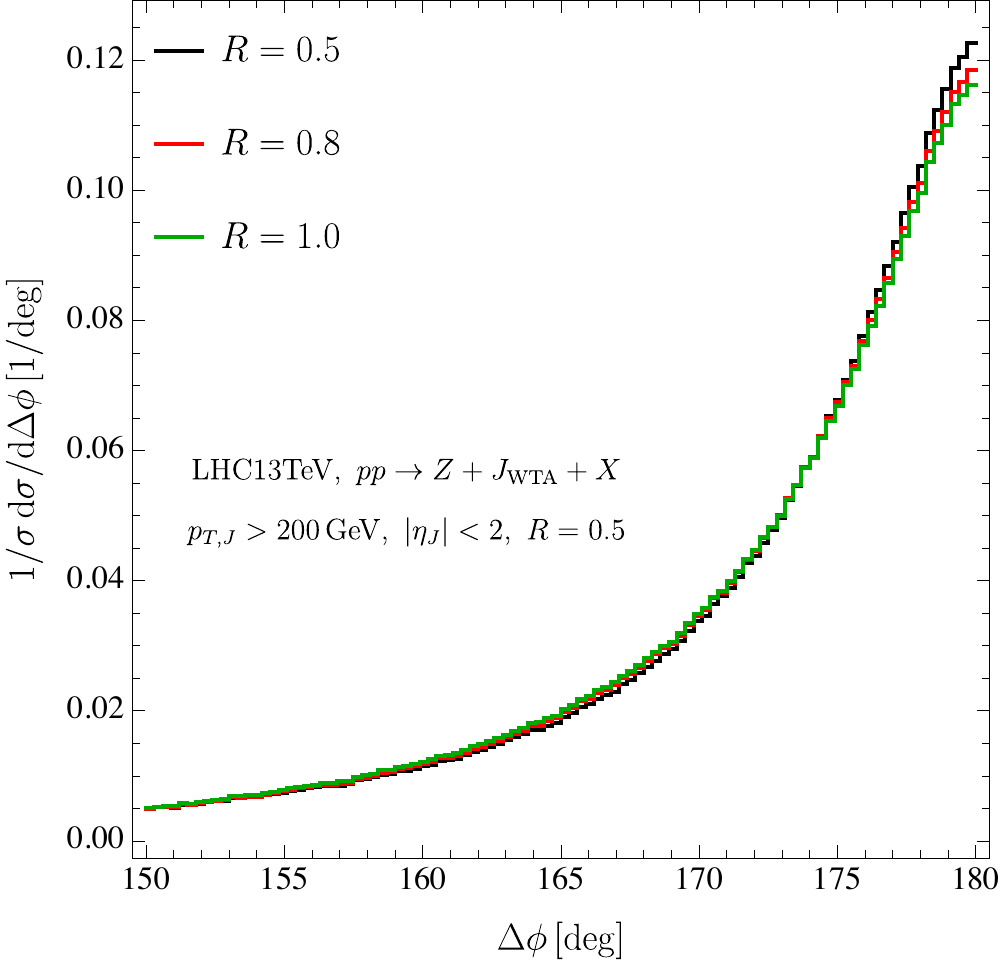}
    \caption{Normalized $\Delta \phi$ distribution for $Z$+jet in \Pythia. Left: for WTA, $p_T^2$ and $p_T$ recombination schemes. Right: for WTA with jet radii $R = 0.5, 0.8$ and $1.0$. Our factorization predicts that the cross section is independent of $R$ if $\delta \phi \ll R$.}
    \label{fig:schemeR}
\end{figure}

To contrast the WTA axis choice, we provide distributions for jets defined using the more general $p_t^n$ recombination scheme in \eq{recomb}. In the left panel of \fig{schemeR} we examine the case where $n=1$ (the standard), $n=2$ and $n\rightarrow \infty$ (WTA). Since the WTA axis is sensitive to momentum-conserving collinear splittings within jets and following the energetic branch, the $\Delta \phi$ distribution is broader near the back-to-back region. The difference between the recoil-free $n=2$ and WTA axis is beyond NLL order, as discussed in sec.~\ref{subsec:recoscheme}, and is indeed small. The difference with the recoil-sensitive case $n=1$ is larger.

We also compare the distributions for jets with different radii as a way to highlight again the insensitivity to soft, typically wide angle, radiation, in the limit where the jet radii are much larger than the azimuthal decorrelation, $\delta \phi \ll R$. The right panel of \fig{schemeR} shows the $\Delta \phi$ distributions for jets reconstructed using different jet radii, namely $R=0.5$, $R=0.8$ and $R=1.0$. The shapes of the distributions are very similar as expected from our factorization analysis. The differences for $\Delta \phi$ in the vicinity of $180^\circ$ arise because these distributions are normalized. However, some normalization is necessary when comparing different jet radii, because the jet radius has a non-negligible (5-10\%) effect on the normalization of the cross section through the cut on the jet transverse momentum.

\subsection{Resummed predictions}
\label{subsec:analytic}

In this section we present numerical results from the resummation formula in \eq{resum}, which we match to \MCFM~\cite{Campbell:2002tg,Campbell:2003hd} using the procedure described in sec.~\ref{subsec:fixed}. 
The electroweak parameters are given by
\begin{align}
    m_Z=91.1876~{\rm GeV},~~~\alpha_{em}=1/132.34\,,~~~ \cos\theta_W=0.88168\,,
\end{align}
and we use CT14nlo \cite{Dulat:2015mca} with $\alpha_s(m_Z)=0.118$ for the collinear PDFs. 

\begin{figure}[t]
    \centering
    \includegraphics[width=0.44\linewidth]{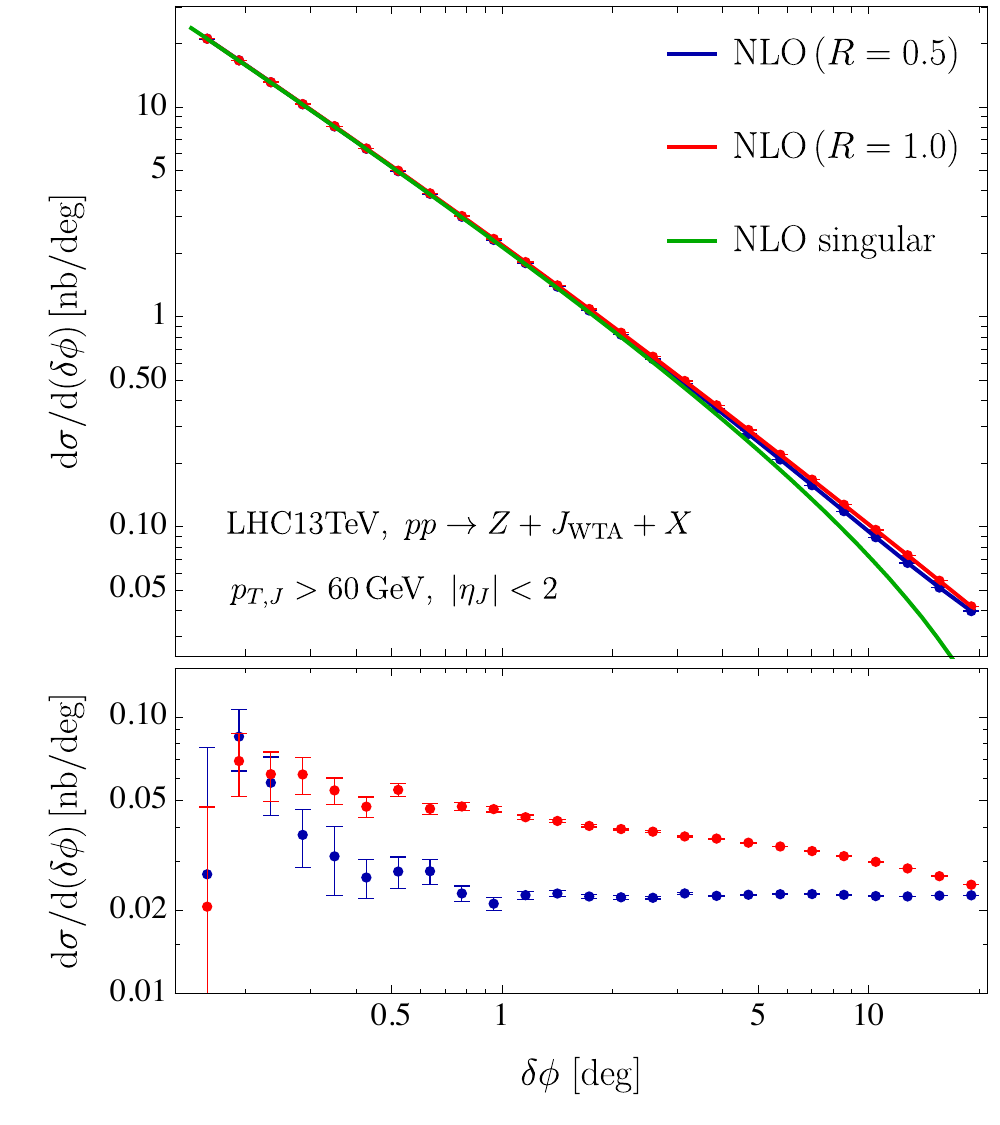} \quad 
    \includegraphics[width=0.44\linewidth]{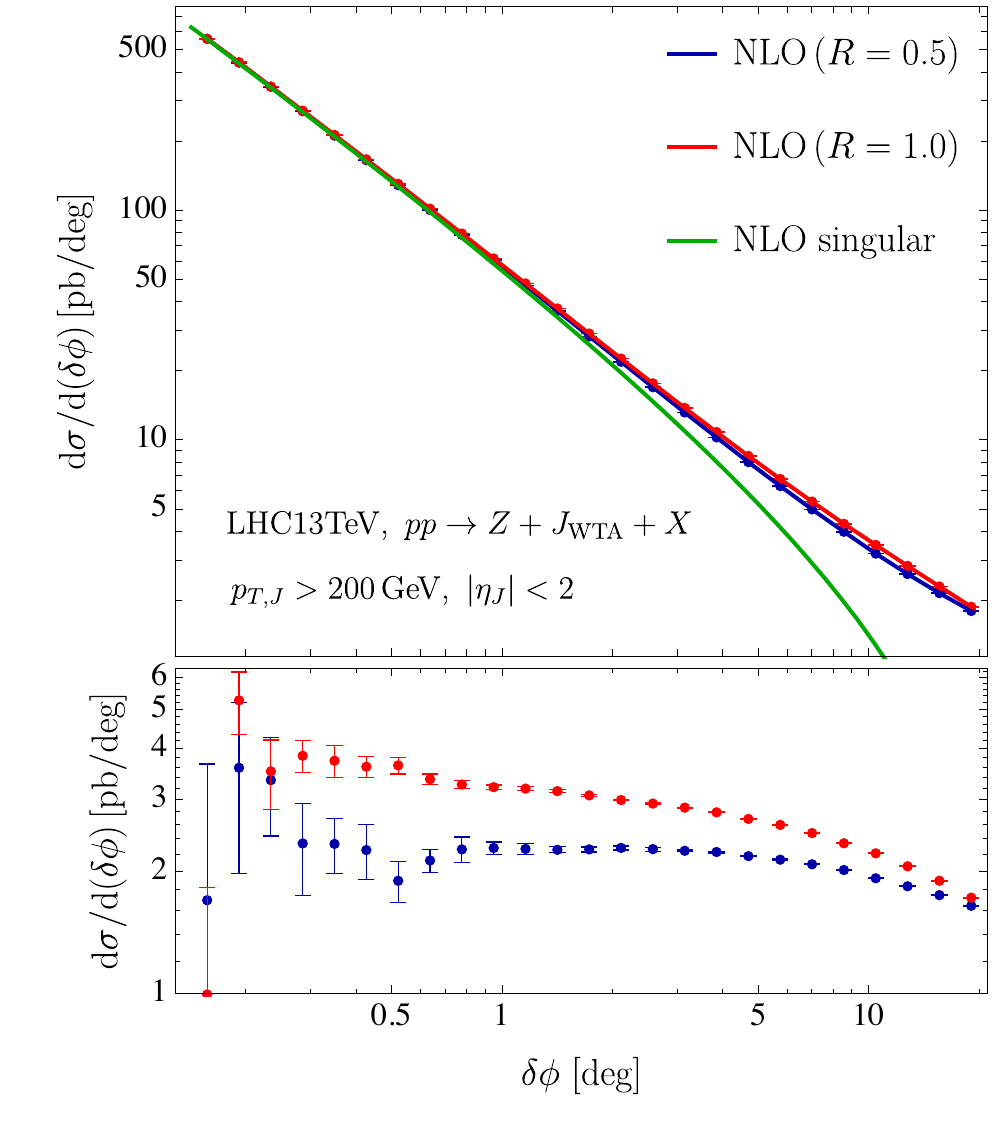}
    \caption{The $\delta\phi$ differential distributions at NLO order for jet transverse momentum $p_{T,J}>60\,{\rm GeV}$ (left) and $p_{T,J}>200\,{\rm GeV}$ (right) with jet radius $R=0.5$ (blue), $1.0$ (red) and the $R$-independent singular terms (green). The non-singular cross section (difference of the NLO and NLO singular) is shown in the lower panels. 
    }
    \label{fig:nonsing-vsnlo}
\end{figure}

\begin{figure}[t]
    \centering
    \includegraphics[width=0.9\linewidth]{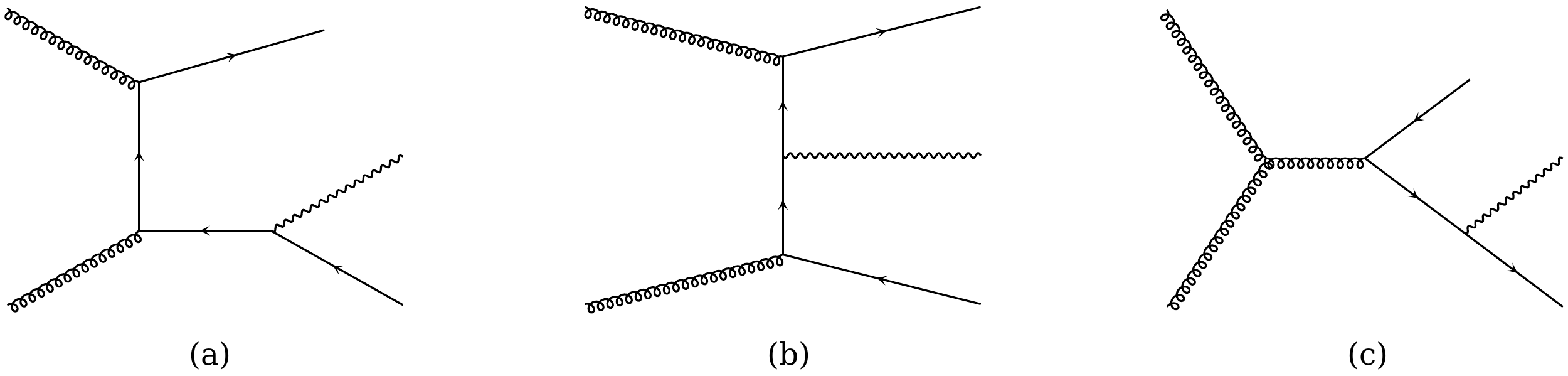}
    \caption{Diagrams for the $gg\to Z q\bar{q}$ contribution to $Z$+jet production (diagrams where the direction of the quark line is reversed are omitted). Our factorization formula only includes the singular contributions in the expansion of diagrams (a) and (b) around the back-to-back limit of the boson and jet. There are power corrections at $\mathcal{O}(\alpha_s(\delta\phi)^0)$ from expanding diagrams (a) and (c) for the emission of a $Z$-boson off dijets, which are included via matching.}
    \label{fig:ggDiags}
\end{figure}

We start in \fig{nonsing-vsnlo} with a detailed comparison of our factorization formula in~\eq{resum}, expanded at NLO in \eq{sing_int}, with predictions from MCFM. As confirmed in the upper panels of this figure, the singular terms agree with the NLO for both $p_{T,J}>60$ GeV and $p_{T,J}>200$ GeV. The lower panels of this figure show the importance of the matching procedure at NNLL because of $\mathcal{O}(\alpha_s (\delta\phi)^0)$ power corrections, which are not contained in our factorization formula. More explicitly, our factorization in \eq{resum} only includes singular $1/(\delta \phi)$ terms from expanding around the back-to-back limit of the boson and jet, missing an important contribution to the power corrections that arises from a dijet configuration where a $Z$ boson is emitted from one of the jets. This contribution enters at the same order in the coupling and is enhanced for $p_{T,J} \ll m_Z$, as will be discussed below.
To illustrate the difference between the contributions in the factorization theorem and this important nonsingular correction, consider the diagrams for the $gg\to q \bar q Z$ process in \fig{ggDiags}: Diagrams (a) and (b) contribute to our factorization theorem in the region of phase-space where a final-state quark is collinear to one of the incoming gluons. However, diagram (a) and the new diagram (c) are also enhanced for the region of phase space describing a $Z$ emission of a dijet configuration. This contribution is included by the matching procedure in sec.~\ref{subsec:fixed} and discussed more below.

\begin{figure}[t]
    \centering
    \includegraphics[width=0.55\textwidth]{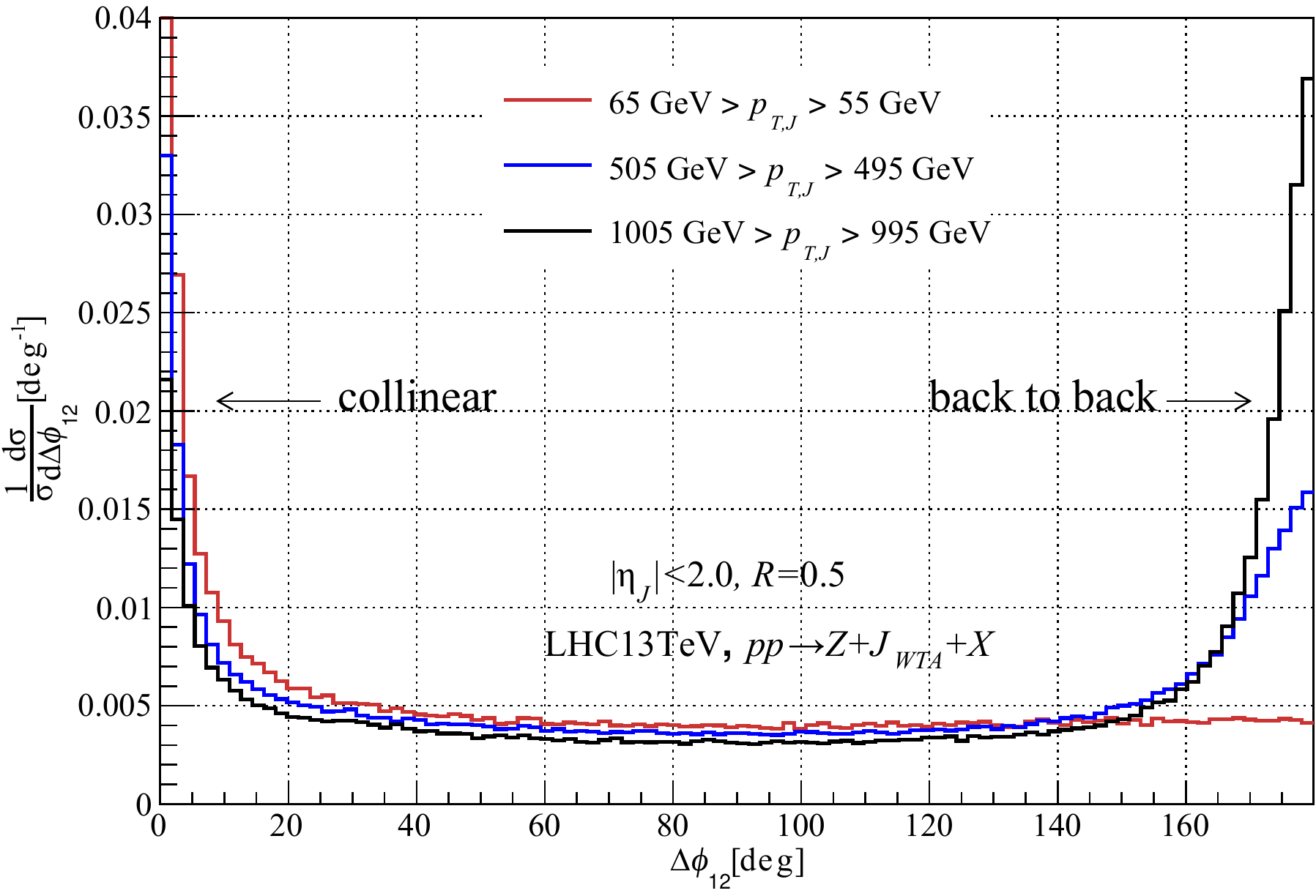}
    \caption{The distribution in the azimuthal angle $\Delta\phi_{12}$ between the two final-state partons at NLO. For $p_{T,J} \lesssim m_V$ there is a collinear singularity at $\Delta\phi_{12} \to 0$, while for $p_{T,J} \gg m_V$ there is also a collinear singularity at $\Delta\phi_{12} \to 180^\circ$ from a $Z$ boson emitted from a dijet partonic configuration.}
    \label{fig:phi_j1j2}
\end{figure}

For $p_{T, J}\ll m_V$, the order $\alpha_s(\delta\phi)^0$ power corrections from the $Z$ emission off dijets are power suppressed in $m_V$. This can be seen by expanding the diagrams in \fig{ggDiags} describing soft-collinear corrections around the back-to-back limit, yielding contributions proportional to $\frac{1}{m_V^2}\frac{1}{(\delta\phi)^2 p_{T,V}^2}$ at the amplitude level\footnote{Including numerator factors it only gives rise to a NLO term $\propto 1/\delta\phi$ in the cross section.}, while the $Z$ emission of dijets is proportional to $\frac{1}{m_V^4}$. Accordingly, at low $p_T$, the $Z$ emission off dijets is expected to be independent of the azimuthal angle based on the above expansion. Indeed, the non-singular in \fig{nonsing-vsnlo} is almost constant at small $\delta \phi$. This is also confirmed by looking at the azimuthal angle $\Delta\phi_{12}$ of the two final-state partons shown in \fig{phi_j1j2}. We find that for $p_{T,J}\lesssim m_V$ (e.g.~the  65~GeV$>p_{T,J}>55$~GeV bin) the $\Delta\phi_{12}$ distribution is flat near $\Delta\phi_{12}=180^\circ$. This flatness is the reason that we do not expect the nonsingular to go to zero at $\Delta \phi \to 180^\circ$, and don't employ a multiplicative matching that would enforce that. Indeed, collinear and soft emissions will smear the nonsingular $\delta \phi$ distribution, but since it is (almost) constant it will not change (much).

Figure~\ref{fig:phi_j1j2} also shows that at high $p_{T,J}$ there are large contributions from the $Z$ emission off back-to-back \emph{dijets}, in addition to soft-collinear QCD radiation for back-to-back boson-jet production. The latter is the focus of this paper, incorporated in the resummation formula \eq{resum}, while the former is included by matching and needed at NNLL and beyond. When $p_{T,J} \gg m_V$, it is insufficient to include this contribution by matching in the region $m_V/p_{T,J} \lesssim \delta \phi \ll 1$, as it now also contains large logarithms of $\delta \phi$ that require resummation.  Alternatively, one can also remove the contribution of a $Z$ boson emitted from a dijet by introducing an isolation cone around the boson (or its decay products).

\begin{figure}[t]
    \centering
    \includegraphics[height=0.4\linewidth]{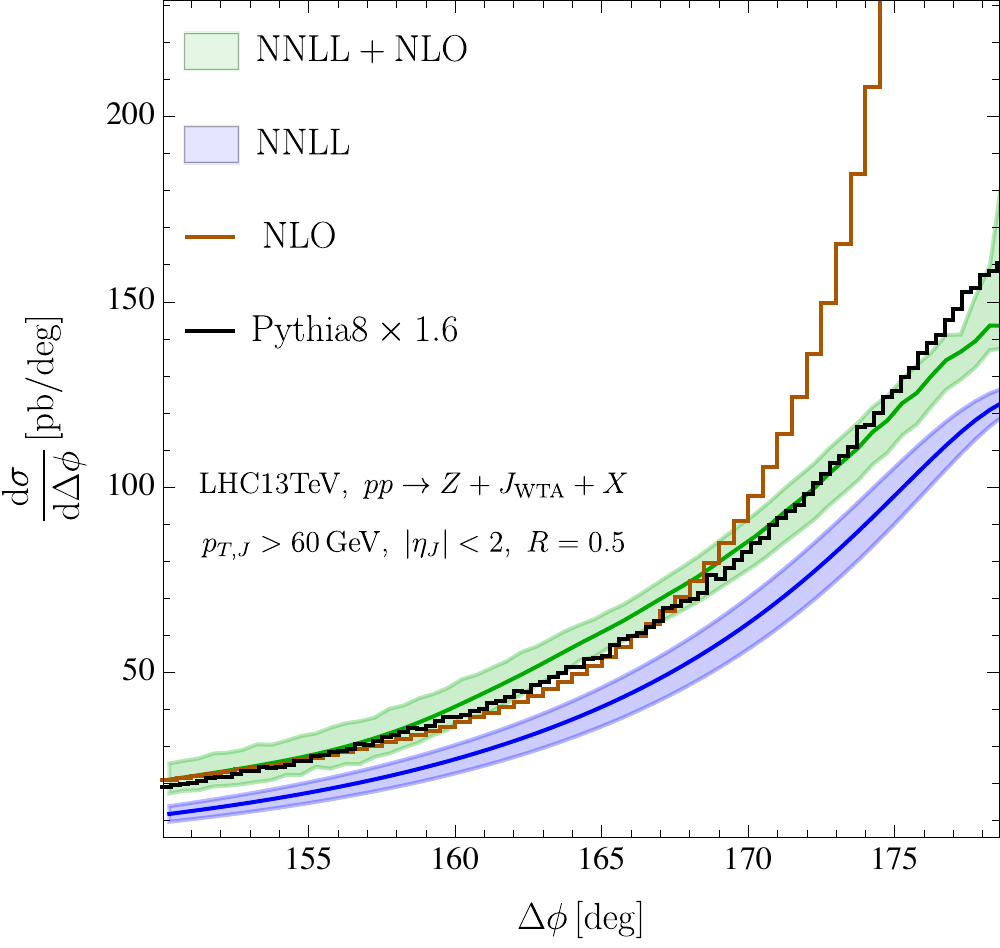}~~
    \includegraphics[height=0.4\linewidth]{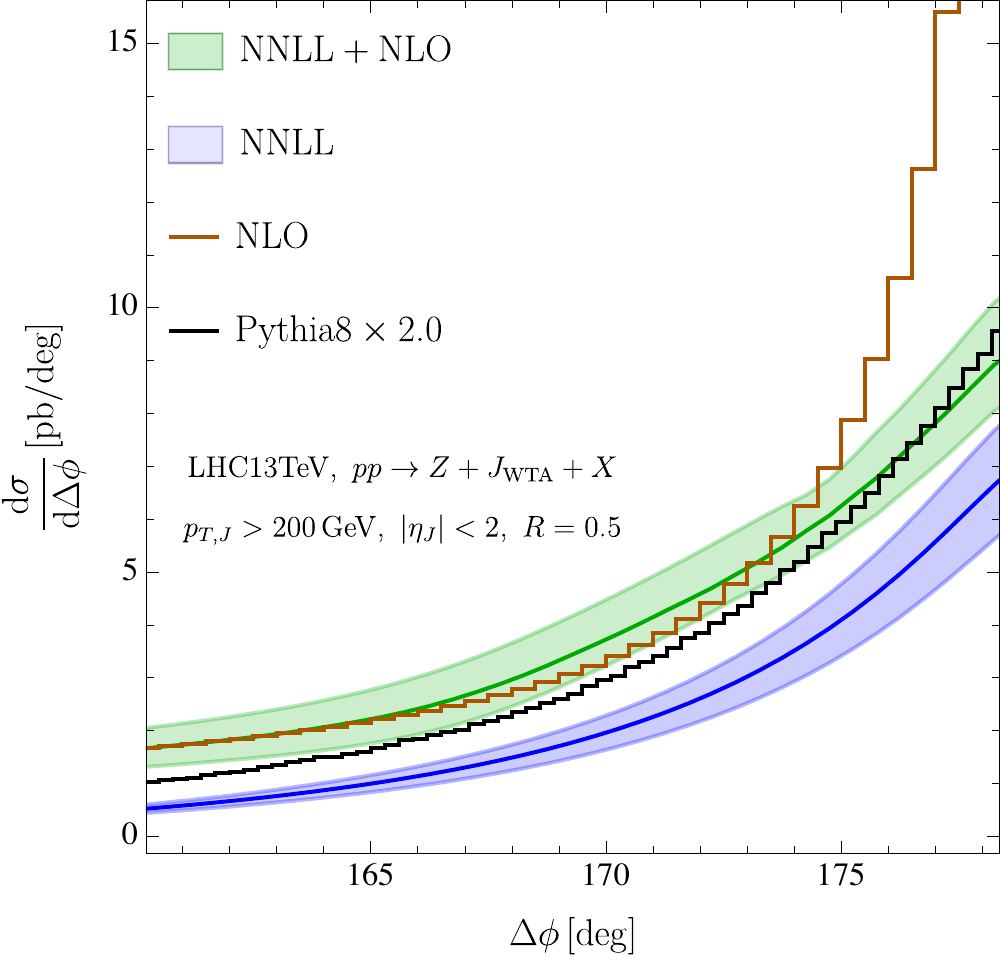}
    \caption{Our resummed predictions (blue) matched to NLO (green) for the $Z$+jet azimuthal decorrelation $\Delta \phi$ for $p_{T,J}>60$ GeV (left) and $p_{T,J}>200$ GeV (right). Our predictions are compared to \Pythia simulations at the hadron level with MPI contributions (black) and the NLO (red). The band represents the perturbative uncertainty, which is estimated by scale variation (see sec.~\ref{subsec:RGE}). The larger nonsingular corrections (from the matching) are discussed in the text.}
    \label{fig:MCcalc}
\end{figure}

We conclude by showing in \fig{MCcalc} our resummed predictions at NNLL+NLO accuracy, and comparing to \Pythia results at hadron level with MPI contributions for the two $p_{T,J}$ regions. We show the resummed predictions with and without matching, in view of the large power corrections we just discussed, and also include the NLO cross section as a separate curve. The NLO cross section divergences as $\Delta \phi \to 180^\circ$, which is remedied by the resummation. Our resummed results without matching agree with the shape of \Pythia simulations. By including the matching, our predictions smoothly approach NLO for smaller values of $\Delta \phi$, where resummation is not needed. The large nonsingular correction (larger than our uncertainty bands) is important to include and cannot be neglected as $\Delta \phi \to 180^\circ$. It is not accounted for in the \Pythia simulations we used. Simply attempting to include it through a $K$-factor does not yield the correct shape, particularly at high $p_{T,J}$.

\section{Conclusions}
\label{sec:conc}

In this paper we describe our calculation of the cross section for a vector boson and jet, differential in the azimuthal decorrelation $\de \phi$, at next-to-next-to-leading logarithmic order. We provide substantially more details than our earlier work~\cite{Chien:2020hzh}, an expanded phenomenological analysis, and also include a \Pythia study. We discuss for the first time \emph{why} the azimuthal angle is simpler than the total transverse momentum $\vec q_T$,  potential Glauber contributions, different recoil-free recombination schemes, the jet radius dependence, and the non-singular matching. 

Our focus is on the region $\de \phi \ll 1$, which dominates the cross section and requires the resummation of logarithms of $\de \phi$ to obtain reliable predictions. We carried out a detailed study of its factorization in SCET, deriving a factorization formula. We investigated potential factorization-violating Glauber contributions, finding that they could first appear at order $\alpha_s^4$. Many of the ingredients in the factorization are available, and we present calculations of the jet functions, including the linearly polarized jet function, as well as jet functions for $p_T^n$-weighted recombination schemes. We find that these different recombination schemes only change the constant term in the jet function, having a minimal effect on the prediction. Furthermore, we verify the independence on the jet radius predicted by the factorization (for $R \gg \delta \phi$) in \Pythia. By using the (rapidity) renormalization group we achieve the resummation of logarithms of $\delta \phi$. 

The key to obtaining predictions beyond NLL is a recoil-free recombination scheme, which reduces the effect of soft radiation to a total recoil of the $V+$jet system, eliminating non-global logarithms. This is in contrast to the standard recombination scheme, or other transverse momentum measurements such as the radial decorrelation, which involve NGLs. 
A recoil-free recombination scheme is also interesting for studying the properties of the medium in heavy-ion collisions, because it is more sensitive to collinear splittings inside the jet and less sensitive to contamination of soft radiation. 
Indeed, we observe negligible effects of hadronization and MPI contributions in \Pythia, as well as compatibility with track-based measurements.
Our resummed predictions are matched to \MCFM in the region where $\de \phi$ is no longer small. Here we find that the non-singular contributions are quite large, particularly for high jet $p_T$, increasing the sensitivity to the details of the matching procedure. It would be interesting to investigate the resummation of these power corrections.

We have established the azimuthal angle between a vector boson and a recoil-free jet, as a robust observable for which high precision is possible. In this paper we achieve NNLL order, but NNNLL is within reach. As the effect of the underlying event, hadronization and from performing the measurement using charged particle tracks is minimal, it is attractive experimentally.

\acknowledgments

We thank Iain Stewart for discussions on Glauber contributions.
R.R.~is supported by the NWO projectruimte 680-91-122. D.Y.S.~is supported by the Shanghai Natural Science Foundation under Grant No.~21ZR1406100. W.W.~is supported by the D-ITP consortium, a program of NWO that is funded by the Dutch Ministry of Education, Culture and Science (OCW). B.W.~is supported by the European Research Council project ERC-2018-ADG-835105 YoctoLHC; by the Maria de Maetzu excellence program under project CEX2020-001035-M; by the Spanish Research State Agency under project PID2020-119632GB-I00; and by Xunta de Galicia (Centro singular de investigaci\'on de Galicia accreditation 2019-2022), by the European Union ERDF.

\newpage
\appendix
\section{Anomalous dimensions}
\label{app:anomdim}
In general, for the (next-to)$^i$
-leading logarithmic (N${}^i$LL) resummation, one needs up to $(i-1)$-loop
fixed-order ingredients, $i$-loop non-cusp anomalous dimensions and
the $(i+1)$-loop cusp anomalous dimension and QCD beta function. In this
appendix, we collect the beta function and all the anomalous dimensions that are needed for this paper.

For the beta function, 
\begin{align}
\frac{\df \alpha_s(\mu)}{\df \ln\mu}=-2 \varepsilon \alpha_s+\beta(\alpha_s),~~~\beta(\alpha_s) = -2\alpha_s \sum_{n=0}^{\infty}\beta_n \left( \frac{\alpha_s}{4\pi} \right)^{n+1}
\end{align}
one has up to three loops~\cite{Tarasov:1980au,Larin:1993tp}
\begin{align}
  \beta_0 &= \frac{11}{3} C_A - \frac{4}{3} T_F n_f \, ,
  \nonumber\\
  \beta_1 &= \frac{34}{3} C_A^2 - \frac{20}{3} C_A T_F n_f - 4 C_F T_F n_f \, ,
  \nonumber\\
  \beta_2 &= \frac{2857}{54} C_A^3 + \left( 2 C_F^2 - \frac{205}{9} C_F C_A -
    \frac{1415}{27} C_A^2 \right) T_F n_f + \left( \frac{44}{9} C_F + \frac{158}{27} C_A
  \right) T_F^2 n_f^2 \, , 
\end{align}
where $n_f$ is the number of active quark flavors.

\smallskip

Identifying standard anomalous dimensions by their associated functions, and
using the universality of the rapidity anomalous dimension, we write the
perturbative expansions of cusp, non-cusp, and rapidity anomalous dimensions as
\begin{equation}
  \Gamma_{\rm
  cusp}=\sum\limits_{n=0}^\infty\left(\frac{\alpha_s}{4\pi}\right)^{n+1}\Gamma_{n}
  \,,\quad
  \gamma_{\mu}^i=\sum\limits_{n=0}^\infty\left(\frac{\alpha_s}{4\pi}\right)^{n+1}\gamma^i_n
  \, , \quad
  \gamma_{\nu}=\sum\limits_{n=0}^\infty\left(\frac{\alpha_s}{4\pi}\right)^{n+1}\gamma^{\nu}_n.
  \end{equation} 
 The cusp anomalous dimension is, up to three loops, given
 by~\cite{Korchemsky:1987wg,Moch:2004pa}
\begin{align}
   \Gamma_0 &= 4 \,, \nonumber\\
   \Gamma_1 &= \left( \frac{268}{9} 
    - \frac{4\pi^2}{3} \right) C_A - \frac{80}{9}\,T_F n_f \,,
    \nonumber\\
   \Gamma_2 &= C_A^2 \left( \frac{490}{3} 
    - \frac{536\pi^2}{27}
    + \frac{44\pi^4}{45} + \frac{88}{3}\,\zeta_3 \right) 
    + C_A T_F n_f  \left( - \frac{1672}{27} + \frac{160\pi^2}{27}
    - \frac{224}{3}\,\zeta_3 \right) \nonumber\\
   & \quad+ C_F T_F n_f \left( - \frac{220}{3} + 64\zeta_3 \right) 
    - \frac{64}{27}\,T_F^2 n_f^2 \,.
\end{align}
The non-cusp anomalous dimension for the hard function can be
extracted~\cite{Idilbi:2006dg,Becher:2006mr,Becher:2009qa} from the massless
quark and gluon form factor, which are known at three loop
order~\cite{Moch:2005tm}. They are given by
\begin{align}
   \gamma_0^q &= -3 C_F \,,\nn \\
   \gamma_1^q &= C_F^2 \left( -\frac{3}{2} + 2\pi^2
    - 24\zeta_3 \right)
    + C_F C_A \left( - \frac{961}{54} - \frac{11\pi^2}{6} 
    + 26\zeta_3 \right)
    + C_F T_F n_f \left( \frac{130}{27} + \frac{2\pi^2}{3} \right),
    \nn \\
    \gamma_0^g &= - \beta_0 
    = - \frac{11}{3}\,C_A + \frac{4}{3}\,T_F n_f \,, \nonumber\\
   \gamma_1^g &= C_A^2 \left( -\frac{692}{27} + \frac{11\pi^2}{18}
    + 2\zeta_3 \right) 
    + C_A T_F n_f \left( \frac{256}{27} - \frac{2\pi^2}{9} \right)
    + 4 C_F T_F n_f \,,
\end{align}
for gluons. The non-cusp anomalous dimensions for the beam and soft functions are given by~\cite{Luebbert:2016itl} 
\begin{align} 
\gamma_0^{B_q}
&= 6C_F
\,, \nn \\
   \gamma_1^{B_q} &= C_F^2 \left( 3 - 4 \pi^2 + 48 \zeta_3 \right)
    + C_F C_A \left( \frac{17}{3} + \frac{44\pi^2}{9} 
    -24\zeta_3 \right)
    + C_F T_F n_f \left( -\frac{4}{3} - \frac{16\pi^2}{9} \right)\,,\nn \\
\gamma_0^{B_g}
&= 2\beta_0
\,, \nn \\
   \gamma_1^{B_g} &= C_A^2 \left( \frac{64}{3} + 24 \zeta_3 \right)
    - \frac{32}{3} C_A T_F n_f - 8 C_F T_F n_f
\,, \end{align}
\begin{align} 
\gamma_0^{S}
&= 0\,, \qquad
\gamma_1^{S}
=  C_A \left(\frac{64}{9} -28\,\zeta_3\right) + \beta_0 \left(\frac{56}{9} - \frac{\pi^2}{3}\right)
\,.\end{align}
Finally, the non-cusp rapidity anomalous dimension is given
by~\cite{Gehrmann:2012ze,Gehrmann:2014yya,Echevarria:2015byo,Luebbert:2016itl}
\begin{align}
      \gamma_0^{\nu} = 0\,, \qquad  \gamma_1^{\nu} = -C_A \left(\frac{128}{9} -56\,\zeta_3\right) - \beta_0 \frac{112}{9}
\,.\end{align}

The most convenient choice of scales in our problem is $\mu=\mu_J=\mu_S=\mu_B$
and $\nu = \mu_B$. With this choice, the simplest path to solve the RG
equations, \eq{RGE}, is shown in figure \ref{fig:pathinmunu}. With $\mu=\mu_B$
fixed, the beam and jet functions run from their natural rapidity scales $\nu_i$
down to $\nu=\mu_B$, the natural rapidity scale of the soft function. Only the
hard function is required to run in $\mu$ from $\mu_H$ to $\mu_B$. Generically,
the hard anomalous dimension takes the form
\begin{align}
\Gamma(\alpha_s)=C_\Gamma \Gamma_\text{cusp}(\alpha_s)\ln\frac{Q_\Gamma^2}{\mu^2}+\gamma (\alpha_s).
\end{align}
The corresponding ordinary RG running boils down to the evaluation of the following functions:
\begin{align} \label{eq:S_and_A}
&S ( \mu_H , \mu ) =  \int_{\mu_H}^\mu \frac{\df \bar \mu}{\bar \mu} \ln
\frac{\mu_{H}}{\bar \mu} \Gamma_{\rm cusp}(\alpha_s(\bar \mu)),\qquad A_{\gamma^i}(\mu_H,\mu) = -\int_{\mu_H}^\mu \frac{\df \bar \mu}{\bar \mu} \gamma^i(\alpha_s(\bar \mu)).
\end{align}
In terms of these two functions, one has
\begin{align}\label{eq:exponential}
e^{\int_{\mu_H}^\mu \frac{\df \bar \mu}{\bar \mu}\Gamma(\bar \mu)}=\left(\frac{Q_\Gamma^2}{\mu_H^2}\right)^{-C_\Gamma A_{\gamma_\text{cusp}}(\mu_H,\mu)}e^{2 C_\Gamma S(\mu_H, \mu)-2A_{\gamma}(\mu_H, \mu) }.
\end{align}
At NNLL, we only keep terms up to $O(\alpha_s)$, that is
\begin{align}
    A_{\gamma^{\text{i}}}(\mu_H,\mu) &= - \int_{\alpha_s(\mu_H)}^{\alpha_s(\mu)} \frac{\df \alpha}{\beta(\alpha)} \gamma^{\text{i}}(\alpha)\notag\\
    &=\frac{1}{2}\frac{\gamma_{0}^\text{i}}{\beta_0}\int_{\alpha_s(\mu_H)}^{\alpha_s(\mu)}\frac{\df \alpha}{\alpha}\frac{1+\sum\limits_{l=1}^\infty\left(\frac{\alpha}{4\pi}\right)^{l}\frac{\gamma_{l}^\text{i}}{\gamma_{l}^\text{0}}}{1+\sum\limits_{l=1}^\infty\left(\frac{\alpha}{4\pi}\right)^{l}\frac{\beta_{l}}{\beta_0}}\notag\\
    &=\frac{1}{2}\frac{\gamma_{0}^\text{i}}{\beta_0}\left[\ln r + \frac{\alpha_s(\mu_H)}{4\pi}\Big(\frac{\beta_1}{\beta_0}-\frac{\gamma_1^i}{\gamma_0^i}\Big)(1-r)\right]
\end{align}
and
\begin{align}
S ( \mu_H , \mu ) =  -\int_{\alpha_s(\mu_H)}^{\alpha_s(\mu)}& \frac{\df \bar{\alpha}}{\beta(\bar{\alpha})}\Gamma_{\rm cusp}(\bar\alpha) \int^{\bar{\alpha}}_{\alpha_s(\mu_H)}\frac{\df \alpha}{\beta(\alpha)}\notag\\
=\frac{\Gamma_0}{4\beta _0^2}\bigg\{&\frac{4\pi}{\alpha_s(\mu_H)}\bigg(\frac{r-1}{r}-\log r\bigg)\notag\\
&+\Big(\frac{\beta_1}{\beta_0}-\frac{\Gamma_1}{\Gamma_0}\Big)(r-1-\log r)+\frac{\beta_1}{2\beta_0}\log^2 r \notag\\
&-\frac{\alpha_s(\mu_H)}{8\pi}\bigg[\frac{\Gamma_2}{\Gamma_0}(r-1)^2 +\frac{\beta_2}{\beta_0} (1-r^2+2 \log r)+\frac{\beta_1^2}{\beta_0^2} (r-1)\notag\\
&\times (r-1+2 \log r)-\frac{\beta_1}{\beta_0} \frac{\Gamma_1}{\Gamma_0} (r^2-4 r+2 r \log r+3)\bigg]\bigg\}
\end{align}
with
\begin{align}
    r\equiv \frac{\alpha_s(\mu)}{\alpha_s(\mu_H)}.
\end{align}

\section{One-loop hard function}
\label{app:oneloophard}

In this appendix we give the expressions of the one-loop hard function. Since we
do not need the one-loop hard function for a linearly-polarized gluon at NNLL accuracy, we only give the results for an unpolarized gluon here. After factorizing the LO hard function, the one-loop hard function has the form as
\begin{align}
    \mathcal{H}^{(1)}_{ij\to V k} = \mathcal{H}^{(0)}_{ij\to V k} C_{ij\to V k}(\hat t, \hat u).
\end{align}
For the $q\bar q \to Vg$ channel, we have 
\begin{align}
    C_{q\bar q\to V g}(t,u)&=  C_{A} \frac{\pi^{2}}{6}+C_{F}\left(-16+\frac{7 \pi^{2}}{3}\right)+2 C_{A} \ln ^{2} \frac{s}{m_{V}^{2}}+C_{A} \ln ^{2} \frac{m_{V}^{2}-t}{m_{V}^{2}} \notag \\
    & +C_{A} \ln ^{2} \frac{m_{V}^{2}-u}{m_{V}^{2}}+\ln \frac{s}{m_{V}^{2}}\left(-6 C_{F}-2 C_{A} \ln \frac{s^{2}}{t u}\right)-C_{A} \ln ^{2} \frac{t u}{m_{V}^{4}}-6 C_{F} \ln \frac{\mu^{2}}{s} \notag\\
&-2 C_{A} \ln \frac{s^{2}}{t u} \ln \frac{\mu^{2}}{s}+\left(-C_{A}-2 C_{F}\right) \ln ^{2} \frac{\mu^{2}}{s} \notag\\
&
+2 C_{A} \operatorname{Li}_{2}\left(\frac{m_{V}^{2}}{m_{V}^{2}-t}\right)+2 C_{A} \operatorname{Li}_{2}\left(\frac{m_{V}^{2}}{m_{V}^{2}-u}\right)
\notag\\
&+ \frac{2}{T_{0}(u, t)}\left\{C_{F}\left(\frac{s}{s+t}+\frac{s+t}{u}+\frac{s}{s+u}+\frac{s+u}{t}\right)\right. \notag\\
&+\left(-C_{A}+2 C_{F}\right)\left[-\frac{m_{V}^{2}\left(t^{2}+u^{2}\right)}{t u(t+u)}+2\left(\frac{s^{2}}{(t+u)^{2}}+\frac{2 s}{t+u}\right) \ln \frac{s}{m_{V}^{2}}\right] \notag\\
&+\left(C_{A} \frac{t}{s+u}+C_{F} \frac{4 s^{2}+2 s t+4 s u+t u}{(s+u)^{2}}\right) \ln \frac{-t}{m_{V}^{2}} \notag\\
&+\left(C_{A} \frac{u}{s+t}+C_{F} \frac{4 s^{2}+4 s t+2 s u+t u}{(s+t)^{2}}\right) \ln \frac{-u}{m_{V}^{2}} \notag\\
&-\left(-C_{A}+2 C_{F}\right)\left[\frac { s ^ { 2 } + ( s + u ) ^ { 2 } } { t u } \left(\frac{1}{2} \ln ^{2} \frac{s}{m_{V}^{2}}-\frac{1}{2} \ln ^{2} \frac{m_{V}^{2}-t}{m_{V}^{2}}+\ln \frac{s}{m_{V}^{2}} \ln \frac{-t}{s-m_{V}^{2}}\right.\right. \notag\\
&\left.+\operatorname{Li}_{2}\left(\frac{m_{V}^{2}}{s}\right)-\operatorname{Li}_{2}\left(\frac{m_{V}^{2}}{m_{V}^{2}-t}\right)\right) \notag\\
&+\frac{s^{2}+(s+t)^{2}}{t u}\left(\frac{1}{2} \ln ^{2} \frac{s}{m_{V}^{2}}-\frac{1}{2} \ln ^{2} \frac{m_{V}^{2}-u}{m_{V}^{2}}+\ln \frac{s}{m_{V}^{2}} \ln \frac{-u}{s-m_{V}^{2}}\right. \notag\\
&\left.\left.+\operatorname{Li}_{2}\left(\frac{m_{V}^{2}}{s}\right)-\operatorname{Li}_{2}\left(\frac{m_{V}^{2}}{m_{V}^{2}-u}\right)\right)\right],
\end{align}
and for the $qg\to Vq$ channel we have 
\begin{align}
    C_{qg\to Vq}(t, u)& = C_{A} \frac{7 \pi^{2}}{6}+C_{F}\left(-16+\frac{\pi^{2}}{3}\right)-6 C_{F} \ln \frac{s}{m_{V}^{2}}-C_{A} \ln ^{2} \frac{-s t}{m_{V}^{4}}+C_{A} \ln ^{2} \frac{m_{V}^{2}-t}{m_{V}^{2}} \notag \\
&+2 C_{A} \ln \frac{\left(s-m_{V}^{2}\right) t}{m_{V}^{2} u} \ln \frac{-u}{m_{V}^{2}}+C_{A} \ln ^{2} \frac{-u}{m_{V}^{2}}-2 C_{A} \ln \frac{\left(m_{V}^{2}-s\right) s t}{m_{V}^{2} u^{2}} \ln \frac{-u}{s} \notag\\
&-2 C_{A} \ln ^{2} \frac{-u}{s}-2 C_{F} \ln ^{2} \frac{-u}{s}+\left(-6 C_{F}+2 C_{A} \ln \frac{t}{u}+4 C_{F} \ln \frac{-u}{s}\right) \ln \frac{\mu^{2}}{s} \notag\\
&-\left(C_{A}+2 C_{F}\right) \ln ^{2} \frac{\mu^{2}}{s}-2 C_{A} \operatorname{Li}_{2}\left(\frac{m_{V}^{2}}{s}\right)+2 C_{A} \operatorname{Li}_{2}\left(\frac{m_{V}^{2}}{m_{V}^{2}-t}\right) \notag\\
&+ \frac{2}{T_{0}(s, t)}\left\{C_{F}\left(\frac{u}{s+u}+\frac{s+u}{t}+\frac{u}{t+u}+\frac{t+u}{s}\right)\right. \notag\\
&+\left(C_{A} \frac{s}{t+u}+C_{F} \frac{s t+2 s u+4 t u+4 u^{2}}{(t+u)^{2}}\right) \ln \frac{s}{m_{V}^{2}} \notag\\
&+\left(C_{A} \frac{t}{s+u}+C_{F} \frac{s t+4 s u+2 t u+4 u^{2}}{(s+u)^{2}}\right) \ln \frac{-t}{m_{V}^{2}} \notag\\
&+\left(-C_{A}+2 C_{F}\right)\left[-\frac{m_{V}^{2}\left(s^{2}+t^{2}\right)}{s t(s+t)}+2\left(\frac{2 u}{s+t}+\frac{u^{2}}{(s+t)^{2}}\right) \ln \frac{-u}{m_{V}^{2}}\right] \notag\\
&-\left(-C_{A}+2 C_{F}\right)\left[\frac { u ^ { 2 } + ( t + u ) ^ { 2 } } { s t } \left(\frac{1}{2} \ln ^{2} \frac{s}{m_{V}^{2}}-\frac{1}{2} \ln ^{2} \frac{m_{V}^{2}-u}{m_{V}^{2}}+\ln \frac{s}{m_{V}^{2}} \ln \frac{-u}{s-m_{V}^{2}}\right.\right. \notag\\
&\left.+\mathrm{Li}_{2}\left(\frac{m_{V}^{2}}{s}\right)-\mathrm{Li}_{2}\left(\frac{m_{V}^{2}}{m_{V}^{2}-u}\right)\right) \notag\\
&+\frac{u^{2}+(s+u)^{2}}{s t}\left(-\frac{\pi^{2}}{2}-\frac{1}{2} \ln ^{2} \frac{m_{V}^{2}-t}{m_{V}^{2}}-\frac{1}{2} \ln ^{2} \frac{m_{V}^{2}-u}{m_{V}^{2}}+\ln \frac{-t}{m_{V}^{2}} \ln \frac{-u}{m_{V}^{2}}\right. \notag\\
&\left.\left.\left.-\mathrm{Li}_{2}\left(\frac{m_{V}^{2}}{m_{V}^{2}-t}\right)-\operatorname{Li}_{2}\left(\frac{m_{V}^{2}}{m_{V}^{2}-u}\right)\right)\right] \right\},
\end{align}
with the function $T_0$ defined as
\begin{align}
    T_{0}(u, t)=\frac{u}{t}+\frac{t}{u}+\frac{2 m_{V}^{2}\left(m_{V}^{2}-t-u\right)}{t u}.
\end{align}

\bibliographystyle{JHEP}
\bibliography{jet}

\end{document}